\documentclass[12pt,preprint]{aastex}
\usepackage{amsmath}

\newcommand{\degree}{\ensuremath{^\circ}}

\begin{document}

\title {Periodic and Aperiodic Variability in the Molecular Cloud $\rho$ Ophiuchus}

\author{J. Robert Parks\altaffilmark{1,2}, Peter Plavchan\altaffilmark{1}, Russel J. White\altaffilmark{2}, $\&$ Alan H. Gee\altaffilmark{1}}
\altaffiltext{1}{Infrared Processing and Analysis Center, California Institute of Technology, Contact: parksj@chara.gsu.edu}
\altaffiltext{2}{Georgia State University, Department of Physics and Astronomy}

\begin{abstract}
  Presented are the results of a near-IR photometric survey on 1678 stars in the direction of the $\rho$ Ophiuchus ($\rho$ Oph) star forming region using data from the 2MASS Calibration Database.  For each target in this sample, up to 1584 individual \emph{J}, \emph{H} and \emph{K$_{s}$} band photometric measurements with a cadence of $\sim$1 day are obtained over 3 observing seasons spanning $\sim$2.5 years; it is the most intensive survey of stars in this region to date.  This survey identifies 101 variable stars with $\Delta$\emph{K$_{s}$} band amplitudes from 0.044 to 2.31 mag and $\Delta$(\emph{J}-\emph{K$_{s}$}) color amplitudes ranging from 0.053 to 1.47 mag.  Of the 72 young $\rho$ Oph star cluster members included in this survey, 79$\%$ are variable; in addition, 22 variable stars are identified as candidate members.  Based on the temporal behavior of the $\emph{K$_{s}$}$ time-series, the variability is distinquished as either periodic, long time-scale or irregular.  This temporal behavior coupled with the behavior of stellar colors is used to assign a dominent variability mechanism.  A new period-searching algorithm finds periodic signals in 32 variable stars with periods between 0.49 to 92 days. The chief mechanism driving periodic variability for 18 stars is rotational modulation of cool starspots while 3 periodically vary due to accretion-induced hot spots.  The time-series for 6 variable stars contain discrete periodic ``eclipse-like'' features with periods ranging from 3 to 8 days..  These features may be asymmetries in the circumstellar disk, potentially sustained or driven by a proto-planet at or near the co-rotation radius.  Aperiodic, long time-scale variations in stellar flux are identified in the time-series for 31 variable stars with time-scales ranging from 64 to 790 days.  The chief mechanism driving long time-scale variability are variable extinction or mass accretion rates.  The majority of the variable stars (40) exhibit sporadic, aperiodic variability over no discernable time-scale.  No chief variability mechanism could be identified for these variable stars.
\end{abstract}


\section{INTRODUCTION}
Photometric variability in young stars is an ubiquitous phenomenon.  Physical mechanisms causing this variability include, but are not limited to, rotational modulation of starspots, evolution of the circumstellar environment, interstellar extinction, transit events and stellar pulsation.  Large sample variability studies such as by \citet{herbst94} indicate these mechanisms often operate concurrently resulting in very complex photometric time-series in T Tauri stars.  Interpretations are primarily limited because broad band photometry acquired by seeing-limited telescopes are unable to spatially resolve the inner regions around young stars, which are on the order of milliarcseconds for low mass stars in nearby star-forming regions.  

High cadence, long temporal baseline photometric surveys, however, can temporally resolve the variations and help identify the dominant mechanism responsible for the variability.  This type of survey work has the prospect of identifying peculiar observational phenomenon that may hold clues to unsolved problems in star and planet formation.  One example is the physical interpretation fo AA Tau-like variability \citep{bouvier03}.  This variability presents as discrete drops in flux from a nearly constant ``continuum'' flux level.  The phenomenon is not unqiue to this star; \citet{morales11} identified 38 stars exhibiting similar variability during an intensive photometric monitoring campaign of young stellar objects (YSOs) in the Orion Nebula Cluster.  They find these dips have durations of 1 to a few days decreasing the flux by several tenths of magnitude in the IRAC bands and as much as 1.5 mag in the \emph{J} band.  These dips either appeared only once in 40 days (the survey's temporal baseline) or were periodic with periods between 2 to 14 days.  The current physical interpretation for these observations is periodic obscuration of the star by a high latitude ``warp'' in the circumstelar disk \citep{bertout00,bouvier03}.  Longer temporal baseline photometric monitoring of additional YSOs will place better constraints on the periodic nature of these events which in turn will confirm/constrain the current physical interpretation. 

Intensive photometric monitoring is most useful when it is conducted at a range of wavelengths.  Near-IR observations probe the very inner regions from $\sim$0.01 to 1 AU for low mass stars \citep{dullemond10}.  As young stars are typically rapidly rotating, this region includes both the co-rotation radius and dust sublimation radius.  For a median disk lifetime of 5 Myr, this corresponds to millions of dynamical times at these orbital radii \citep{haisch01}.  Therefore any indication of assymmetries in the circumstellar disk must be driven by some phenomenon (i.e. inclined magnetic dipole, planetary formation). Understanding the physics in this region is of particular interest to protoplanetary formation and migration, as well as how mass accretes onto the star.    

$\rho$ Ophiuchus ($\rho$ Oph) makes an excellent laboratory to test the ability of high cadence, long temporal near-IR observations to distinquish between variability mechanisms in young stars.  $\rho$ Oph is a dense star-forming region containing a few hundred known young stellar objects (YSOs) with ages ranging from 0.3 to 3 Myr.  The region is rich with variable stars; previous surveys having identified more than 100 photometrically variable stars \citep{greene92,barsony97,barsony05,bontemps01,wilking05,oliveira08}.  Photometric surveys are limited to the near- to far-IR due to large amounts of visual extinction ranging from 5 to 25 A$_{V}$ in the cloud core \citep{cambresy99}.  This complex interstellar environment could itself be responsible for detected photometric variability.  

\citet[hereafter P08]{plavchan08a} carried out a pilot study of 57 stars in the $\rho$ Oph field using photometry collected by the Two Micron All-Sky Survey Calibration Point Source Working Database (2MASS Cal-PSWDB).  That study identified periodic variability in two YSOs given a sample of candidate M stars.  The study presented here expands on the initial pilot study performed in P08 and will include the full $\rho$ Oph field data set from the 2MASS Cal-PSWDB to better understand the variability of young stars in this cloud.

In $\S$2, the details concerning the observations and source selection for this survey are discussed.  The variability analysis, along with a discussion on advantages to high cadence observations, is presented in $\S$3.  The methods used to find both periodic and long time-scale variability are found in $\S$4.  In $\S$5, the characteristics of the variability catalog, as well as a discussion of variability mechanisms is presented.  This section also discusses 6 stars where two variability mechanisms are estimated to operate concurrently.  Finally $\S$6 contains a summary of the findings reported by this work.

\section{OBSERVATIONS AND SAMPLE SELECTION}

\subsection{Observations}
The Two Micron All-Sky Survey \citep[2MASS]{skrutskie06} imaged nearly the entire sky via simultaneous drift scanning in three near-infrared bands (\emph{J}\emph{H}\emph{K$_{s}$}) between 1997 and 2001.  Observations were taken at the northern Mt. Hopkins Observatory and the southern CTIO facility.  Photometric calibration for 2MASS required hourly observations of 35 calibration fields split evenly between the northern and southern hemispheres.  Each calibration field is 1$\degree$ in length and 8.5$\arcmin$ wide.  One calibration field lies in the direction of the Ophiuchus constellation.  This field is centered at $\alpha$ = 16$^{h}$27$^{m}$15.6$^{s}$ and $\delta$ = -24$\degree$41$\arcmin$23$\arcsec$ (J2000) and covers part of the $\rho$ Oph L1688 cloud core \citep{bok56}.  These data have an observing cadence of $\sim$1 epoch per day.  A complete observation is comprised of six consecutive 1.3 second scans in declination with a nearly constant right ascension.  Each scan is offset by 5$\arcsec$ in right ascension to minimize errors from pixel effects.  The six scans, or "scan group", are finally co-added to minimize short time-scale and systematic variations.  A complete scan group is obtained in approximately 8 minutes \citep[$\S$III.2b]{cutri06}.  The maximum number of scans for a single star is 1584 divided by 6 or 264 scan groups.

Photometry is extracted from the calibration field via the 2MASS Point Source Catalog (2MASS PSC) automated processing system.  Details of the system's implementation are described in \citet{cutri06}; here a brief summary is given.  Photometry for sources fainter than \emph{J} = 9, \emph{H} = 8.5 and \emph{K$_{s}$} = 8 mag, are extracted by profile-fitting.  Profile-fitting compares the source flux to a pre-generated point spread function (PSF) via $\chi$$^{2}$ minimization. The PSFs are selected from a look-up table with respect to a dimensionless seeing index that is updated regularly during each scan.  The seeing index characterizes the atmospheric seeing during specific observations.  The library of PSFs is generated by empirically fitting the 50 brightest stars in a single 2MASS calibration scan with a specific average seeing index.  This scan is not necessarily of the $\rho$ Oph field, but a calibration field containing a different slice of the sky.  An error at the few-percent level may be present in the resulting photometry due to mismatched PSFs arising from rapid seeing variations.  

For the few sources brighter than the above cut off magnitudes, photometry are extracted using a 4$\arcsec$ fixed aperture corrected using a curve-of-growth.  Atmospheric seeing conditions can place as much as 15$\%$ of the flux from a point source outside this fixed aperture.  A curve-of-growth correction is a constant factor added to measured photometry to simulate measurements taken using an ``infinite'' aperture.  The benefit of this method is avoiding decreased signal-to-noise and potential source confusion arising from large aperture photometry. However, curve-of-growth corrections assume the sources are unresolved single stars that can be approximated by a PSF.  Therefore photometry for extended sources (i.e. stars embedded in bright nebular emission) or multiple systems are not properly characterized with this method.  All the data scans are compiled in the 2MASS Cal-PSWDB.
  
\subsection{Source Identification}
The source selection in the $\rho$ Oph field is similar to that described in P08, which is summarized here.  A \emph{parent sample} catalog of 7815 sources is constructed from a co-added deep image of the field \citep{cutri06}.  For each target in the parent sample, the 2MASS Cal-PSWDB is searched for detections within a 2$\arcsec$ matching radius.  This radius is several $\sigma$ larger than the 2MASS Point Source Catalog (PSC) astrometric precision and astrometric bias between the PSWDB and PSC \citep{zacharias05,skrutskie06}.  This ensures confidence all Cal-PSWDB detections for the parent 7815 sources are found within the PSC astrometric precision.

Of the 7815 stars identified in the parent sample, 1678 stars have a sufficient number of detections for variability and periodic analysis.  This sample of 1678 stars is henceforth referred to as the \emph{target sample}.  A 'sufficient number' is defined as stars detected in $\geq$10$\%$ of the observations in either \emph{J}, \emph{H} or \emph{K$_{s}$} and $\geq$ 50 detections in the \emph{J} band.  The first constraint ensures a sufficient number of data points for a robust periodogram computation.  The 10$\%$ limit is an \emph{ad hoc} limit chosen to reduce the noise present in the variability statistics.  The second constraint removes sources near the FOV edges that are not present in most scans.  

Finally, despite the success of the 2MASS prescription to produce high-quality photometric measurements, occasionally photometry affected by latent image artifacts, spurious detections and poor quality detections still persist in the database.  The reader is referred to P08 for a full treatment on how sources with poor photometry are characterized and excluded.  \citet{cutri06} describes the different varieties of latent image artifacts arising from a number of phenomenon associated with the optical system.  These artifacts are identified and removed via visual inspection.  Multiple simultaneous detections found within the 2$\arcsec$ search radius of a target, which are typically spurious byproducts of the source extraction pipeline, are eliminated.  Simultaneous detections are when two (or more) detected sources are identified with a single source in the 2MASS Cal-PSWDB.  Secondary detections are typically $\sim$0.5 - 1.5 magnitudes fainter than the primary detection.  In addition, they are typically detected in only one passband and only in one of the six scans.  Unaccounted for spurious detections can give the appearance of variability and introduce systematic noise into any underlying periodic signals.  Photometric measurements with poor spatial fits to the model PSF are also excluded from our analysis.  A poor spatial fit occurs when the $\chi$$^{2}$ value between the observed stellar profile and a model PSF is $>$ 10.  This is flagged as 'E' quality photometry within the Cal-PSWDB.  Image saturation, cosmic rays, hot pixels, extended emission or partially resolved doubles could account for this poor quality fit to the photometry \citep{cutri06}.  Photometry with poor spatial fits are systematically brighter by a few tenths of a magnitude, and this can falsely trigger the identification of variability.

Table 1, available only on-line, lists the 1678 stars analyzed for variability.  The magnitudes and errors listed are extracted from a co-added image of all calibration scans in that particular band.

  \renewcommand{\thetable}{\arabic{table}} 
  \setcounter{table}{0} 
\begin{deluxetable}{c c c c c c c c}
    \tablecolumns{8}
    \tablewidth{0pt}
    \tablecaption{Catalog of 2MASS stars in $\rho$ Oph}
    \tabletypesize{\scriptsize}
    \tablehead{
      \colhead{RA} & \colhead{Dec} & \colhead{N$_{\emph{J}obs}$} & \colhead{N$_{\emph{H}obs}$} & \colhead{N$_{\emph{K}obs}$} &\colhead{$\emph{J}$\tablenotemark{a}} & \colhead{$\emph{H}$\tablenotemark{a}} & \colhead{$K_{s}$\tablenotemark{a}} \\
      \colhead{(degrees)} & \colhead{(degrees)} & \colhead{} & \colhead{} & \colhead{} & \colhead{(mag)} & \colhead{(mag)} & \colhead{(mag)}
    }
    \startdata
    246.730759&-25.067987&262&261&244&16.972$\pm$0.011&16.174$\pm$0.012&15.659$\pm$0.028\\
    246.731262&-25.062096&258&258&258&17.048$\pm$0.016&16.163$\pm$0.014&15.631$\pm$0.029\\
    246.731293&-24.955822&6&70&72&16.508$\pm$0.005&15.796$\pm$0.006&15.431$\pm$0.009\\
    246.731308&-25.015520&263&263&262&13.980$\pm$0.001&13.048$\pm$0.001&12.747$\pm$0.001\\
    246.731323&-24.472427&6&41&32&17.272$\pm$0.000&16.440$\pm$0.043&15.327$\pm$0.014\\
    246.731323&-25.130241&198&174&67&16.973$\pm$0.010&16.211$\pm$0.012&15.720$\pm$0.026\\
    246.731750&-25.146633&187&139&47&16.987$\pm$0.010&16.227$\pm$0.013&15.726$\pm$0.028\\
    246.731796&-24.863796&102&105&36&15.868$\pm$0.003&15.030$\pm$0.003&14.694$\pm$0.004\\
\enddata
    \tablenotetext{a}{Unweighted mean apparent magnitude of Cal-PSWDB photometry}
  \end{deluxetable}

\subsubsection{Detection and Completeness Limits}
For non-variable stars, the photometric measurement uncertainty is characterized by the standard deviation of all photometric measurements in a particular band.  P08 showed that this photometric standard deviation as a function of apparent magnitude, for 2MASS photometry, follows the form of two distinct power laws.  One power law describes brighter sources, where Poisson statistics dominate the uncertainty, while the second describes the dimmer sources, where the uncertainty is dominated by instrumental noise.  The point of intersection between these two power laws, or ``break point'', designates the survey completeness limit where source detection drops below 100$\%$.  This power law model is used to predict the photometric scatter for a star, and any star that has a dispersion significantly ($>$ 5$\sigma$) above this is identified as a candidate variable. The model, as a function of apparent magnitude $\emph{m}$, is given by the following expression:

\begin{equation}
  10^{[\sigma_{m,model} \pm \nu_{m,model}(m)]} = b_{m,l} \pm \sigma_{b_{m,l}} + (a_{m,l} \pm \sigma_{a_{m,l}})10^{0.4m}
\end{equation}

\noindent where $\emph{a$_{m,l}$}$, $\emph{b$_{m,l}$}$, $\sigma$$_{a,m,l}$, and $\sigma$$_{b,m,l}$ represent the slope, intercept and respective errors for each fit in each band over magnitude region $\emph{l}$.  This model is first applied to our sample of 1678 stars using coefficients derived by P08 from the entire 2MASS Calibration Field data set.  These coefficients, however, yield a relatively poor fit to the $\rho$ Oph calibration field.  The lower noise in the $\rho$ Oph data is attributed to better average seeing conditions during these observations.  As a result, the model is re-fit on the $\rho$ Oph data set alone to derive a new set of coefficients.  The new coefficients with errors are listed in Table 2.  Fig 1 shows the best-fit model along with the observed photometric scatter in each band.

\begin{deluxetable}{llrr}
  \tablecolumns{4}
  \tablewidth{0pc}
  \tablecaption{Model Fit Parameters for Observed Photometric Scatter\tablenotemark{a}}
  \tablehead{
    \colhead{Band}  &
    \colhead{Range\tablenotemark{b}} & \colhead{$a_{m,l}\pm\sigma_{a_{m,l}}$} & \colhead{$b_{m,l}\pm\sigma_{b_{m,l}}$}
  }
  \startdata
  \emph{J} & $<$16.63 & (3.046$\pm$0.043)$\times$10$^{-8}$ & 1.01326$\pm$0.00087 \\
  \emph{J} & $>$16.63 & (1.484$\pm$0.083)$\times$10$^{-8}$ & 1.08343$\pm$0.00046 \\
  \emph{H} & $<$15.75 & (6.467$\pm$0.055)$\times$10$^{-8}$ & 1.01444$\pm$0.00041 \\
  \emph{H} & $>$15.75 & (4.03$\pm$0.16)$\times$10$^{-8}$ & 1.0628$\pm$0.0059 \\
  \emph{K$_{s}$} & $<$15.10 & (1.2247$\pm$0.0094)$\times$10$^{-7}$ & 1.0134$\pm$0.0036 \\
  \emph{K$_{s}$} & $>$15.10 & (4.98$\pm$0.25)$\times$10$^{-8}$ & 1.0934$\pm$0.0042 \\
  \enddata
  \tablenotetext{a}{See ${\S}$2.2.1 for explanation of parameters.}
  \tablenotetext{b}{Range in apparent magnitude; apparent magnitudes of
    $<$ 6 are excluded from this model.}
\end{deluxetable}

\begin{figure}
  \plotone{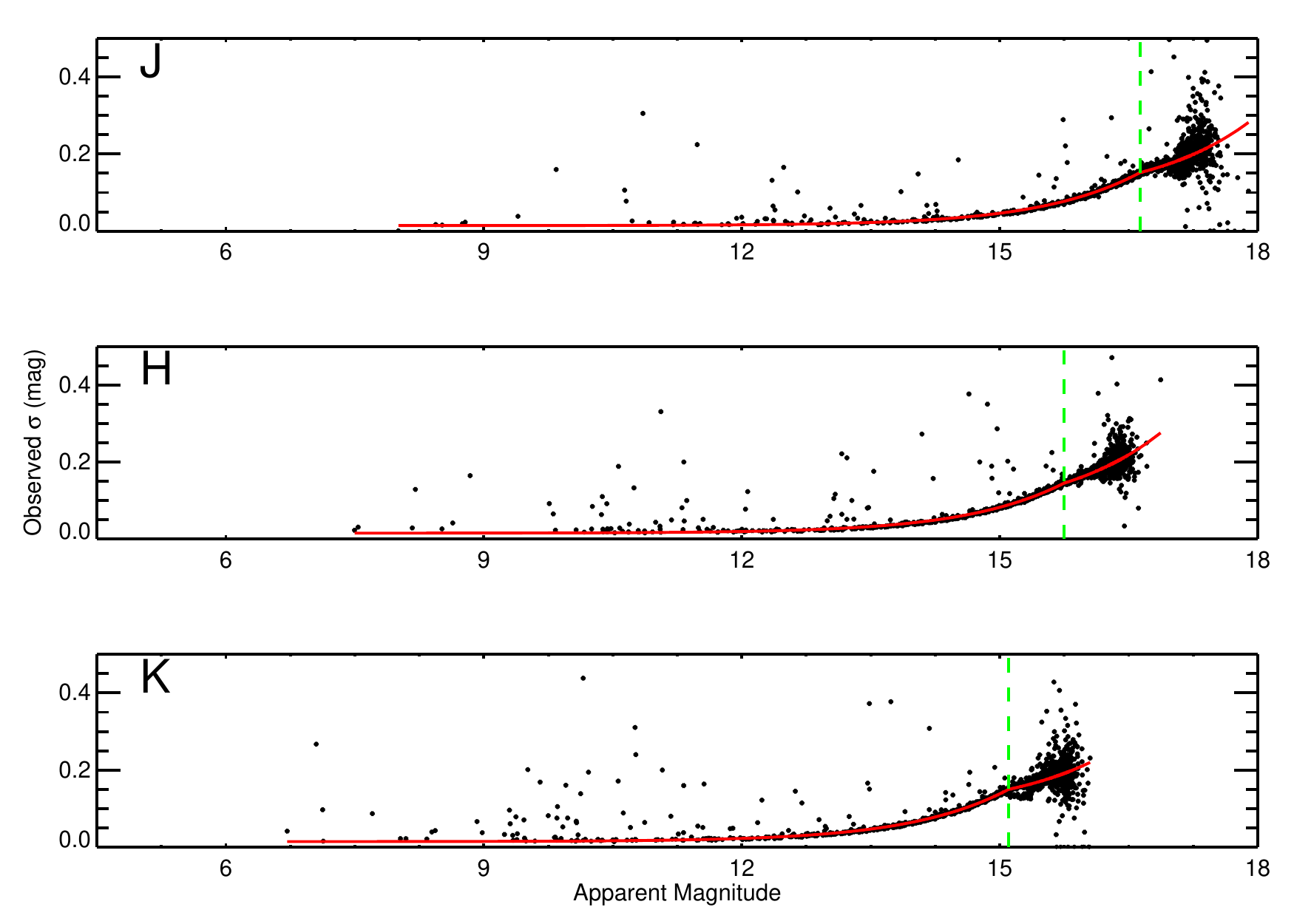}
  \caption{Photometric standard deviation versus apparent magnitude derived from up to 1584 observations of each sample star; there are 1678 sample stars in total.  The solid red line corresponds to the photometric model fit to this sample.  The dashed green line marks the break magnitude, where the detection rate drops below 100$\%$, in each band.  The break magnitudes are \emph{J} = 16.63, \emph{H} = 15.75, and K$_s$ = 15.10 mag.}
\end{figure}

The model yields completeness limits for this survey of 16.63, 15.75 and 15.10 mag in \emph{J}, \emph{H} and \emph{K$_{s}$}, respectively.  These are significantly fainter limits than the 2MASS PSC as a whole, which are 15.8, 15.1 and 14.3 mag in \emph{J}, \emph{H} and \emph{K$_{s}$}, respectively.  The approximate detection limits for this study, found by averaging the apparent magnitudes for the 10 faintest objects meeting our detection criteria, are 17.7, 16.7 and 16.0 mag in \emph{J}, \emph{H} and \emph{K$_{s}$} respectively.
  
\section{VARIABILITY ANALYSIS}
  
\subsection{Selection Criteria for Variability}
Numerous surveys have used time-series analysis on multi-wavelength photometry to characterize young star variability \citep[and references therein]{mathieu97,carpenter01,carpenter02,grankin07,grankin08,morales11,findeisen13,wolk13}.  The methods for identifing stellar variability are nearly as numerous as the variability studies themselves.  These include, but are not limited to, the Stetson index, excess photometric dispersion, $\chi$$^{2}$ statistic, cross-correlation and Fourier analysis \citep{stetson96,carpenter01,barsony05,oliveira08}.

Variable stars are identified in this work through 3 complementary methods that are sensitive to different types of variability.  A full description of these techniques are presented in P08.  The same terminology used in P08 is adopted in this work.  Here a summary is presented along with specifics regarding this sample.  The first and second methods, ``flickering'' and ``excursive'', identify variability in each band individually.  The third method uses the Stetson index to identify correlated variability between bands.
  
\subsubsection{Flickering Variability}
Flickering variability describes when the star's photometric scatter significantly differs from the predicted scatter.  Flickering variability is sensitive to continuous variability, as consistent, substantial variations are needed to significantly increase the observed photometric dispersion above the expected non-variable value.  To identify flickering variables, an observed dispersion is calculated for all scan measurements of a star prior to conbining as a scan group.  This is then compared to the star's expected dispersion with associated uncertainty, $\sigma$, calculated using the noise model described in $\S$ 2.2.1 (Eqn 1 and Fig 1).  If the observed dispersion exceeds the expected dispersion by more than 5$\sigma$, the star is a candidate variable.  This search is done separately for each of the 3 bands (\emph{J},\emph{H},\emph{K$_{s}$}); a star can thus be flagged as a flickering variable in 1, 2 or all 3 bands.  Following this criterion, 17 stars flag in only a single band, 23 flag in two bands and 54 flag in all three bands.  If variability is intrinsic to the star, the expectation is the flickering will occur in more than one band.  Low signal-to-noise photometry (the number of observations, N$_{obs}$, is typically less than 500) might likely account for the 9 candidate variables that flicker in the K band only.  It might also account for the 11 candidate variables that flicker in both the \emph{H} and K bands.  However there is no obvious explanation for the 17 candidate variable stars that flicker only in the \emph{J} or \emph{H}, \emph{J} and \emph{H}, or \emph{J} and K bands.  The average dispersions for these variable stars in \emph{J}, \emph{H} and \emph{K$_{s}$} are 0.12 $\pm$ 0.46, 0.12 $\pm$ 0.43 and 0.11 $\pm$ 0.35 mag, respectively.  The listed errors are the standard deviation of the average dispersion.  These values represent the dispersion intrinsic to the source, or specifically the dispersion after the predicted non-variable measurement dispersion is subtracted in quadrature from the observed dispersion.

\subsubsection{Excursive Variability}
Excursive variability describes when the average magnitude of a individual scan group is significantly deviant from the mean of all the star's scan groups.  Excursive variability is sensitive to short time-scale variations such as a single eclipse event or flare.  Excursive candidate variables are identified if the average magnitude for a single scan group exceeds the global mean by more than 5$\sigma$, where here $\sigma$ is the co-added uncertainty in the scan group photometry.  As with flickering variability, this search is done separately for each of the 3 bands.  From the final variable catalog, 21 stars flag in only a single band, 19 flag in two bands and 41 flag in all three bands.  Low signal-to-noise photometry (N$_{obs}$ typically less than 500) might account for the 10 candidate variables that are excursive in the K band only.  It might also account for the 12 candidate variables that are excursive in both the \emph{H} and K bands.  However there is no obvious explanation for the 19 candidate variable stars that are excursive only in the \emph{J} or \emph{H}, \emph{J} and \emph{H}, or \emph{J} and K bands. The average number of deviant scan groups per star in our variable catalog is 24, 42 and 57 in \emph{J}, \emph{H} and \emph{K$_{s}$} bands, respectively.
  
\subsubsection{Stetson Index}
The Stetson index describes the correlation in a star's photometric variation between different bands.  The Stetson index is sensitive to variability whose amplitude is not significantly different between photometric bands.  For example, the Stetson index is not sensitive to a strong increase in the Br$\gamma$ emission line strength that might only affect 1 band.  This index has been previously used on other molecular cloud 2MASS variability surveys in Orion A and Chameleon I \citep{carpenter01,carpenter02}.  The Stetson index is computed for all 1678 stars; a star is considered a candidate variable if this index is $>$ 0.2.  P08 determined this criterion based on 18 of 23 periodic variables, in that work, having indices above this value.  The same index is adopted here since the observing methodology is identical in both works. This index is smaller than those adopted for the Orion A (0.55) and Chameleon I (1.00) surveys.  The Orion A survey contained 29 epochs over a 36 day temporal baseline and the Chameleon I survey contained 15 epochs over 5 months.  The smaller number of observed epochs in each case causes these surveys to be less sensitive to variability and thus in need of a higher index.  A Stetson index of zero indicates random noise or no correlation between the photometry in different bands.  A positive index indicates correlation between the photometry in two bands.  The higher the index, the greater the correlation between the photometry.  Using the Stetson index 57 stars flag as variable.

\subsubsection{Excluding Seeing Induced Variables}
A common way in which a non-variable star is misidentified as variable is from photometric variations caused by changing atmospheric seeing.  Both photometric techniques used here (PSF fitting, fixed sums) are susceptible to this, especially in regions that are crowded or where there is bright nebular emission.  Seeing estimates, corresponding to the average FWHM for each calibration scan, are provided for the Cal-PSWDB photometry.  The typical seeing values range between 2.5'' to 2.7'' over the entire observing season \citep{cutri06}.

The possibility of changes in brightness being correlated with changes in the seeing are first investigated.  This is done by computing the Pearson r-correlation statistic for each star, \emph{n}.  The statistic is given by the following:

\begin{equation}
  r_n =  \frac{\Sigma_{t=1}^{N_{m,n}}(m_{n,t} - \overline{m}_n)(S_{m,t} - \overline{S_m})}{\sqrt{\Sigma_{t=1}^{N_{n,t}}(m_{n,t} - \overline{m}_n)^{2}}\sqrt{\Sigma_{t=1}^{N_{n,t}}(S_{m,t} - \overline{S_m})^{2}}}
\end{equation}

\noindent where \emph{m} is the band, \emph{S$_{m,t}$} is the \emph{m}-band seeing FWHM in arcseconds at epoch \emph{t} and \emph{$\overline{S}$} is the average seeing in \emph{m}-band.  The separate quantities are summed over all \emph{N$_{\emph{m,n}}$} \emph{m}-band observations for star, \emph{n}.  This statistic spans the range from -1 to 1 with negative values indicating inversely correlated variations and positive values corresponding to directly correlated variations.  An inverse correlation means as the seeing worsens the star gets brighter.  A direct correlation refers to the opposite effect.  Since in Eqn 2, the photometry comparison (numerator) is computed in magnitudes and the photometric standard deviation (denominator) is computed first in flux units then converted to magnitudes, this can result in \emph{r} values slightly outside the -1 to 1 range.
A slight trend exists in the sample of 1678 stars toward an inverse seeing correlation in each band.  The average \emph{r} statistics in \emph{J}, \emph{H} and \emph{K$_{s}$} are -0.12, -0.11 and -0.05, respectively.  Inverse correlation is likely caused by crowded fields where as the seeing worsens, flux from surrounding stars may encroach into the measured star's aperture or spatial profile.  While these correlations are not very significant in most cases, it is noted the seeing in one band is slightly correlated with the seeing in another band.  This is consistent with multi-band photometry taken simultaneously. To look for correlations between bands, the Pearson index for \emph{J} and \emph{H} are plotted in Fig 2.  To characterize and flag seeing induced variability, a single seeing test is constructed to provide an estimate of seeing effects on measured photometry.  Each correlation statistic (\emph{r$_{\emph{J}}$}, \emph{r$_{H}$}, \emph{r$_{K}$}) is considered a component of a single ``seeing vector''.  This vector is rotated and transformed from cartesian to cylindrical coordinates so the \emph{z}-axis corresponds to \emph{r$_{J}$} = \emph{r$_{H}$} = \emph{r$_{K}$}.  This representation causes the seeing correlation to be axisymmetric about the \emph{z}-axis, thus reducing the characterization of multi-band seeing correlation by one dimension.  A ``seeing ellipse'' is described by

\begin{equation}
  \frac{z_n^{2}}{\sigma_z^{2}} + \frac{\rho_n^{2}}{\sigma_\rho^{2}} = 1
\end{equation}

\noindent where $\emph{z$_{n}$}$ is the component of the seeing vector for star \emph{n}, with standard deviation $\sigma$$_{z}$, along the \emph{z}-axis. $\rho$$_{\emph{n}}$ is the component for star \emph{n} along the $\rho$-axis, with the standard deviation $\sigma$$_{\rho}$.  Both $\sigma$$_{z}$ and $\sigma$$_{\rho}$ are determined from the distribution of the ensemble 1678 stars.  A candidate variable is flagged as seeing correlated when the seeing vector length is larger than the seeing ellipse for the ensemble.  This is the case when the left-hand side of Eqn 3 is greater than unity.  This test excludes 19 candidate variables; the variability of these stars is likely solely caused by fluctuations in atmospheric seeing. 

\begin{figure}
  \plotone{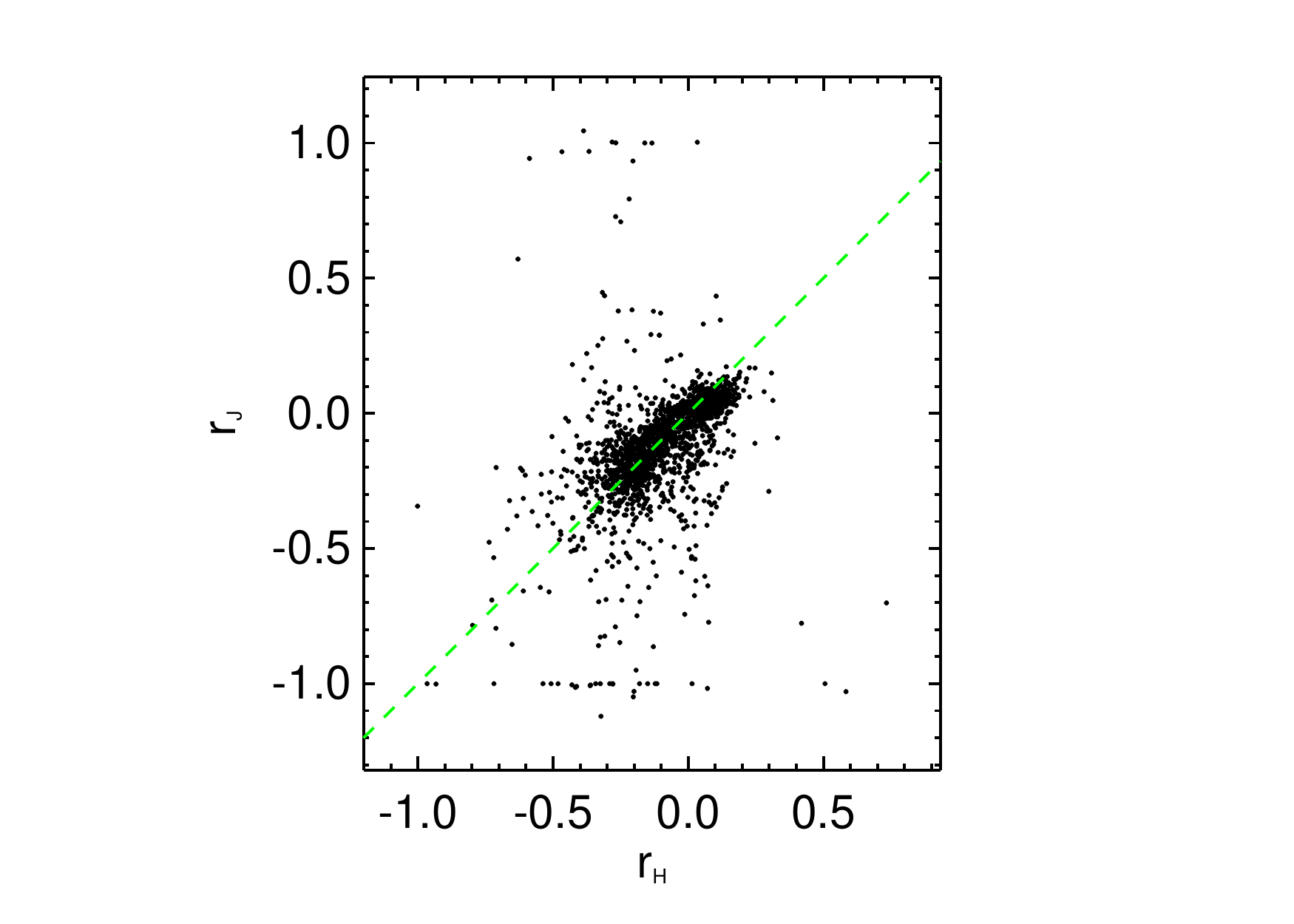}
  \caption{Seeing correlation between the \emph{J} and \emph{H} bands for the 1678 sample stars.  The dashed green line is a 1:1 correlation between the seeing correlation in the \emph{J} band, r$_\emph{J}$,  and \emph{H} band, r$_\emph{H}$.  The line also corresponds to the projected z-axis as described in the text just prior to Eqn 3}
\end{figure}

\subsubsection{Final Variable Catalog}
From the target sample of 1678 stars, 101 stars (6$\%$) are identified as variable.  These variable stars are referred to as the \emph{variable catalog}.  The variable catalog is listed in Table 3.  The full set of light curves, color curves and color-color plots for all variables stars is only available on-line.  For inclusion into the variable catalog, a star must not exhibit seeing correlated photometry (see $\S$ 3.1.4) and must meet 2 of the 7 variability criteria (see $\S$3.1.1 - 3.1.3).  In addition, the 2 criteria must be met in different bands or in a single band along with the Stetson criterion.  This last condition is imposed in order to prevent identifying variability due to poor quality or spurious photometry that is missed by the previous filters.  The amplitudes of variability for stars within the variable catalog span a wide range.  The range in $\Delta$K$_{s}$ spans 0.04 to 2.31 mag and $\Delta$(\emph{H}-\emph{K$_{s}$}) varies from 0.01 to 1.62 mag.  The variable catalog contains 47 stars with $\Delta$K$_{s}$ $>$ 0.25 mag and 66 stars with $\Delta$(\emph{H}-\emph{K$_{s}$}) $>$ 0.1 mag.

\begin{deluxetable}{c c c c c c c c c}
  \tablecolumns{9}
  \tablewidth{0pc}
  \tablecaption{Catalog of Variable Stars in $\rho$ Oph}
  \tabletypesize{\scriptsize}
  \tablehead{
    \colhead{RA} & \colhead{Dec} & \colhead{Varability Flags\tablenotemark{a}} & \colhead{$\Delta$K$_{s}$} & \colhead{$\Delta$({\emph{J}}-\emph{H})} & \colhead{$\Delta$(\emph{H}-\emph{K$_{s}$})} & \colhead{Type} & \colhead{YSO Class\tablenotemark{b}} & \colhead{'On Cloud'}\\
    \colhead{(degrees)} & \colhead{(degrees)} & \colhead{} & \colhead{(mag)} & \colhead{(mag)} & \colhead{(mag)} & \colhead{} & \colhead{} & \colhead{}
  }
  \startdata
  246.732269&-25.137508&1101010&0.114&0.150&0.123&irregular&---&no\\
  246.736557&-24.230934&1111111&0.476&0.235&0.173&periodic&---&yes\\
  246.737396&-24.922001&1000010&0.167&0.153&0.180&irregular&---&no\\
  246.7388&-24.594057&0010010&0.166&---&---&irregular&II&yes\\
  246.739273&-24.88303&0001010&0.160&0.105&0.139&irregular&---&no\\
  246.73941&-24.880718&1111110&0.854&---&---&irregular&---&no\\
  246.741165&-24.964417&0001100&0.108&0.093&0.124&irregular&---&no\\
  246.743301&-24.358301&1111111&0.292&0.140&0.257&periodic&II&yes\\
  246.743774&-24.76016&1111111&0.500&0.415&0.270&LTV&II&yes\\
  246.744202&-24.76741&0010100&0.131&0.198&0.168&irregular&---&yes\\
  246.744324&-24.309591&1111111&0.294&0.241&0.180&irregular&II&yes\\
  246.746017&-24.599096&1111111&0.218&0.185&0.380&LTV&II&yes\\
  246.746155&-24.787884&1111110&1.109&---&---&irregular&---&yes\\
  246.74649&-24.582909&0000010&0.631&---&---&irregular&I&yes\\
  246.748657&-24.261997&1111100&0.078&0.124&0.073&irregular&---&yes\\
  246.752335&-24.273695&1100000&0.061&0.082&0.066&irregular&---&yes\\
  246.752975&-24.774199&1111101&0.061&0.091&0.084&LTV&---&yes\\
  246.75676&-24.360228&1111111&0.078&0.074&0.180&LTV&III&yes\\
  246.75972&-24.624172&1111100&0.082&0.083&0.176&irregular&---&yes\\
  246.761093&-24.776232&1101001&0.049&0.061&0.122&LTV&---&yes\\
  246.76149&-25.045506&1100000&0.053&0.062&0.083&irregular&---&no\\
  246.761902&-24.31514&0001110&0.060&0.086&0.064&irregular&---&yes\\
  246.764984&-24.33481&1110111&0.894&---&0.707&LTV&II&yes\\
  246.76709&-24.474903&1110111&0.294&0.275&0.523&LTV&II&yes\\
  246.768814&-24.716524&1111111&0.084&0.051&0.065&periodic&III&yes\\
  246.768997&-24.70384&1111111&0.086&0.047&0.050&periodic&III&yes\\
  246.769058&-24.454285&1111111&0.411&0.232&0.329&periodic&II&yes\\
  246.770874&-25.146795&1100000&0.052&0.084&0.077&irregular&---&no\\
  246.771545&-24.335421&1110000&0.049&0.079&0.078&irregular&---&yes\\
  246.773605&-24.670246&0010010&0.337&---&---&periodic&II&yes\\
  246.774628&-24.99383&1100000&0.044&0.078&0.055&irregular&---&no\\
  246.774902&-24.476698&0000110&0.182&---&0.157&irregular&II&yes\\
  246.777481&-24.696856&1111111&0.224&0.128&0.099&periodic&II&yes\\
  246.778244&-24.637451&0110111&1.125&---&1.065&LTV&I&yes\\
  246.784149&-24.707903&1001001&0.069&0.061&0.195&irregular&---&yes\\
  246.787857&-24.200172&1111111&0.508&0.242&0.204&periodic&II&yes\\
  246.787964&-24.56889&1111111&0.330&0.137&0.102&periodic&II&yes\\
  246.788971&-24.672836&0110110&0.195&---&0.119&LTV&II&yes\\
  246.78923&-24.621819&1111111&1.636&---&0.628&periodic&I&yes\\
  246.791824&-24.486958&1111111&0.334&---&0.366&irregular&II&yes\\
  246.792862&-24.320118&1110010&0.198&1.272&0.098&LTV&II&yes\\
  246.7957&-24.758245&1000001&0.056&0.073&0.074&irregular&---&yes\\
  246.79657&-24.679569&1110111&0.984&---&1.123&LTV&II&yes\\
  246.798706&-24.394924&1111001&0.058&0.067&0.122&LTV&III&yes\\
  246.798828&-24.642199&0110111&1.012&---&0.784&LTV&II&yes\\
  246.79892&-24.786337&1010001&0.069&0.079&0.074&irregular&---&yes\\
  246.800537&-24.58028&1111111&0.729&0.346&0.402&periodic&II&yes\\
  246.803055&-25.067175&1111111&0.339&0.157&0.112&periodic&---&no\\
  246.807236&-24.304626&1111111&0.155&0.085&0.090&irregular&II&yes\\
  246.807388&-25.095842&1110000&0.059&0.065&0.091&irregular&---&no\\
  246.807602&-24.725399&1111111&0.560&0.249&0.133&LTV&II&yes\\
  246.807755&-24.262215&0110000&0.205&---&0.291&irregular&---&yes\\
  246.808533&-24.252649&1110000&0.109&0.200&0.149&irregular&---&yes\\
  246.813034&-24.860764&1111111&0.330&0.213&0.111&periodic&---&no\\
  246.813858&-24.264278&1010000&0.069&0.099&0.088&irregular&---&yes\\
  246.814423&-24.444342&0110111&0.562&---&0.850&LTV&II&yes\\
  246.814636&-24.514885&0010010&0.445&---&---&LTV&II&yes\\
  246.815674&-24.645327&1111111&0.299&0.262&0.297&LTV&II&yes\\
  246.816071&-24.645321&1111111&0.213&0.258&0.305&periodic/LTV&II&yes\\
  246.816208&-24.420513&0110010&0.356&---&---&periodic&II&yes\\
  246.816925&-24.250999&0110000&0.352&---&0.534&irregular&---&yes\\
  246.816971&-24.271143&1010000&0.391&---&0.517&irregular&---&yes\\
  246.821976&-24.374475&0110010&0.163&---&1.137&LTV&0&yes\\
  246.822739&-24.218828&0100000&0.155&0.217&0.315&irregular&---&yes\\
  246.823074&-24.4823&0110111&0.728&---&0.813&LTV&I&yes\\
  246.826492&-24.914923&1111111&0.807&0.354&0.327&periodic&---&no\\
  246.826584&-24.407238&0110111&0.135&---&0.371&periodic/LTV&III&yes\\
  246.826599&-24.654037&1110111&0.528&---&0.009&periodic&I&yes\\
  246.827026&-24.484921&1111111&0.640&0.161&0.186&periodic&II&yes\\
  246.831299&-24.694487&0001111&0.092&0.057&0.071&periodic&III&yes\\
  246.839462&-24.69525&1111111&0.215&0.276&0.158&irregular&II&yes\\
  246.840378&-24.363819&0100000&0.050&0.184&0.069&irregular&III&yes\\
  246.840836&-24.498091&0110111&1.199&---&1.256&LTV&I&yes\\
  246.840942&-24.726538&0010000&0.062&---&0.056&periodic&III&yes\\
  246.843735&-25.126837&1111111&0.700&---&0.751&irregular&---&no\\
  246.84552&-24.299223&1111111&0.130&0.044&0.071&periodic&III&yes\\
  246.845718&-24.801941&1111111&0.134&0.097&0.062&LTV&---&yes\\
  246.846909&-24.809896&1001001&0.074&0.099&0.094&irregular&---&yes\\
  246.848297&-24.207954&1111001&0.052&0.066&0.080&LTV&---&yes\\
  246.852676&-24.684278&0010010&0.749&---&---&LTV&I&yes\\
  246.852737&-24.493141&0010010&0.094&---&0.327&periodic&---&yes\\
  246.854782&-24.775953&0100101&0.065&0.089&0.173&LTV&---&yes\\
  246.85556&-25.105873&1111111&0.402&0.272&0.295&periodic&---&no\\
  246.859329&-24.323023&0111111&0.326&0.432&0.175&irregular&II&yes\\
  246.859558&-24.712914&0010010&0.098&---&0.200&periodic&I&yes\\
  246.860397&-24.65638&1111111&0.318&0.185&0.207&periodic&II&yes\\
  246.860779&-24.431711&1111111&0.211&0.099&0.096&periodic&II&yes\\
  246.862335&-24.680904&0110110&0.393&---&0.927&LTV&I&yes\\
  246.862778&-24.538191&0010010&0.090&---&0.478&periodic&---&yes\\
  246.864105&-24.521235&1111111&0.305&0.121&0.082&periodic&II&yes\\
  246.866684&-24.65926&0110111&0.784&---&0.549&periodic&I&yes\\
  246.872681&-24.654474&1111111&2.312&---&1.318&LTV&I&yes\\
  246.875793&-24.462006&1111111&0.155&0.110&0.233&LTV&II&yes\\
  246.877228&-24.542961&1100000&0.070&0.075&0.082&irregular&---&yes\\
  246.87851&-24.790745&1110101&0.067&0.057&0.057&periodic&III&yes\\
  246.878571&-24.415533&1111111&0.282&0.106&0.160&LTV&II&yes\\
  246.878784&-24.459188&0010000&0.926&---&---&LTV&I&yes\\
  246.879166&-25.065256&1111110&1.057&---&---&irregular&---&no\\
  246.879456&-24.567505&1111001&0.058&0.054&0.071&periodic&III&yes\\
  246.880157&-25.071445&1111110&0.566&0.386&0.473&irregular&---&no\\
  246.883682&-25.148535&1110000&0.613&0.318&0.679&irregular&---&no\\
  \enddata
  \tablenotetext{a}{First three flags correspond to flickering variability.  The second three flags correspond to excursive variability.  The seventh flag corresponds to the Stetson index.  Flag is set to 1 when true; 0 otherwise}
  \tablenotetext{b}{\citep{bontemps01,gutermuth09}}
\end{deluxetable}

Fig 3 contains the co-added calibration field in the direction of $\rho$ Oph; the target sample of 1678 stars and the variable catalog of 101 stars are plotted to show their spatial distribution.  It is clear that target stars are not evenly distributed in the field.  A demarcation line at $\delta$ = -24$\degree$51' is set as an ad hoc determination of cloud membership.  North of this limit is considered ``on cloud'' while anything south is classified as ``off cloud''.  This demarcation corresponds roughly to where A$_{V}$ = 5 mag \citep{cambresy99}.  Comparing the variability north and south of this demarcation, the ``on-cloud'' variable fraction increases to 15$\%$ while the variable fraction for the ``field'' drops to a mere 1$\%$.  This is consistent with the expectation young stars are are more often found spatially close to molecular clouds and are more variable than field stars.

\begin{figure}
  \plotone{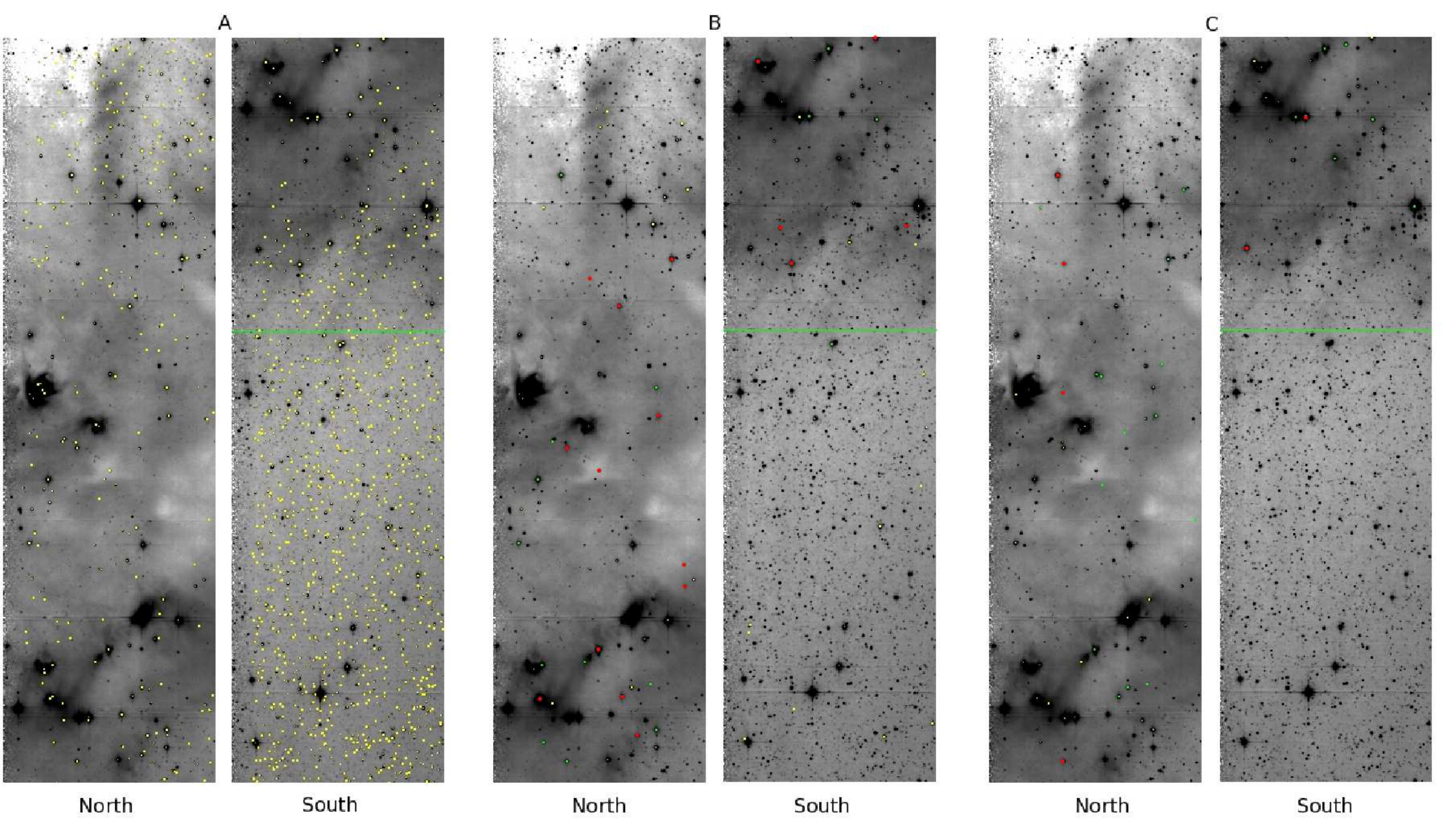}
  \caption{$\rho$ Ophiuchus field. $\emph{A:}$ The field is split into ``North'' and ``South'' panels.  The 1678 source sample is overlaid in yellow.  $\emph{B:}$ The same field overlaid with all variable sources.  Green - periodic variables ($\S$5.1).  Red - time-scale variables ($\S$5.2).  Yellow - irregular variables ($\S$5.3).  $\emph{C:}$ Same field overlaid with all classified YSO sources ($\S$3.3).  Yellow - Class I.  Green - Class II.  Red - Class III.  The green line in all ``South'' panels represents a demarcation at $\delta$: -24$\degree$ 51$'$ where A$_V$ = 5 mag \protect\citep{cambresy99}.  North of this demarcation contains higher visual extinction.}
\end{figure}

\subsection{Known Young Stars in the $\rho$ Oph Field}
Clues to the formation and evolution of young stars may be revealed by relating the variability to the stars' evolutionary states.  As originally proposed by \citet{lada87}, young stars are classified into four evolutionary stages or \emph{classes} (Class 0, Class I, Class II and Class III). Class assignment is typically based on photometry through the infrared slope index in the wavelength range from 2 to 25 $\mu$m.  Class 0 stars represent cloud cores undergoing the initial stages of protostellar collapse.  Class I stars are heavily embedded protostars with in-falling material from a circumstellar envelope forming an accretion disk.  Class II stars are fully assembled stars with accretion primarily from the circumstellar disk channeled onto the star along magnetic field lines; classical T Tauri (CTTS) stars are another name for Class II stars.  The last stage, Class III, represents stars yet to reach the main-sequence with depleted or no accretion disks due to mass accretion onto the star, photo-evaporation, or planet formation.  They may nevertheless retain debris disks or disks with depleted inner holes.  These stars are also known as weak-lined T Tauri stars (WTTS).

To identify if any of the 1678 sample stars have a previously assigned evolutionary class, the sample is cross-referenced with the $\rho$ Oph L1688 cloud core mid-IR surveys by \citet[hereafter B01]{bontemps01} and \citet[hereafter G09]{gutermuth09}.  The measurements obtained by B01 were taken with the \emph{ISO} ISOCAM LW2 and LW3 broad band cameras centered on 6.7 $\mu$m and 14.7 $\mu$m, respectively.   The G09 survey obtained measurements in the \emph{Spitzer} IRAC 3.6, 4.5, 5.8 and 8.0 $\mu$m bands complemented by \emph{J}, \emph{H} and \emph{K$_{s}$} 2MASS data for stars are used when there are no detections in either 5.8 or 8.0 $\mu$m.  In addition, 24 $\mu$m \emph{Spitzer} MIPS data is also used to verify YSO classifications in cases with high SNR ($\sigma$ $<$ 0.2 mag) and star luminosity ([24] $<$ 7 mag).  

The B01 survey provides YSO classifications for 54 of the 1678 target sample stars, while G09 provides classifications for 58 stars.  However, overlapping targets between these surveys results in 40 stars classified by both B01 and G09, yielding YSO classifications for only 72 target sample stars.  For 5 stars classified by both B01 and G09, the two surveys disagree on the classification.  G09 classifies these 5 stars as belonging in an earlier evolutionary stage by one class than B01 (i.e. WL 22 is classified as Class I by G09 and a Class II by B01).  In these cases, the classification by G09 is adopted because of the broader wavelength coverage utilized.  Assuming the B01 survey identified all the young stellar objects in the $\rho$ Oph region (425 YSOs), this survey contains $\sim$17$\%$ of these YSOs.  Of the 72 stars with YSO classifications, 79$\%$ are identified as variable stars.  As a function of YSO class, 92$\%$ of both Class I (12 of 13) and Class III (11 of 12) are variable stars.  The variable fraction decreases to 72$\%$ (34 of 47) for Class II stars.  The majority (14 stars) of the non-variable YSOs are Class II while ISO-Oph 99 is Class I.  All of these stars are located ``on cloud''.  As a YSO evolves in time the median brightness and color variability amplitudes decrease.  The median peak-to-trough $\Delta$\emph{K$_{s}$} amplitude for Class I, II, and III stars are 0.77, 0.31, and 0.08 mag, respectively.  The median peak-to-trough $\Delta$(\emph{H}-\emph{K$_{s}$}) color amplitudes are 0.81, 0.21, and 0.07 mag for each class respectively.

\subsection{Advantages of High Cadence Variability Studies}
In this section, the advantages of high cadence, long temporal baseline observations in variability studies are investigated.  The results of this work are compared to the \citet[hereafter AC08]{oliveira08} survey of the $\rho$ Oph central cloud core.  The AC08 survey searched for variability in thousands of target stars within a $\sim$0.8 deg$^{2}$ field of view.  These stars were observed in the \emph{H} and \emph{K$_{s}$} bands during 14 epochs spanning May, June and July 2005 and 2006.  The magnitudes of target stars fell within 11 to 19 mag in \emph{H} and 10 to 18 mag in \emph{K$_{s}$}.

This survey and AC08 have 464 stars in common.  The prescription for identifying variables in AC08 is based on $\chi$$^{2}$ fitting and cross-correlations between the \emph{H} and \emph{K$_{s}$} photometry.  Comparing the number of variables detected from the 464 stars, AC08 identifies 32 (7$\%$) variables while this work identifies 82 (18$\%$).  The larger fraction of detected variables by this survey could be attributed to the higher sampling over a longer temporal baseline or from different sensitivities in the adopted variability criteria.  To determine which explanation is more probable, histograms of the $\Delta$K$_{s}$ peak-to-trough amplitudes for the variables identified by both this work and AC08 within the joint 464 star sample are computed. Fig 4 contains these histograms as well as the histograms for the $\Delta$(\emph{H}-\emph{K$_{s}$}) peak-to-trough color amplitudes.  It is clear from Fig 4 the fraction of variables with $\Delta$K$_{s}$ $<$ 0.5 mag detected by each survey is nearly identical.  The same is true for variables with $\Delta$(\emph{H}-\emph{K$_{s}$}) $<$ 0.55 mag.  Therefore, the higher fraction of variables detected, as compared to AC08, is most likely a consequence of the higher observing cadence.  It is worth noting that 7 stars within the joint sample are identified as variable by AC08, but are not in this work.  This work identified 5 of these stars as having photometry correlated with seeing.  Therefore these stars may have been intrinsically variable within the observing window, however this variability could not be confidently confirmed.

\begin{figure}
  \plotone{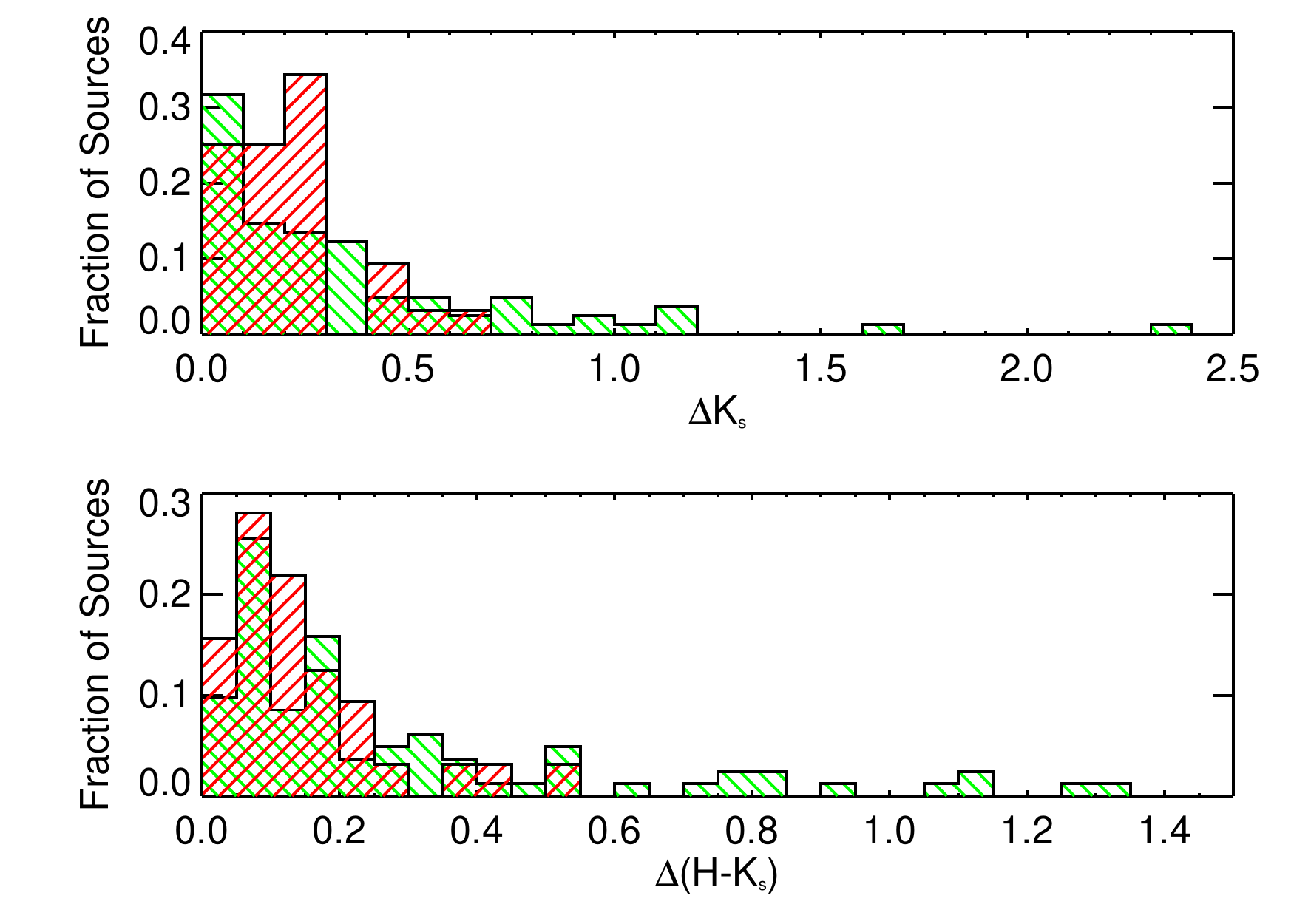}
  \caption{\emph{Top:} Histograms of $\Delta$\emph{K$_{s}$} for AC08 (red forward hatching) and this work (green backward hatching).  The two distributions are statistically indistinguishable.  \emph{Bottom:} Histograms of $\Delta$(\emph{H}-\emph{K$_{s}$}) for AC08 and this work.  Each survey is represented the same as the top plot.  This work detects larger amplitudes in both $\Delta$\emph{K$_{s}$} and $\Delta$(\emph{H}-\emph{K$_{s}$}) than AC08.}
\end{figure}

While the detection fraction of low amplitude variables is nearly identical between surveys, the detection fraction of high amplitude variability stars is not.  AC08 does not detect variables with $\Delta$\emph{K$_{s}$} $>$ 0.7 mag or $\Delta$(\emph{H}-\emph{K$_{s}$}) $>$ 0.55 mag.  This work finds 5.25$\%$ of detected variables have $\Delta$\emph{K$_{s}$} amplitudes greater than these upper limits.  In addition, 6$\%$ of detected variables have $\Delta$(\emph{H}-\emph{K$_{s}$}) color amplitudes greater than these upper limits.  Within the 464 star joint sample, 25 stars are identified as variable in both this work and AC08.  Strong correlations exist between the difference in amplitudes measured between surveys and the amplitudes measured in this work (see Fig 5).  Sparsely sampled photometry will underestimate the amplitude of variability in both magnitude and color.
 
\begin{figure}
  \plotone{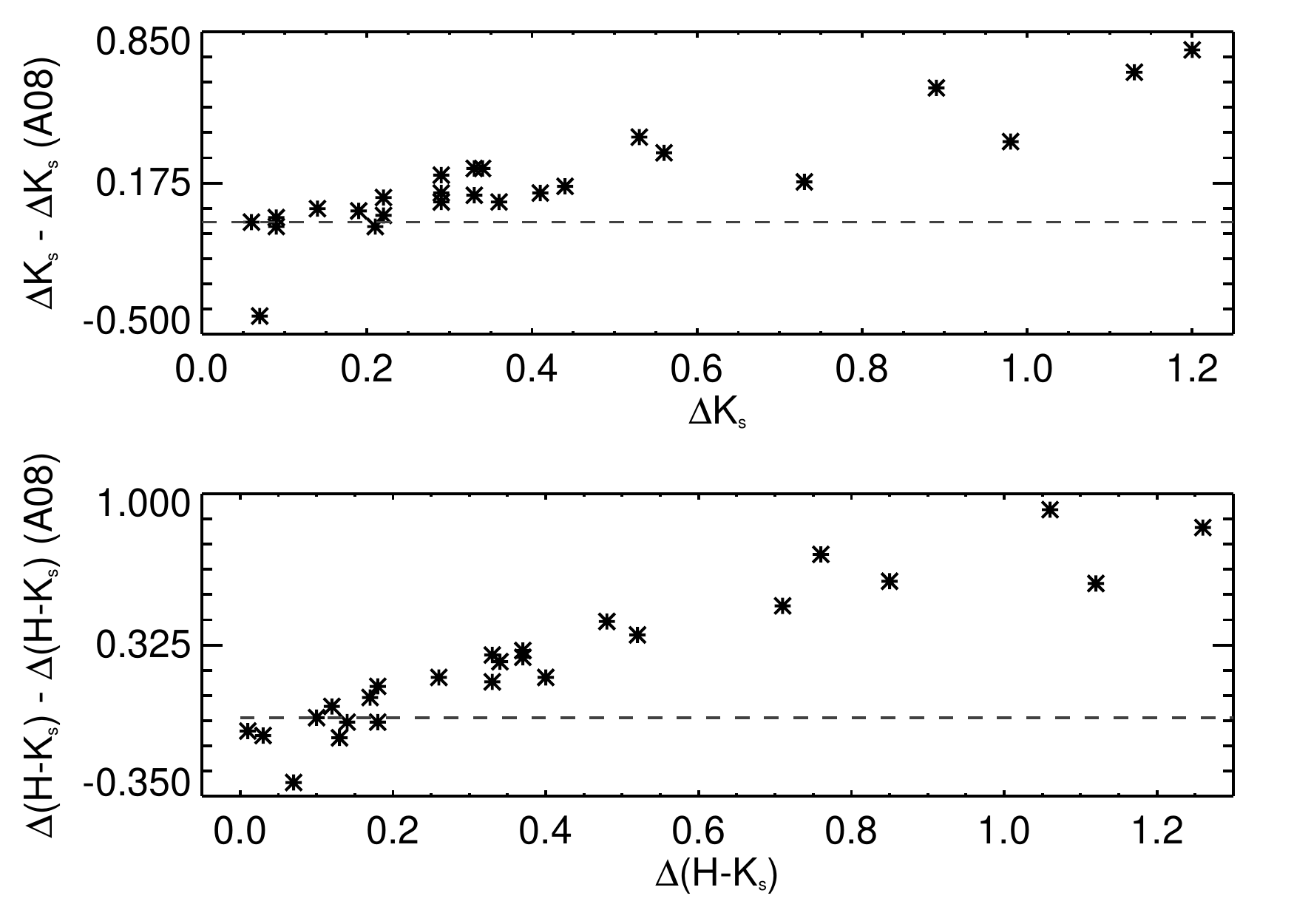}
  \caption{The difference between amplitudes measured by AC08 and this work.  \emph{Top:} The comparison between measured \emph{K$_{s}$} variability.  \emph{Bottom:} The comparison between measured (\emph{H}-\emph{K$_{s}$}) color variability.  In both cases, there is good agreement between the surveys for low amplitude variability.  However as the amplitude increases, AC08 underestimates the variability.  The dashed line in both plots indicates a difference of zero.}
\end{figure}

High cadence, long temporal baseline observations are vital for fully characterizing the variability of young stars.  It increases the detection fraction of the survey allowing for more accurate statistics, such as the incidence of variable stars and distribution of variability amplitudes. In addition, this strategy is needed to sample the full amplitude of variability.

\subsection{New Candidate $\rho$ Oph Members}
As photometric variability is an ubiquitous characteristic of young stars, it is a useful tool for assessing youth and potential membership in the $\rho$ Oph star forming region.  However, variability alone is not sufficient evidence for identifying potential members and additional constraints are needed, such as spatial location and location on a color-magnitude diagram. Candidate $\rho$ Oph membership is first determined by cross-referencing the final variable catalog with previous surveys to identify previously known $\rho$ Oph members \citep{strom95,barsony97,barsony05,grosso00,ozawa05,wilking05,pillitteri10}.  These are the same surveys used by AC08 to assign membership to their variable stars.  This identifies 62 of the 101 variable stars as confirmed members of $\rho$ Oph, which are plotted on a \emph{K$_{s}$} versus (\emph{H}-\emph{K$_{s}$}) color-magnitude diagram in Fig 6.  For comparison, the 53 variable stars determined as $\rho$ Oph members by AC08 are also plotted.   Eleven stars are identified as $\rho$ Oph in both surveys.  A dashed line connects the data for these stars as observed by AC08 and this work.  The solid black line indicates a 3 Myr isochrone constructed using NextGen models for masses between 0.02 to 1.4 M$_{\sun}$ at a distance of 129 pc \citep{baraffe98}.  The distance is the weighted average between previous measurements \citep{loinard08,mamajek08}.  A star is classified as a new candidate member if it is located ``on cloud'' (see $\S$3.1.5) and is brighter and redder than the 3 Myr isochrone (see Fig 6).  Table 4 contains the 22 stars identified as candidate $\rho$ Oph members from the previously unassociated 39 stars.  Candidate member 2MASS J16270597-2428363 is classified as a Class II YSO thereby increasing the likelihood of membership.  Follow up spectroscopic observations in the mid-IR for the remaining candidates to determine whether these stars are YSOs will provide additional evidence for membership.

\begin{deluxetable}{c c c c c c c c}
  \tablecolumns{8}
  \tablewidth{0pc}
  \tablecaption{Candidate $\rho$ Ophiuchus Members}
  \tabletypesize{\scriptsize}
  \tablehead{
    \colhead{RA} & \colhead{Dec} & \colhead{Catalog ID\tablenotemark{a}} & \colhead{$\emph{J}$\tablenotemark{b}} & \colhead{$H$\tablenotemark{b}} & \colhead{$K_{s}$\tablenotemark{b}} & \colhead{$(\emph{J}-\emph{H})$} & \colhead{$(H-K_{s})$}\\
    \colhead{(degrees)} & \colhead{(degrees)} & \colhead{} & \colhead{(mag)} & \colhead{(mag)} & \colhead{(mag)} & \colhead{(mag)} & \colhead{(mag)}
  }
  \startdata
  246.744202&-24.76741&65861-2446029&15.392$\pm$0.001&14.000$\pm$0.001&13.364$\pm$0.001&1.392&0.636\\
  246.746155&-24.787884&70054-2446444&16.935$\pm$0.032&16.255$\pm$0.011&15.567$\pm$0.012&0.680&0.688\\
  246.748657&-24.261997&65967-2415433&14.229$\pm$0.001&12.404$\pm$0.001&11.642$\pm$0.001&1.825&0.762\\
  246.752335&-24.273695&70055-2416255&13.432$\pm$0.001&11.840$\pm$0.001&10.997$\pm$0.001&1.592&0.842\\
  246.752975&-24.774199&70072-2446272&13.670$\pm$0.001&12.002$\pm$0.001&11.233$\pm$0.001&1.668&0.769\\
  246.761093&-24.776232&70266-2446345&13.348$\pm$0.001&11.596$\pm$0.001&10.665$\pm$0.001&1.752&0.930\\
  246.761902&-24.31514&70285-2418546&13.090$\pm$0.001&11.049$\pm$0.001&10.096$\pm$0.001&2.041&0.953\\
  246.771545&-24.335421&70516-2420077&12.700$\pm$0.001&10.440$\pm$0.001&9.341$\pm$0.001&2.260&1.099\\
  246.774902&-24.476698&70597-2428363&16.905$\pm$0.005&14.467$\pm$0.001&13.029$\pm$0.001&2.438&1.439\\
  246.784149&-24.707903&60819-2442286&15.365$\pm$0.001&12.252$\pm$0.001&10.723$\pm$0.001&3.113&1.529\\
  246.795700&-24.758245&71096-2445298&13.011$\pm$0.001&11.056$\pm$0.001&10.156$\pm$0.001&1.955&0.900\\
  246.807755&-24.262215&71384-2415441&16.760$\pm$0.005&15.065$\pm$0.002&14.251$\pm$0.002&1.696&0.814\\
  246.808533&-24.252649&71404-2415096&15.356$\pm$0.002&13.899$\pm$0.001&13.274$\pm$0.001&1.457&0.625\\
  246.813858&-24.264278&71531-2415515&13.990$\pm$0.001&12.532$\pm$0.001&11.865$\pm$0.001&1.458&0.667\\
  246.816925&-24.250999&71605-2415039&16.829$\pm$0.005&15.461$\pm$0.003&14.768$\pm$0.003&1.368&0.693\\
  246.816971&-24.271143&71604-2416163&17.056$\pm$0.007&15.708$\pm$0.004&15.007$\pm$0.004&1.348&0.701\\
  246.821976&-24.374475&71726-2422283&17.341$\pm$0.124&15.567$\pm$0.003&13.418$\pm$0.001&1.774&2.149\\
  246.822739&-24.218828&71744-2413079&15.993$\pm$0.002&14.378$\pm$0.002&13.707$\pm$0.001&1.614&0.671\\
  246.845718&-24.801941&72297-2448071&10.922$\pm$0.001&9.832$\pm$0.001&9.336$\pm$0.001&1.089&0.496\\
  246.846909&-24.809896&72325-2448357&14.124$\pm$0.001&12.613$\pm$0.001&11.982$\pm$0.001&1.511&0.631\\
  246.848297&-24.207954&72357-2412288&12.805$\pm$0.001&10.796$\pm$0.001&9.851$\pm$0.001&2.009&0.945\\
  246.854782&-24.775953&72514-2446335&15.527$\pm$0.001&12.957$\pm$0.001&11.683$\pm$0.001&2.569&1.275\\
  \enddata
  \tablenotetext{a}{The catalog ID has been truncated by 2MASS J162 for 2MASS catalog stars}
  \tablenotetext{b}{Unweighted mean apparent magnitude of Cal-PSWDB photometry}
\end{deluxetable}

\begin{figure}
  \plotone{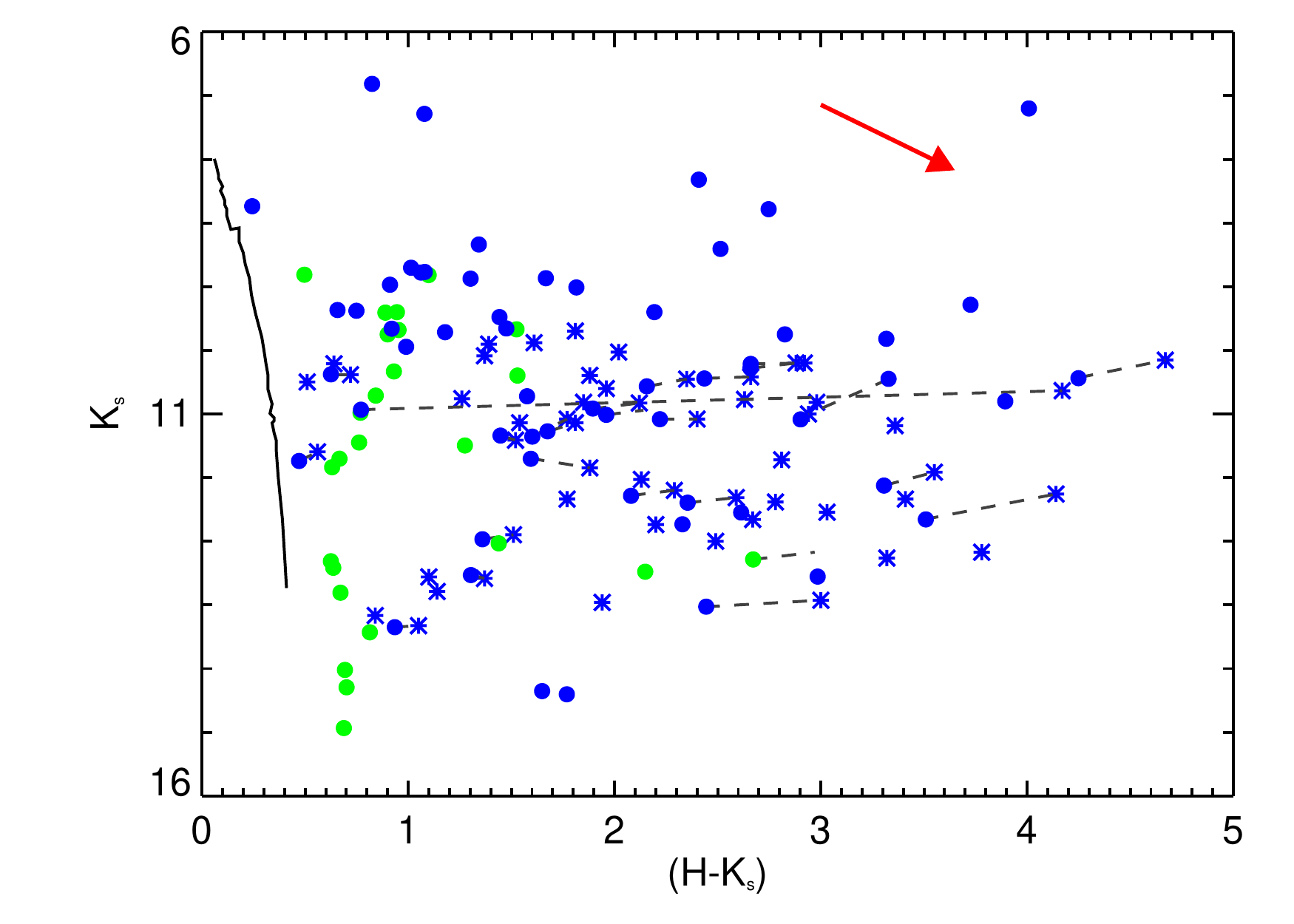}
  \caption{Candidate $\rho$ Oph membership.  The filled blue circles indicate variables previously identified as $\rho$ Oph members.  The green filled circles mark the 22 new candidate $\rho$ Oph members.  The blue asterisks indicate the AC08 measured \emph{K$_{s}$} and (\emph{H}-\emph{K$_{s}$}) for variables previously identified as $\rho$ Oph members.  The dashed lines connect the same variable as it is detected in this work.  The solid black line is a 3 Myr isochrone constructed using NextGen models for masses between 0.02 and 1.4 M$_{\sun}$ for a distance of 129 pc \citep{baraffe98}.  The red arrow corresponds to a reddening vector A$_{V}$ = 10 mag.}
\end{figure}

\section{TIME-SERIES ANALYSIS}
Characterizing the amplitude, time-scale, and form (e.g. periodic vs. aperiodic) of variability provides valuable insights into the underlying physical mechanism(s) causing the brightness variations.  Period-searching algorithms have been very helpful in this regard (e.g. \citep{lomb76,scargle82}).  In this section, two separate methods for measuring the time-scales of variability are discussed.

\subsection{Periodicity Analysis via the Plavchan Algorithm}
A novel period-searching algorithm, henceforth called the Plavchan algorithm (PA), is implemented to detect periodicity in identified variable stars.  The algorithm described below is a more mature version than the one used in \citet{plavchan08a}.  The version of the algorithm is used in the NASA Exoplanet Archive periodogram tool \citep{braun09,ramirez09}.  Tens of thousands of test periods are investigated by the PA algorithm with a uniform frequency sampling between 0.1 and 1000 days.  For each trial period, \emph{P$_{j}$}, the PA starts by generating a phase-folded light curve from the time-series photometry.  A phase is defined as the time (\emph{t$_{i}$}) modulo the test period (\emph{P$_{j}$}).  This light curve is smoothed via boxcar smoothing with a phase width, \emph{p} = 0.06.  This smoothed light curve is designated as the prior, or reference curve.  When the measured photometry for a periodic source is folded to the test period, the photometry is assumed to be approximately continuous and smoothly varying over the phased cycle.  The difference between the measured photometry and the prior is computed for every photometric measurement, \emph{m$_{i}$}.  This difference is compared to the difference between the measured photometry and a ``non-variable'' straight line, defined by the photometric mean (see Fig 7).  A poor fit results when these two differences are equal or nearly equal to each other.  A good fit results when the difference between the data and the smoothed prior is smallest.  This normalization removes the dependence on the absolute value and dispersion in \emph{m$_{i}$}.  A quality of fit, $\chi$$_{n_{0}}^{2}$, is computed by Eqn 4 only over the 40 data points with the poorest fits (\emph{n$_{0}$} = 40) (i.e. the epochs with the largest difference between the data, \emph{m$_{i}$}, and the prior (average) in the denominator (numerator)):

\begin{equation}
  \chi_{n_0}^{2} = \frac{\Sigma_{i=1}^{n_0}(m_i-\overline{m})^{2}}{\Sigma_{i=1}^{n_0}(m_i - m_{prior_i})^{2}}
\end{equation}

\noindent where the prior term, \emph{m$_{prior_{t}}$}, is the mean of \emph{m$_{i}$} if \emph{m$_{i}$} is within the boxcar smoothing window.  The summations in the numerator and denominator in Eqn 4 are over independent sets of poorly fit measurements, since the poorest fit measurements by the prior might not be the same as the measurements that deviate the most from the mean.  The best-fits periods have the largest $\chi$$_{n_{0}}^{2}$ value.  In other words, $\chi$$_{n_{0}}^{2}$ represents the power of the periodic signal.  The power indicates, for the PA, the relative improvements of the prior compared to a straight line for a given test period \emph{P$_{j}$}.

\begin{figure}
  \plotone{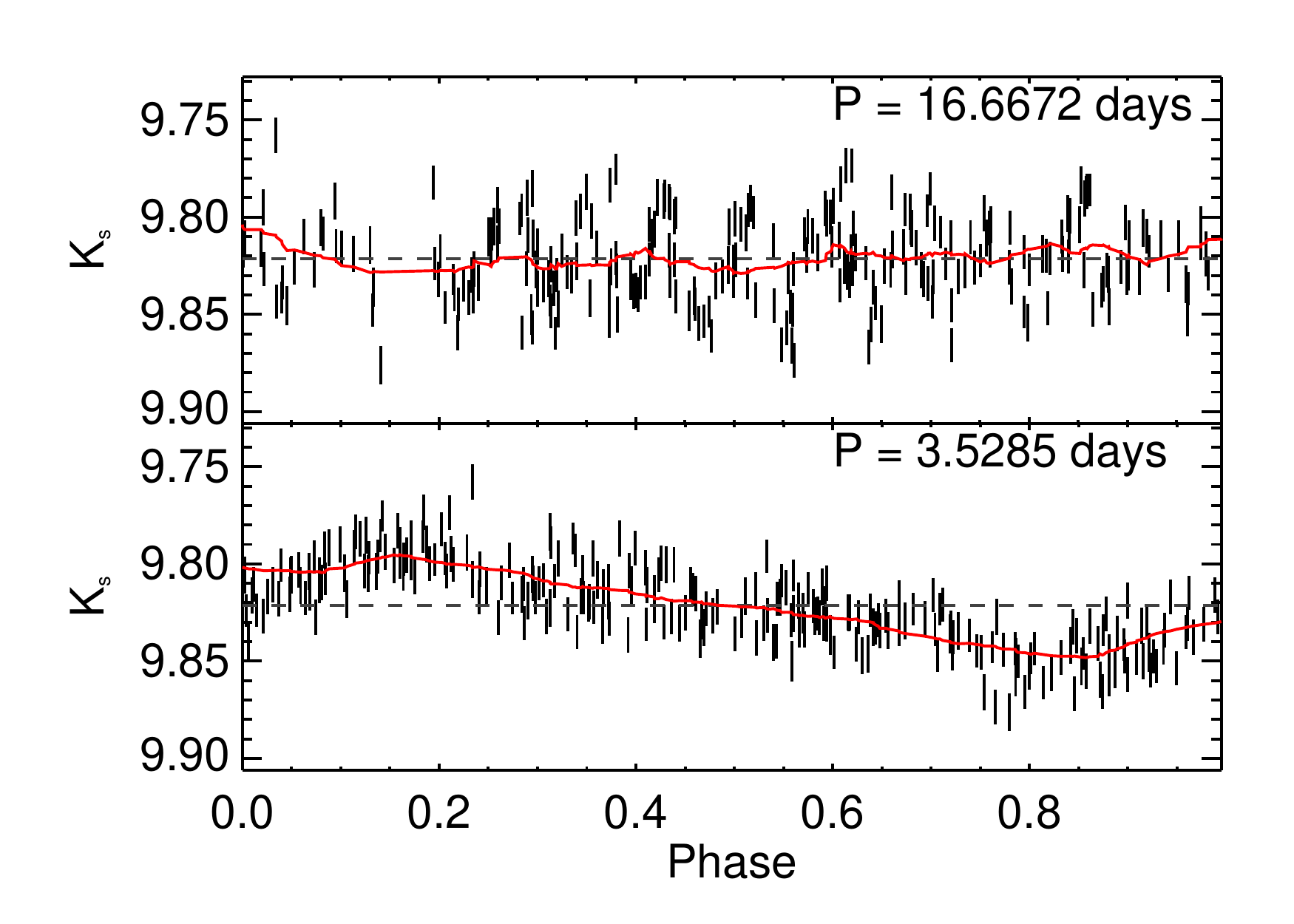}
  \caption{Demonstration of Plavchan Algorithm on ISO-Oph 96.  $\emph{Top:}$ Light curve phased to a period of 16.6672 days  This period is considered insignificant.  $\emph{BOTTOM:}$ phased to a period of 3.5285 days. This is the most significant period from the periodogram.   The dotted line indicates the mean magnitude for this star.  The red lines in the middle and bottom panels is the prior generated for each period.  Computing the $\chi$$_{40}^{2}$ for the 3.5285 and 16.6672 day periods indicates the power value for the former is $\sim$9x larger, implying a much larger statistical significance.}
\end{figure}

To evaluate the statistical significance of the power value for a peak period in the periodogram, or in other words to compute a false-alarm probability (FAP), there are several possible quantitative methodologies to arrive at an appropriate probability distribution.  The approaches include - one, an analytic derivation from first principles; two, a monte carlo of periodograms generated by randomly swapping measurement values at each epoch; three, the distributions of power values at other periods in the same (adequately sampled) periodogram; and four, the distribution of maximum power values for all sources in an ensemble (mostly non-variable) survey.  The first approach is rarely used in the literature, with the noted exception of the Lomb-Scargle periodogram \citep{scargle82}.  In the case of the Lomb-Scargle periodogram and typical radial velocity surveys, however, systematic errors in the velocity measurements can invalidate the assumptions in the first approach.  The second monte carlo approach is often used as a more reliable method for Lomb-Scargle periodograms \citep{marcy98}, and is equally applicable to the PA periodogram.  In this section, the third method to evaluate a period's statistical significance is discussed.  This third method is readily applicable to most time-series and is the method used in this work for computing the FAP for found periods.  In the Appendix, the fourth method is discussed.  The fourth method is survey dependent, but provides the insight that the PA periodogram is ``well-behaved'' with respect to changes in data values, number of observations, and algorithm parameters \emph{p} and \emph{n$_{0}$}.

The distribution of power values in an adequately sampled PA periodogram for a non-variable source is best described by a log-normal distribution.  In this instance, adequately sampled means covering a broad dynamic range of periods and sampling the periodogram at a large number of periods representative of the expected frequency resolution dictated by the cadence.  Fig 8 contains the periodogram for the non-periodic star 2MASS J16265576-2508150.  The power values vary about a mean value, or a ``significance floor''.  The distribution is slightly asymmetric with a slight bias towards power values greater than the mean, consistent with a normal distribution in log-space.  Fig 9 shows this distribution is very similar to the periodogram power value distribution for the boxcar least-squares (BLS) periodogram applied to the same source \citep{kovacs02}, albeit with a different mean and standard deviation.  The BLS periodogram traditionally makes the assumption of a normal distribution for evaluating the statistical significance of a peak period in the distribution of power values from an adequately sampled periodogram.  However, again, a log-normal distribution is a more appropriate prescription for the BLS distribution \citep{braun09,ramirez09}.  While the assumption of a normal distribution of power values is probably adequate for both algorithms, a normal distribution will ascribe a greater statistical significance (i.e. a smaller FAP) to a peak period than a log-normal distribution.  Therefore, the more conservative log-normal distribution is adopted in evaluating the statistical significance of peak periods in both the BLS and PA periodograms. 
  
\begin{figure}
  \plotone{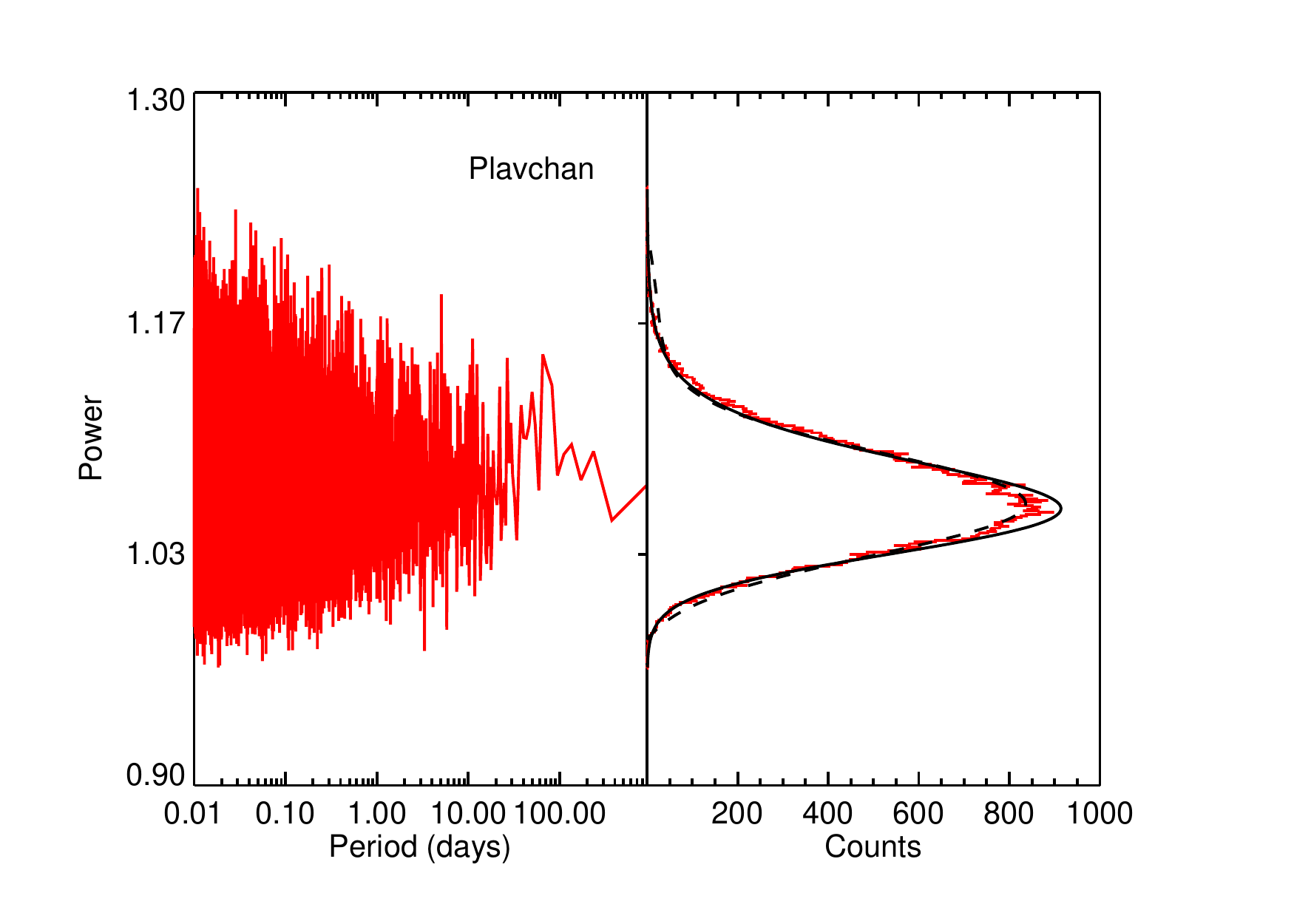}
  \caption{\emph{Left:} Periodogram for the irregular variable 2MASS J16265576-2508150 using the PA. \emph{Right:} The histogram of periodogram power values used to determine the significance of calculated periods.  The solid line indicates a log-normal distribution fit to the histogram values and the dashed line indicates a normal distribution fit.  The log-normal fit is used as it results in a more conservative higher false alarm probability.}
\end{figure}

\begin{figure}
  \plotone{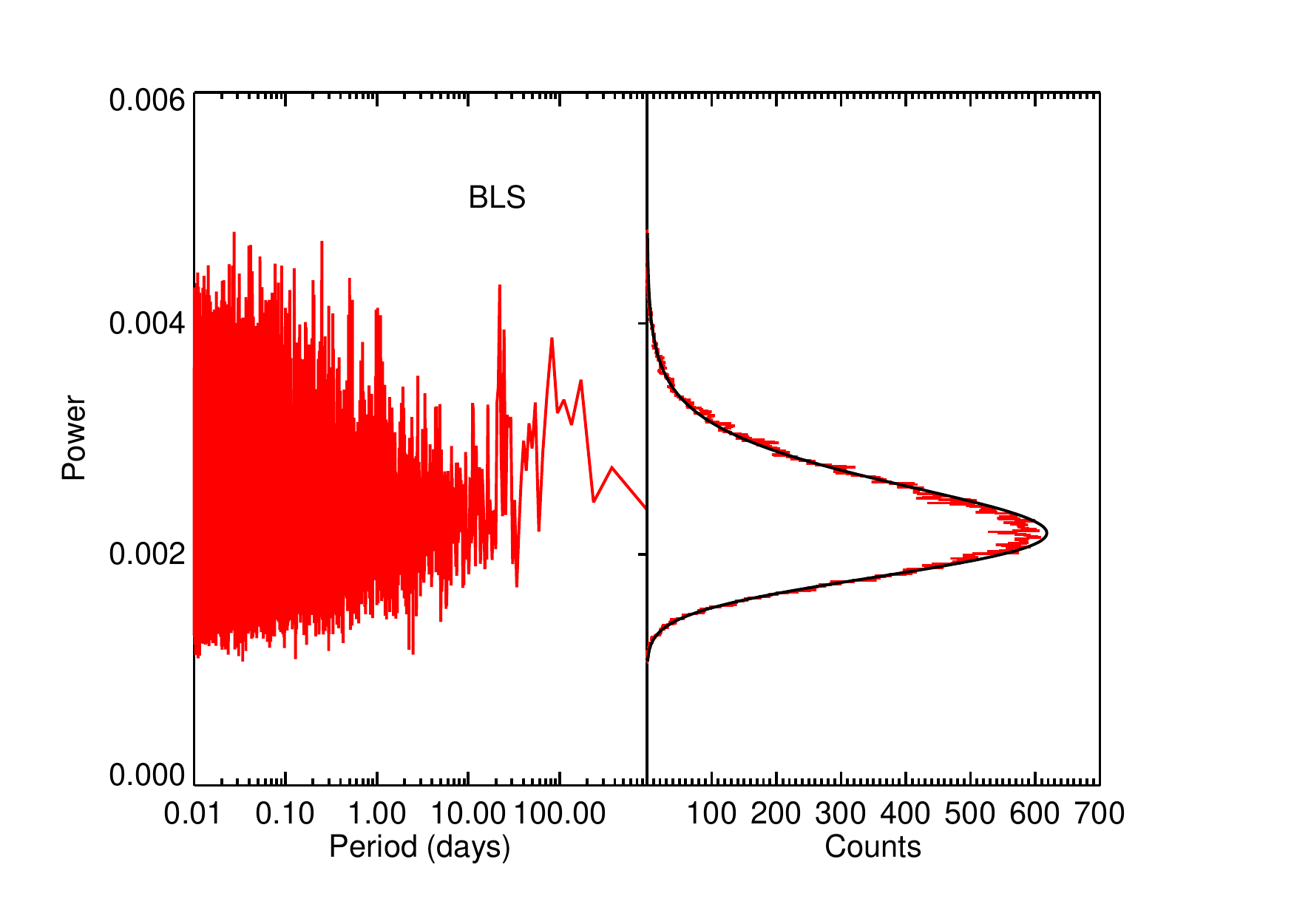}
  \caption{Same as Figure 8 however using the BLS algorithm}
\end{figure}

To determine if a period is statistically significant for a given source is this survey, the log of power values from the PA periodogram are computed, as well as the mean and standard deviation of the log-distribution.  Power values that are 5$\sigma$ outliers in the periodogram are identified as statistically significant periods with low FAP.  Each of these significant periods are investigated via visual inspection of the photometry folded to the period in question.  Finally the statistical significance of the derived period is confirmed by either the Lomb-Scargle or BLS algorithms, depending on the folded light curve shape.  The Lomb-Scargle algorithm is optimized to identify sinusoidal-like periodic variations, while the BLS algorithm is better equipped in identifying eclipse-like periodic variations.   Thus, the PA periodogram excels at identifying periodic signatures from both sinusoidal-like and eclipse-like time-series periodic variations \citep{plavchan08a}. The period error is derived from the 1$\sigma$ width of a Gaussian fit to the period's peak in the periodogram.  In order to avoid confusion in the fit from other peaks, only periods within $\pm$3$\%$ of the most significant peak are fit.  An upper bound to confident periods is placed at 200 days.  Stars are rejected as truly periodic with larger periods since the star will complete at most 3 cycles within the observing baseline.  These ``periods'' are reported as timescales and described in $\S$5.2.1.

From the 101 variables, 32 stars (32$\%$) are identified to exhibit periodic variability with periods ranging from 0.49 to 92 days.  Table 5 contains the list of periodic variables.

\subsubsection{Detecting Secondary or Masked Periodic Variability}
The PA found two statistically distinct ($>$20$\sigma$) periods for YLW 1C.  The time-series folded to the shorter period (5.7792 days) exhibits a sinusoidal-like shape.  The time-series folded to the longer period (5.9514 day) exhibits an ``eclipse-like'' shape where the star periodically dims from a near constant continuum flux.  This prompted a search for secondary periods in the other 5 stars that exhibit eclipse-like periodic variability.  We found 3 stars (YLW 1C, 2MASS J16272658-2425543, YLW 10C) to vary periodcally at two distinctly different periods; sinusoidal-like variability at one period and eclipse-like variability at the other.  Initially the secondary period is not statistically significant; it is only discovered when the time-series of the eclipse event is removed.  The PA is run only on the time-series preceding each eclipse ingress and after each eclipse egress.  A small number ($\sim$10) of sharp drops outside the eclipse events in the time-series for 2MASS J16272658-2425543 and YLW 10C are also omitted from the PA analysis.  Errors in the secondary periods are determined in the same manner as the primary periods.

Since multiple variability mechanisms may be common in variable stars \citep{herbst94,morales11}, we attempted to search for periodic variability in stars where the variability was complex.  For 6 variable stars, the stellar brightess fluxuates about a mean level for one or two consecutive years.  During the remaining time, a large amplitude variation is observed lasting longer than 50 days.  The PA is run on the nearly constant time-series omitting the large amplitude variation event.  In 2 stars (WL 20W and ISO-Oph 126), the PA found a significant period in the ``whitened'' time-series.  The time-series folded to the appropriate period results in sinusoidal-like variability with an amplitude $\sim$50$\%$ smaller than the large amplitude variation.  This larger amplitude variation effectively masked the smaller amplitude periodic signal.  For each star, the periodic variability could not be recovered during the large amplitude variation.  Fig 10 contains the \emph{K$_{s}$} light curves for WL 20W and ISO-Oph 126, as well as the \emph{K$_{s}$} light curves folded to the identified periods.  For WL 20W, 93 out of 262 scan groups were removed before the PA analysis.  For ISO-Oph 126, 149 out of 262 scan groups were removed.  Fig 11 shows the periodograms for both stars using the full time-series and the whitened time-series.

\begin{figure}
  \plotone{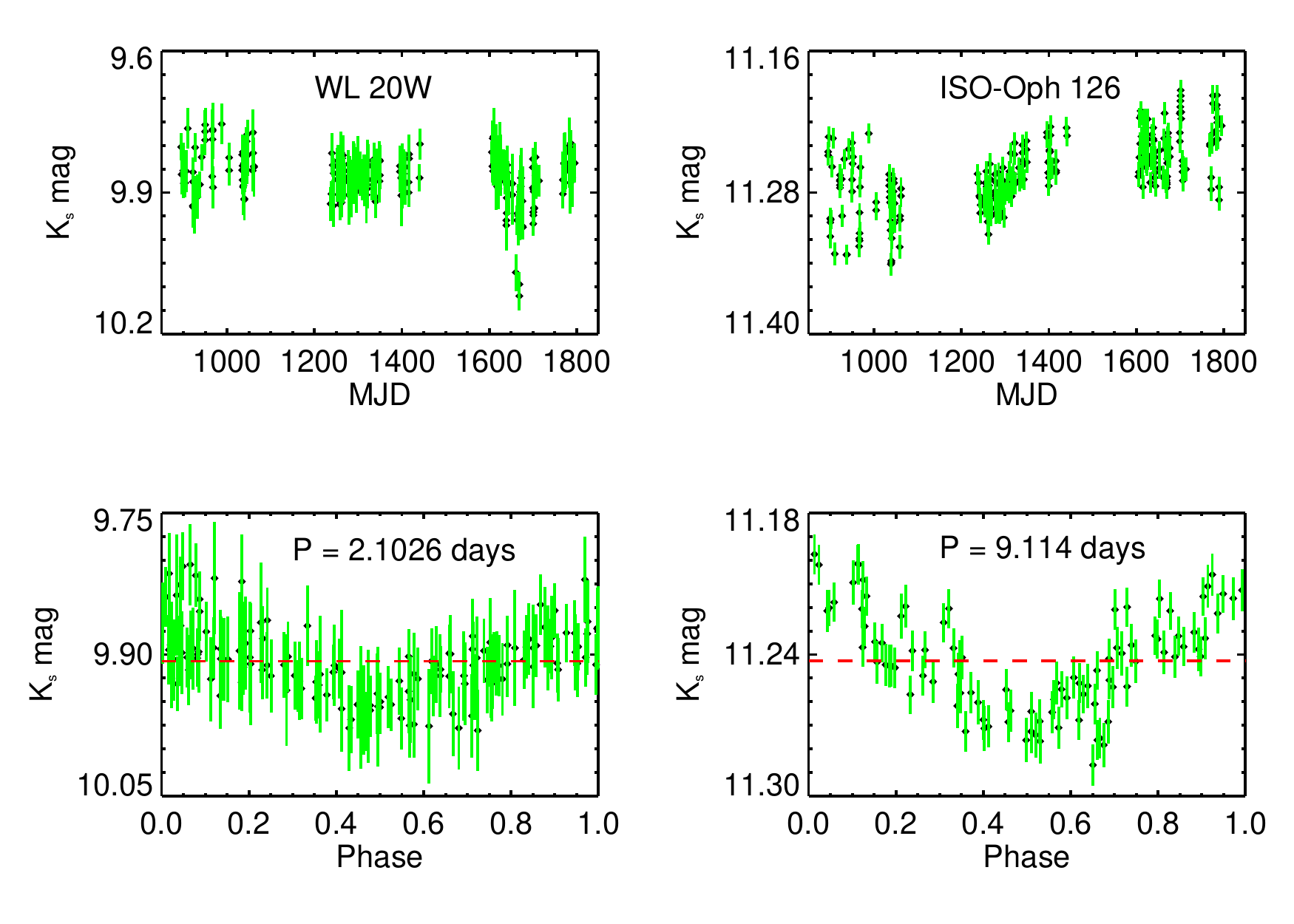}
  \caption{\emph{TOP:} The \emph{K$_{s}$} light curves for WL 20W and ISO-Oph 126.  Both light curves display a large amplitude long time-scale variation.  \emph{BOTTOM:} The folded \emph{K$_{s}$} light curves for WL 20W (P = 2.1026 $\pm$ 0.0060 days) and ISO-Oph 126 (P = 9.114 $\pm$ 0.90 days).  The periods are only detected once the photometry affected by the large amplitude variation is removed.  This data is not included in the folded light curves.}
\end{figure}

\begin{figure}
  \plotone{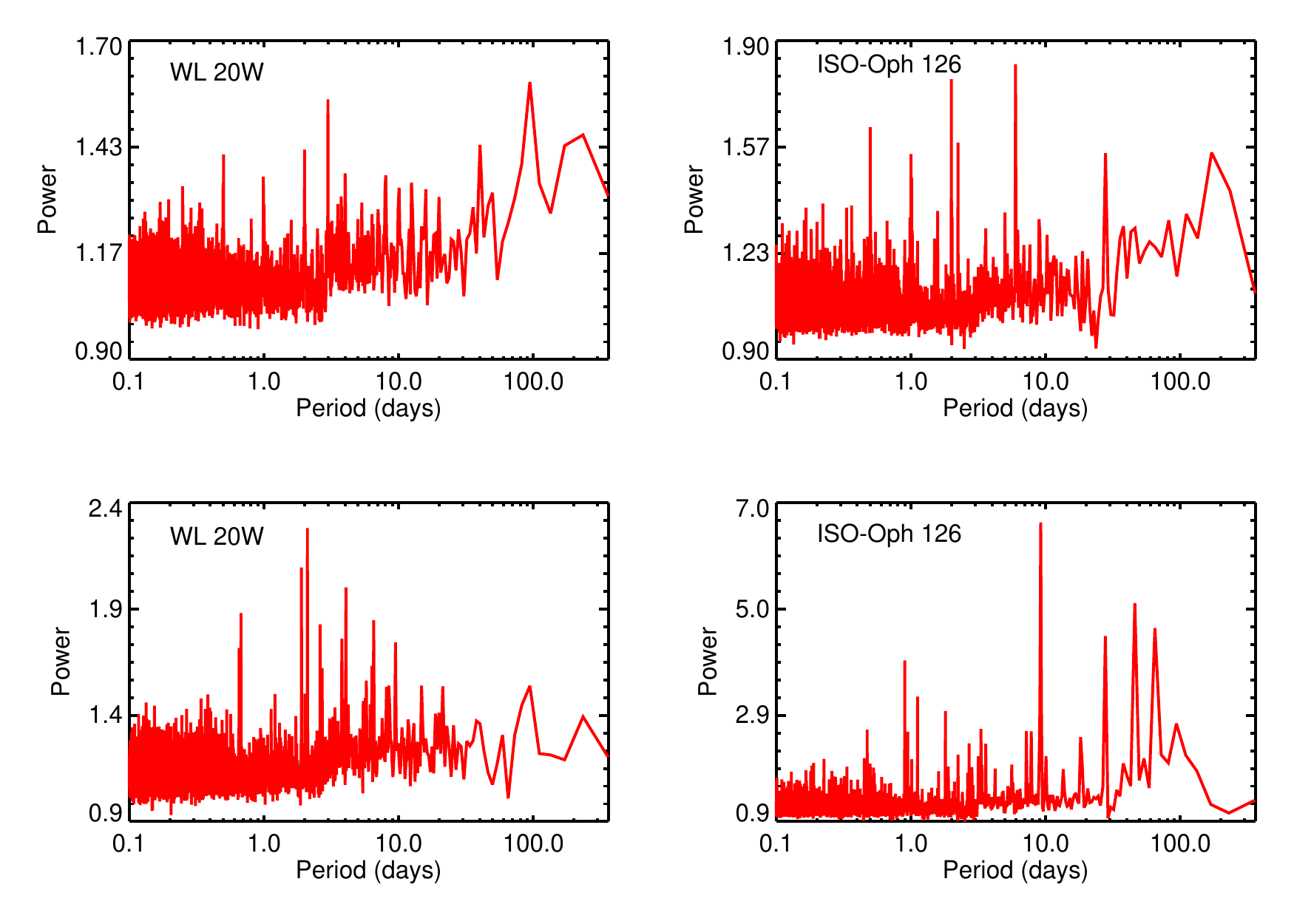}
  \caption{Periodograms for WL 20W and ISO-Oph 126 including and excluding the long time-scale event photometry.  \emph{TOP:} The periodograms from running the PA on the full data set including the large amplitude long time-scale variation.  \emph{BOTTOM:} The periodograms from the PA only on the photometry not affected by the large amplitude variation.  The 2.1026 day period is only seen and significant in the lower periodogram for WL 20.  The same is true for the 9.114 day period of ISO-Oph 126.  Additionally, this period peak power value is nearly 3x more significant than any power value detected using the complete set of photometry.} 
\end{figure} 

\subsection{Measuring Long Time-Scale Variability}
The long temporal baseline of the photometric time-series allows for the analysis of variability on month and year time-scales which are time-scales not well explored for young stars.  Long time-scale ($>$50 days) variability differs from periodic or irregular variability in that the mean flux value may not remain nearly constant from season to season.  In addition, the photometry in one season may systematically brighten or dim while remaining constant in the other two seasons.  Examples of these two phenomenon in the time-series for WL 20W and ISO-Oph 126 are shown in Fig 10.  The intention in this section is to measure the time-scale of the single largest amplitude aperiodic or irregular variation.

Two criteria are used to identify stars exhibiting long time-scale variability.  The first criteria is the difference between the photometric mean magnitude from one season to either of the remaining two seasons must be greater than 3$\sigma$, where $\sigma$ is the average photometric error of the data over the entire temporal baseline (see Fig 23, WL 6).  The second criteria is that the slope in the photometry in at least one season must be greater than $\pm$5$\degree$.  The quality of the line fit determining the slope is assessed by visual inspection.  The motivation for the second criterion is illustrated by WL 14 (see Fig 21).  An obvious decreasing trend in the photometry is seen in the third season, however the sharp flux drop in the second season causes the mean flux between the two seasons to not satisfy the first criterion.  Of the 101 variables, 31 stars (31$\%$) satisfy at least one of these criteria and are designated long time-scale variables (LTVs).

A differencing technique is employed to measure the time-scale over which a LTV changes from one extreme in flux to the other.  Fig 12 provides a visual demonstration of this method.  In the top panel of Fig 12, a gradual dimming over the entire data set is observed.  This global trend is seen in the time-series of 68$\%$ of the LTVs.  Two different types of variability are believed responsible for the global trend and the long time-scale variation.  Removal of the global trends provides an unbiased analysis of the shorter time-scale variation in the time-series superimposed on these trends.  The global trend is a sustained, but small amplitude effect superimposed over the time-series including the larger amplitude, long time-scale variation.  LTVs with these global trends are split evenly with 50$\%$ dimming over time and 50$\%$ brightening.  The amplitude of the global trends range from 7.5 to 330 millimag/year, with a median value of 26 millimag/year.  The median value corresponds to a change in the stellar flux of $\sim$60 millimag over the temporal baseline.  

An accurate time-scale measurement for the largest amplitude variation can be complicated by the presence of small time-scale variability.  The middle panel of Fig 12 shows how the light curve is smoothed with a 50 day moving median filter.  The length of 50 days is chosen by visual inspection of the smoothed light curves; this timescale suppresses the smaller amplitude, shorter time-scale variability while preserving the shape of the long time-scale variation.   The time-scale for the long time-scale variation is set to be the time difference between when the LTV is at one extreme in flux (i.e. brightest state) to the opposite extreme (i.e. dimmest state).  This time-scale is determined by subtracting the smoothed magnitude found at time \emph{i} with the smoothed magnitude found at time \emph{j} using the following:

\begin{equation}
  M_{i,j} = \sum_{i=1}^{N_{obs}}\sum_{j=1}^{N_{obs}}(m_j^* - m_i^*)
\end{equation}

\noindent where \emph{M$_{i,j}$} is the $\Delta$mag between time \emph{j} and time \emph{i}, \emph{m$_{j}$$^{*}$} is the magnitude at time \emph{j}, \emph{m$_{i}$$^{*}$} is the magnitude at time \emph{i} and \emph{N$_{obs}$} is the total number of observations.  The time between the largest $\Delta$mag is recorded as the time-scale.   In many cases, the full time-scale of the variation cannot be measured due to the data sampling.  The bottom panel of Fig 13 shows the quantity \emph{M$_{i,j}$} as a function of times between measurements \emph{i} and \emph{j}.  The ``landscape'' shows multiple peaks each corresponding to various time-scales of variability.  The highest peak is only considered as only the time-scale associated with the greatest change in magnitude (i.e. largest \emph{M$_{i,j}$}) is sought.  Either extreme flux state may fall within a gap in the photometry or outside the date range of observations.  Therefore these time-scales should be treated as lower bounds. The variability time-scales range from 64 to 790 days.  Not all LTVs display only one discrete long time-scale variation.  ISO-Oph 119 clearly shows two distinct long time-scale variations.  For ISO-Oph 119 and similar cases, only the time-scale for the largest amplitude variation is measured.  Figs 27 to 30 contain the \emph{K$_{s}$} light curves for these LTVs.  

\begin{figure}
  \plotone{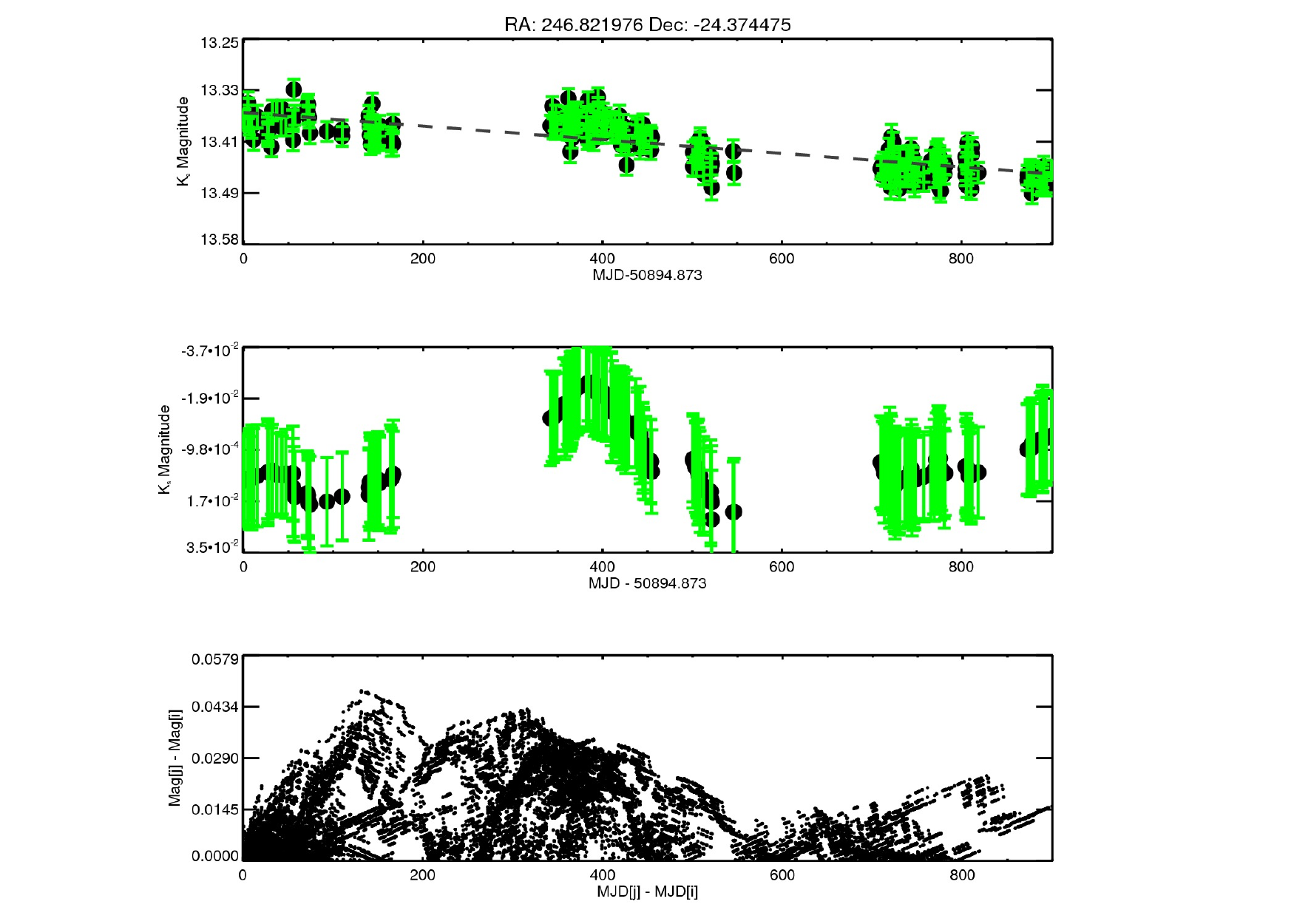}
  \caption{A demonstration of the method used to estimate the variability time-scale of LTVs.  $\emph{Top:}$ The \emph{K$_{s}$} light curves for 2MASS J16271726-2422283.  The gray dashed line is a linear least-squares fit to the data.  $\emph{Middle:}$ The same light curve after the data are smoothed and the linear fit is removed.  The smoothing is done using a moving median filter with a 50 day width.  $\emph{Bottom:}$ This shows the $\Delta$mag as a function of the time between individual photometric measurements, \emph{m$_{j}$$^{*}$} and \emph{m$_{i}$$^{*}$}.  The recorded 132 day time-scale corresponds to highest peak, or largest $\Delta$ mag occurring in the middle plot.  This time-scale describes the star flux decrease from $\sim$400 to $\sim$525 2MASS MJD.}   
\end{figure}

Despite observations spanning $\sim$2.5 years, in most cases it is not possible to conclude whether or not long time-scale variability is periodic.  However, 6 LTVs have photometry suggestive of periodic behavior based on visual inspection of the stars' folded light curves correpsonding to periods ranging from 207 to 589 days.  The light curves are folded to the most significant period found by the PA.  These candidate periodic stars are identified in the first column of Table 6.  These sources are not included with the periodic variables as the found periods are greater than the 200 day confidence limit (see $\S$4.1).

\section{DISCUSSION}
The observational goal of this study is to measure the amplitudes and timescales of stellar variability, particularly in young stars.  This information, in turn, places constraints on the physical mechanisms responsible for the variability.  Empirical methods based on correlations between observed magnitudes and color have been employed to characterize stellar variability of young stars \citep{carpenter01,carpenter02,oliveira08}.  These methods consider variability due to rotational modulation of hot or cool starspots, variable extinction, variable mass accretion and structure changes in the circumstellar environment.  Cool starspots are believed to be caused by localized magnetic inhibition of convection energy transport.  Hot starspots, on the other hand, result from either surface flaring or heating by mass accretion onto the surface along magnetic field lines.  Extinction may occur from asymmetries in an accretion disk or even from isolated dense regions of the parent molecular cloud passing through the line of sight.  Variable mass accretion rates can cause the star brightness to vary through the clearing of the inner circumstellar disk.  In addition, variability may be caused by energy released as material in an accretion disk moves toward a star by viscous processes.  Finally, these mechanisms are not mutually exclusive and are often seen to exist simultaneously \citep{herbst94}.

Each of the above variability mechanisms can be distinguished based on the temporal nature of the variability and correlations between color variability to stellar brightness.\footnote{The correlations between stellar color and brightness are based on models in \citet[and references herein]{carpenter01}.}  The following set of qualitative observables are developed to classify the observed variability and to connect these variations to physical mechanisms.

\begin{itemize}
  
\item Long-lived cool starspots result in periodic variability with periods consistent with the rotational periods of young stars ($\lesssim$ 14 days) \citep{rebull01}.  This variability is often sinusoidal in shape.  At the temperature range of most YSOs, the near-IR wavelength regime samples the Rayleigh-Jeans tail of the stellar energy distribution where the contrast between the starspot and surrounding photosphere is small (e.g. \citet{vrba85}).  Therefore, the (\emph{J}-\emph{H}) and (\emph{H}-\emph{K$_{s}$}) colors should remain constant (within photometric errors) as the brightness varies.
  
\item Variability by hot starspots can either result in periodic or irregular variability.  Long-lived hot starspots caused by accretion onto the stellar surface may result in periodic variability.  However it should be noted that accretion induced hot spots may display aperiodic behavior due to a stochastic accretion rate.  Variability caused by flares will be aperiodic and will have time-scales on the order of hours to days.  As with cool starspots, the period of variability will be consistent with the rotational periods of young stars.  In both cases the affected photosphere should be hotter than the surrounding surface resulting in the star becoming bluer as the star brightens \citep{rodono88,panagi95,yu06}.
  
\item Variable extinction can result in either periodic or long time-scale variability.  Variability caused by asymmetries in the inner circumstellar disk, if present, may be periodic with periods from days to weeks.  Unlike variability caused by starspots, periodic variable extinction need not appear sinusoidal but present more likely as eclipse-like features.  These eclipse-like features are sharp drops or ``dips'' in the stellar flux with a regularity dependent on the observing cadence.  Variability caused by asymmetries in the outer circumstellar disk ($>$ 1 AU) will not be periodic within the temporal baseline of this study due to long period of revolution around the host star.  This variability and variable extinction from inter-cloud material can occur on long time-scales, however as the time-scale depends on the system geometry, there is no expectation as to its duration.  Variable extinction causes the star to redden as the star dims.
  
\item Variability caused by a variable accretion rate within the circumstellar disk is not expected to be periodic.  The time-scale of variability does place constraints on the physics causing this rate change (e.g. disk viscosity, time variable magnetic field) \citep{armitage95,mahdavi98,lai99,terquem00,carpenter01}.  During times of lower accretion rates, the inner disk cools and the inner hole becomes larger.  This, in turn, decreases the contribution of dust reradiation, particularly in the \emph{K$_{s}$} band, to the overall energy budget of the star and circumstellar disk system.  Therefore while the total system flux drops, a larger percentage of emitted radiation is from the star causing the system to become bluer as the system dims.  However, if the inner circumstellar disk edge is dominated by the dust sublimation temperature, a observationally similar effect will result.  In this case, an increased accretion rate raises the star's effective temperature in turn increasing both the distance to the circumstellar disk inner rim and disk vertical height.  The result would be that the system would become brighter as it reddens.  Both physical scenarios produce a qualitatively identical result to the observed correlation between brightness and color. 
  
\end{itemize}

In an attempt to identify the dominant variability mechanism, stars in the variable catalog are placed into subclasses based on the observed shape and time-scale of variability. These subclasses are: periodic, long time-scale and irregular. These classifications along with the above criterion identified the likely dominant variability mechanism for 53 of the 101 stars in the variable catalog.  The type of variability associated with each star is listed in Table 3 and each sub-class is described in the following subsections.  The periodic sub-class accounts for 32$\%$ of the variable catalog with the majority (88$\%$) lying ``on cloud''.  Long time-scale variables make up 31$\%$ of the variable catalog.  All LTVs reside ``on cloud''.  The irregular subclass contains the most members comprising 40$\%$ of the variable catalog.  Only 68$\%$ of irregular variables lie ``on cloud''.  These subclasses are rough descriptions and are by no means mutually exclusive.  For instance, WL 20W and ISO-Oph 126 are placed into both the periodic and long time-scale subclasses.

These criteria do not always allow for the dominant variability mechanism to be identified.  The main reasons preventing an estimate of the mechanism are: the time-scale/period or color correlation is contrary to the above diagnostics, no dominant amplitude variability is clearly evident, or the photometry in \emph{J} and H is below the completeness limits in each band resulting in no useful color information.  Mechanisms appended with a question mark in Table 3 either possess a marginal color correlation via visual inspection, or the diagnostics did not definitively differentiate between proposed mechanisms.

\subsection{Periodic Variables}
The PA identifies 32 of 101 stars (32$\%$) within the variable catalog as periodic with periods ranging from 0.49 to 92.28 days.  Table 4 contains the list of periodic variables.  The light curves for certain subsets of periodic variables are very similar in form when phased to the identified period.  This allows for periodic variables to be separated into two sub-categories: sinusoidal-like and eclipse-like.  Assignment to a particular sub-category is based upon visual inspection of the folded light curve in the band with the highest signal-to-noise.  

\begin{deluxetable}{c c c c c c c c}
  \tablecolumns{8}
  \tablewidth{0pc}
  \tablecaption{Periodic Variables}
  \tabletypesize{\scriptsize}
  \tablehead{
    \colhead{Catalog ID\tablenotemark{a}} & \colhead{Period\tablenotemark{b}} & \colhead{$\Delta$K$_{s}$} & \colhead{$\Delta$(\emph{J}-\emph{H})} & \colhead{$\Delta$(H-K$_s$)} & \colhead{YSO Class} & \colhead{Sub-Category} & \colhead{Var. Mech.\tablenotemark{c}}\\
    \colhead{} & \colhead{(days)} & \colhead{(mag)} & \colhead{(mag)} & \colhead{(mag)} & \colhead{} & \colhead{} & \colhead{}
  }
  \startdata
  ISO-Oph 83&25.554$\pm$0.071&0.476&0.235&0.173&---&Sinusoidal&Extinction\\
  YLW 1C&5.7753$\pm$0.0085&0.292&0.140&0.257&II&Sinusoidal&Hot Starspot(s)\\
  &5.9514$\pm$0.0014&0.29&---&---&&Eclipse&Extinction\\
  ISO-Oph 96&3.5285$\pm$0.0032&0.084&0.051&0.065&III&Sinusoidal&Cool Starspot(s)\\
  ISO-Oph 97&14.520$\pm$0.088&0.086&0.047&0.050&III&Sinusoidal&Cool Starspot(s)\\
  ISO-Oph 98&5.9301$\pm$0.0092&0.411&0.232&0.329&II&Sinusoidal&Cool Starspot(s)\\
  ISO-Oph 100&3.682$\pm$0.002&0.337&---&---&II&Sinusoidal&Unknown\\
  ISO-Oph 102&3.02173$\pm$0.00044&0.224&0.128&0.099&II&Eclipse&Cool Starspot(s)?\\
  ISO-Oph 106&3.4370$\pm$0.0012&0.508&0.242&0.204&II&Eclipse&Extinction\\
  WL 10&2.4149$\pm$0.0027&0.330&0.137&0.102&II&Sinusoidal&Cool Starspot(s)?\\
  WL 15&19.412$\pm$0.085&1.636&---&0.628&I&Sinusoidal&Unknown\\
  WL 11&3.0437$\pm$0.0038&0.729&0.346&0.402&II&Sinusoidal&Hot Starspot(s)?\\
  65744-2504017&0.83141$\pm$0.00030&0.339&0.157&0.112&---&Sinusoidal&Hot Starspot(s)?\\
  71513-2451388&8.004$\pm$0.046&0.330&0.213&0.111&---&Eclipse&Extinction\\
  WL 20W&2.1026$\pm$0.0060&0.213&0.258&0.305&II&Sinusoidal&Cool Starspot(s)\\
  YLW 10C&2.9468$\pm$0.0029&0.356&---&---&II&Eclipse&Extinction?\\
  &3.0779$\pm$0.0025&0.28&---&---&&Sinusoidal&Cool Starspot(s)?\\
  71836-2454537&2.7917$\pm$0.0017&0.807&0.354&0.327&---&Sinusoidal&Hot Starspot(s)?\\
  ISO-Oph 126&9.114$\pm$0.090&0.135&---&0.371&III&Sinusoidal&Cool Starspot(s)\\
  ISO-Oph 127&6.365$\pm$0.014&0.528&---&0.009&I&Sinusoidal&Cool Starspot(s)\\
  WL 4&65.61$\pm$0.40&0.640&0.161&0.186&II&Inverse Eclipse&Circumbinary Disk\\
  YLW 13A&7.0270$\pm$0.0056&0.092&0.057&0.071&III&Sinusoidal&Cool Starspot(s)\\
  ISO-Oph 133&6.354$\pm$0.011&0.062&---&0.056&III&Sinusoidal&Cool Starspot(s)\\
  ISO-Oph 135&5.536$\pm$0.019&0.130&0.044&0.071&III&Sinusoidal&Cool Starspot(s)\\
  72463-2429353&6.581$\pm$0.012&0.094&---&0.327&0&Sinusoidal&Cool Starspot(s)\\
  72533-2506211&0.485143$\pm$0.000050&0.402&0.272&0.295&---&Sinusoidal&Unknown\\
  ISO-Oph 139&3.7202$\pm$0.0041&0.098&---&0.200&I&Sinusoidal&Cool Starspot(s)\\
  YLW 16C&1.14182$\pm$0.00043&0.318&0.185&0.207&II&Sinusoidal&Cool Starspot(s)\\
  72658-2425543&2.9602$\pm$0.0013&0.211&0.099&0.096&II&Eclipse&Extinction\\
  &1.52921$\pm$0.00065&0.17&---&---&&Sinusoidal&Cool Starspot(s)\\
  72706-2432175&18.779$\pm$0.099&0.090&---&0.478&---&Sinusoidal&Cool Starspot(s)\\
  WL 13&23.476$\pm$0.077&0.305&0.121&0.082&II&Sinusoidal&Cool Starspot(s)\\
  YLW 16A&92.28$\pm$0.84&0.784&---&0.549&I&Inverse Eclipse&Circumbinary Disk\\
  ISO-Oph 149&1.24505$\pm$0.00039&0.067&0.057&0.057&III&Sinusoidal&Cool Starspot(s)\\
  ISO-Oph 148&3.5548$\pm$0.0039&0.058&0.054&0.071&III&Sinusoidal&Cool Starspot(s)\\
  \enddata
  \tablenotetext{a}{The catalog ID has been truncated by 2MASS J162 for 2MASS catalog stars}
  \tablenotetext{b}{The FAP for all periods are $<$1$\%$}
  \tablenotetext{c}{A question mark denotes variability mechanisms that are uncertain due to insufficient color information}
\end{deluxetable}

\subsubsection{Sinusoidal-like Periodic Variables}
Figs 13 to 16 contain the \emph{K$_{s}$} light curves for sinusoidal-like periodic variables.  This sub-category of sinusoidal-like periodic variables includes the most periodic variables (25 stars) with periods ranging from 0.49 to 25.55 days.  The peak-to-trough $\Delta$K$_{s}$ amplitudes range from 0.06 to 1.64 mag, with a median value of 0.29 mag.  The peak-to-trough $\Delta$(\emph{H}-\emph{K$_{s}$}) color amplitudes range from 0.01 to 0.63 mag, with a median value of 0.19 mag.  Typically, the light curve folded to the most significant period shows only one sinusoidal cycle.  However, four stars (ISO-Oph 100, WL 10, WL 13, YLW 13A) show what could be interpreted as a second cycle at half the frequency (i.e. double the period).

\begin{figure}
  \plotone{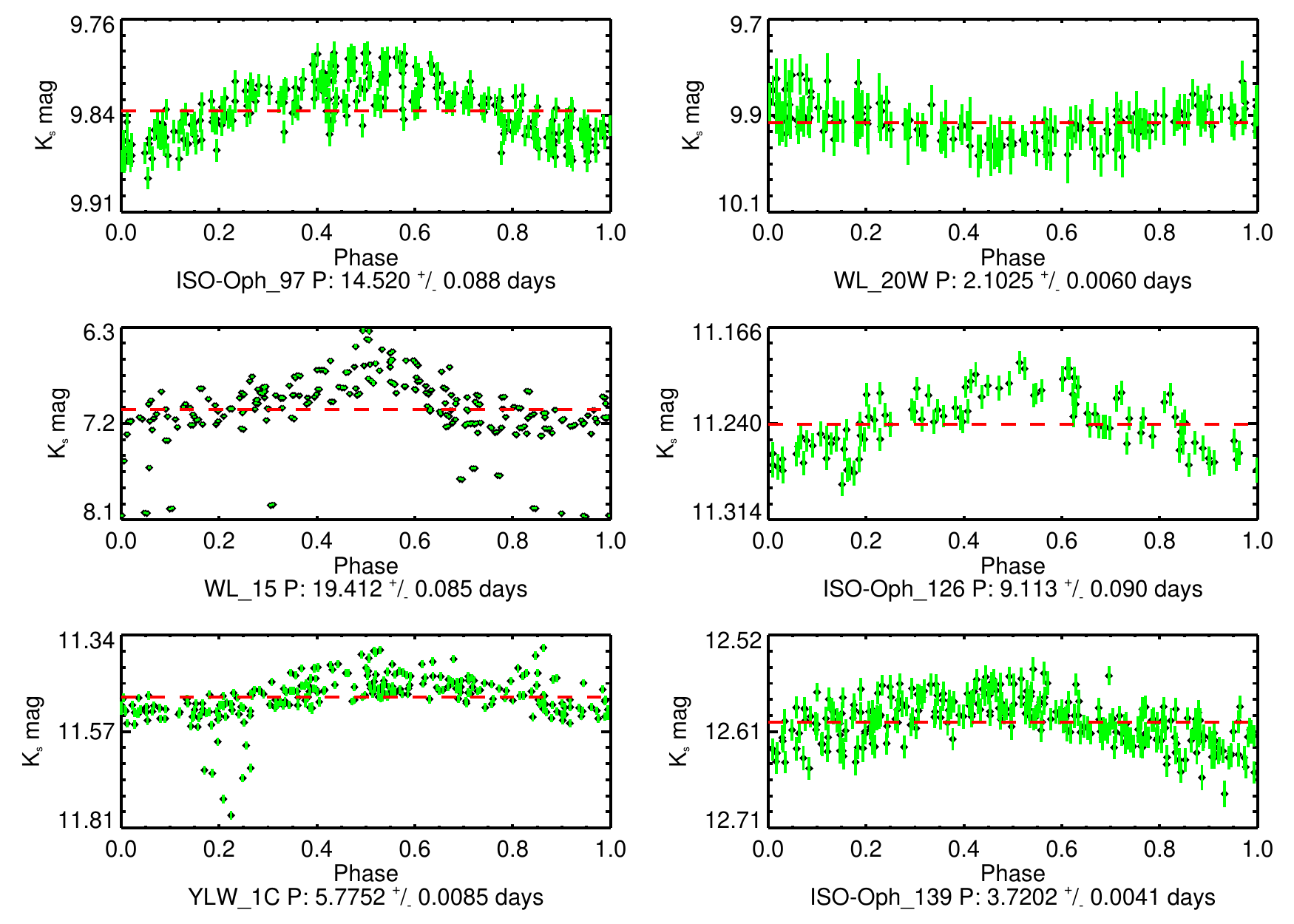}
  \caption{The folded \emph{K$_{s}$} light curves for 6 sinusoidal-like periodic variables.  The red line indicates the star's mean magnitude.}
\end{figure}

\begin{figure}
  \plotone{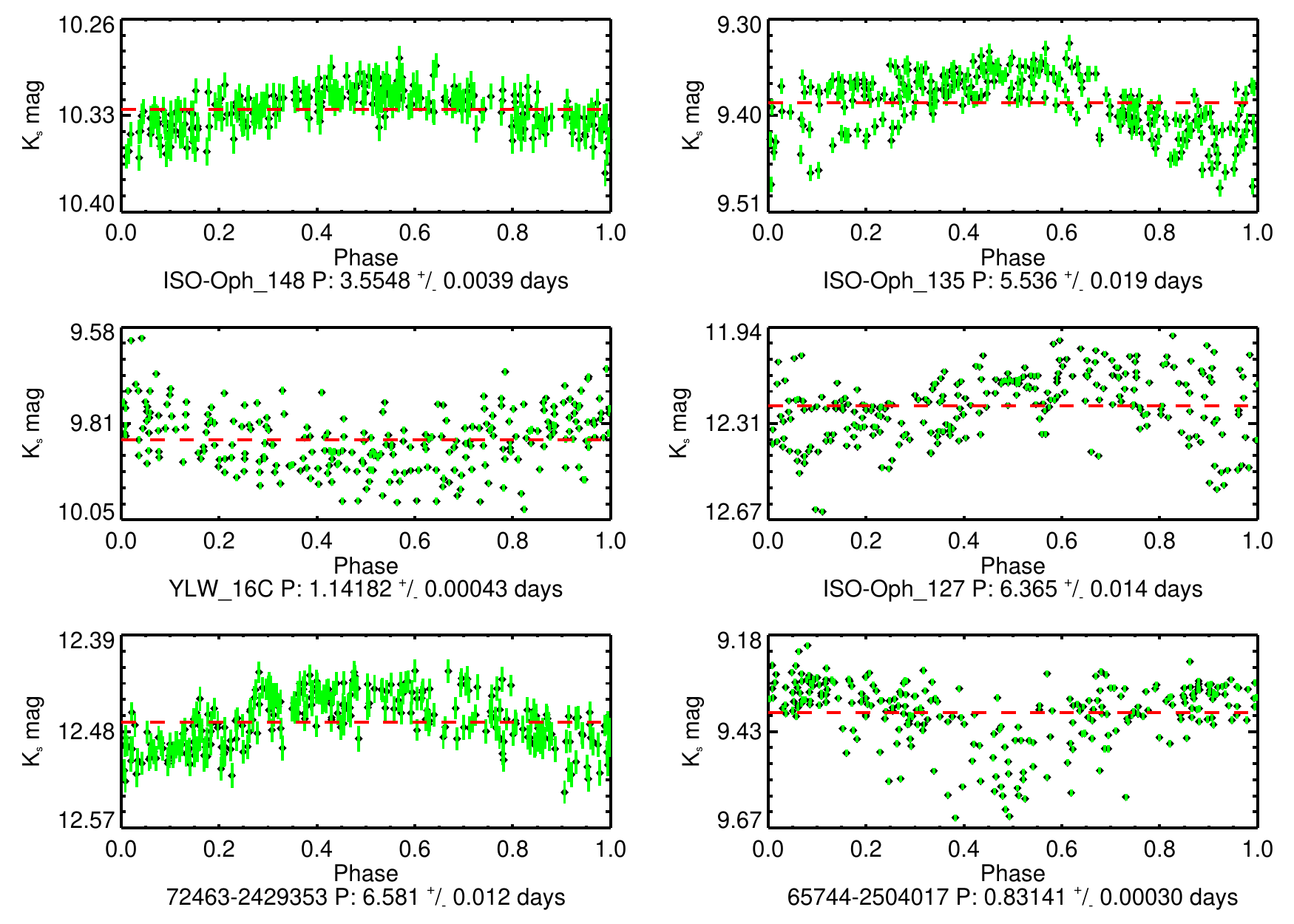}
  \caption{The folded \emph{K$_{s}$} light curves for 6 sinusoidal-like periodic variables.  The catalog name for stars labeled with a 2MASS designation have been truncated by 2MASS J162.  The red line indicates the star's mean magnitude.}
\end{figure}

\begin{figure}
  \plotone{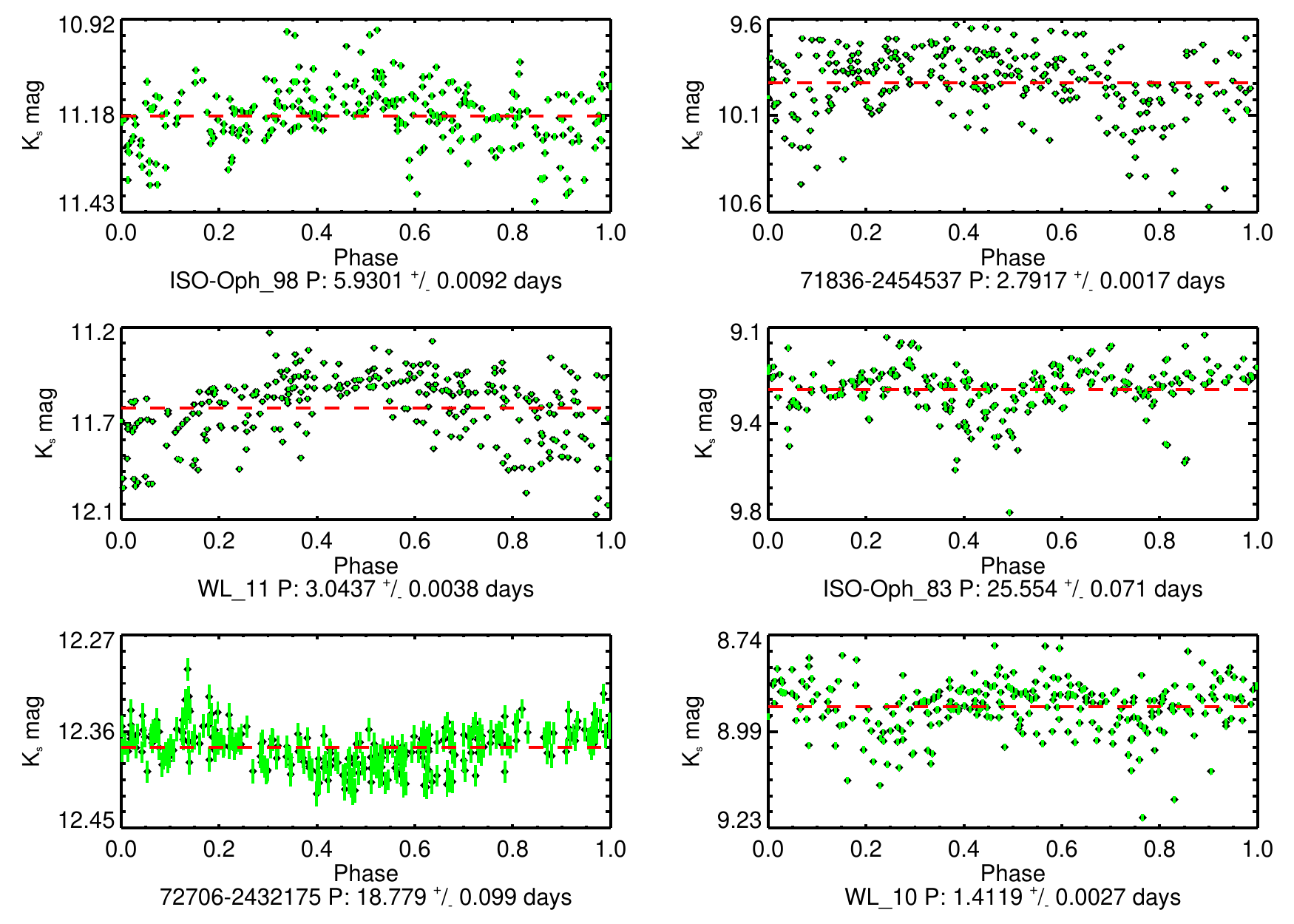}
  \caption{The folded \emph{K$_{s}$} light curves for 6 sinusoidal-like periodic variables.  The catalog name for stars labeled with a 2MASS designation have been truncated by 2MASS J162.  The red line indicates the star's mean magnitude.}
\end{figure}

\begin{figure}
  \plotone{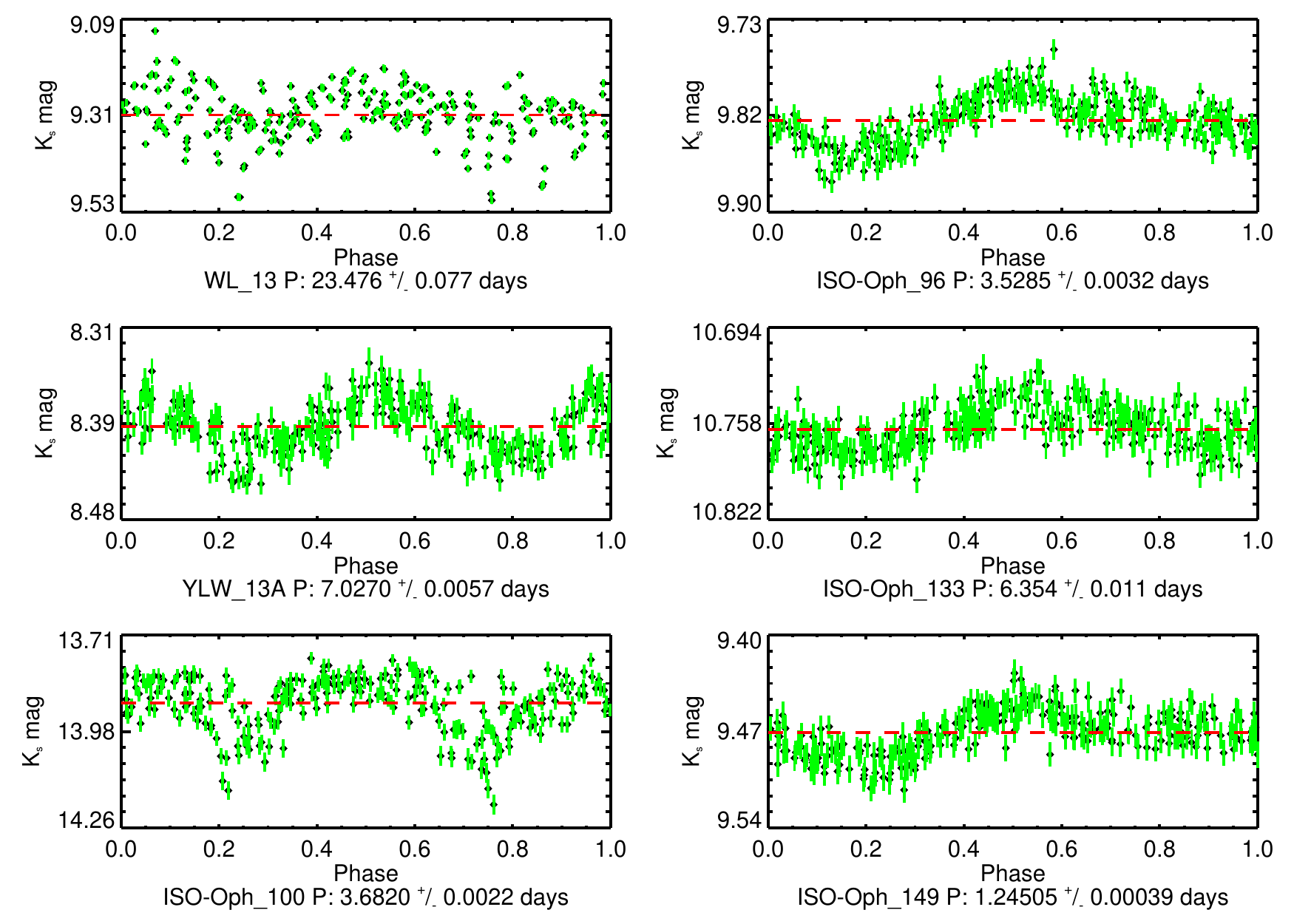}
  \caption{The folded \emph{K$_{s}$} light curves for 6 sinusoidal-like periodic variables.  The red line indicates the star's mean magnitude.}
\end{figure}

Probable variability mechanisms are identified for these stars by applying the criteria discussed in $\S$5.  Correlations are qualitatively examined between the ensemble \emph{K$_{s}$} photometry and stellar colors when folded to the star's period.  Fig 17 illustrates examples of these interpretations for both cool and hot starspots.  For 18 of the sinusoidal-like periodic variables (72$\%$), the variability in the (\emph{J}-\emph{H}) and (\emph{H}-\emph{K$_{s}$}) colors does not correlate with the \emph{K$_{s}$} variability.  This favors rotational modulation by cool starspots as the dominant variability mechanism.  For 3  sinusoidal-like periodic variables (12$\%$), the (\emph{J}-\emph{H}) and (\emph{H}-\emph{K$_{s}$}) colors become bluer as the star brightens.  This is consistent with the behavior expected from rotational modulation by an accretion induced hot starspot.  The (\emph{J}-\emph{H}) and (\emph{H}-\emph{K$_{s}$}) colors become redder as the star dims for 1 (4$\%$) sinusoidal-like periodic variable.  This favors variable extinction as the dominant variability mechanism.  Finally for the remaining 3 sinusoidal-like periodic variables (12$\%$), no dominant variability mechanism could be assigned using the adopted criteria.

\begin{figure}
  \plotone{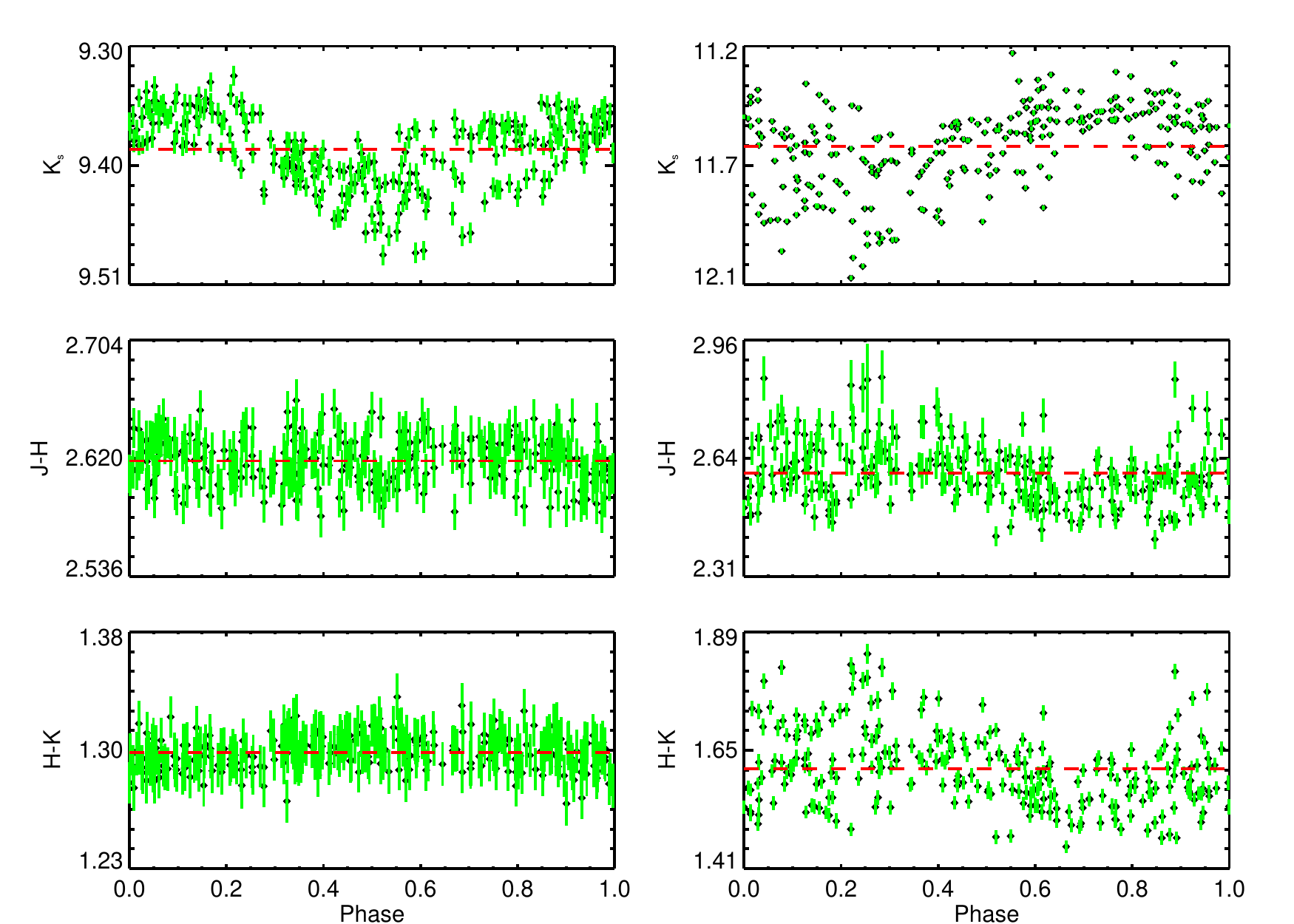}
  \caption{\emph{Left}: The folded \emph{K$_{s}$} and color curves for the sinusoidal-like periodic variable ISO-Oph 135. The lack of near-IR color changes with brightness changes favors cool starspots as the variability mechanism.  \emph{Right}: The folded \emph{K$_{s}$} and color curves for the sinusoidal-like periodic variable WL 11.  The stellar color becomes bluer as the \emph{K$_{s}$} photometry becomes brighter.  This favors rotational modulation of accretion induced hot starspots as the dominant mechanism.  The red line in each plot indicates the mean value.}
\end{figure}

 All sinusoidal-like periodic variables except 3 (2MASS J1625744-2504017, 2MASS J16271836-2454537 and 2MASS J16272533-2506211) are located ``on cloud''.  This sub-category contains 19 stars with a YSO classification: 3 Class I (25$\%$), 8 Class II (24$\%$) and 8 Class III (73$\%$).\footnote{The percentages indicate the percentage of variable stars in each class that are sinusoidal-like periodic variables.}  The variability mechanism for 2 of the Class I stars is cool starspots, while the mechanism could not be identified for the third.  Of the Class II stars, 5 vary due to cool starspots, 1 from an accretion-induced hot starspot and 1 is unknown.  The variability of all Class III stars is caused by cool starspots.

Two key points can be made by analyzing the 18 sinusoidal-like periodic variables where cool starspots is the believed variability mechanism.  First, cool starspots on young stars persist on preferential longitudes on year timescales.  This is evidenced by a lack of phase drift in the folded light curves.  This phenomenon of preferential or \emph{active} longitudes have been associated with a number of chromospherically active stars (e.g. RS CVns, FK Com) \citep{strassmeier88,zeilik88,henry95,jetsu96}.  Second the variability amplitude due to cool starspots changes on much shorter timescales as evidenced by the significant scatter within the folded light curves.  This amplitude variability is likely caused by an evolving starspot covering factor and/or starspot temperature.  The covering factor is defined as the area of the observed stellar disk covered by the starspot(s).

ISO-Oph 96, ISO-Oph 133, ISO-Oph 149 and 2MASS J16272533-2506211 differ from the remaining sinusoidal-like periodic variables as the light curves for these 4 stars are asymmetric (i.e. they have a sharp increase in flux then decrease more slowly).  The (\emph{J}-\emph{H}) and (\emph{H}-\emph{K$_{s}$}) color variability are not correlated to the \emph{K$_{s}$} variability for the first three asymmetric sinusoidal-like periodic variables.  This favors a dominant variability mechanism of rotational modulation by cool starspots.  Asymmetric light curves have been observed for both WTTS and chromospherically active dwarf stars \citep{cutispoto01,cutispoto03,grankin08,frasca09}.  In both cases, the variability is believed to be caused by magnetically generated cool starspots. ``Reverse'' asymmetric light curves with a slow rise in source flux followed by a steep drop are also observed.  \citet{frasca09} is able to closely model the \emph{RIJH} asymmetric light curves of the WTTS V1529 Ori by rotating a stellar surface with two cool starspots of unequal areas separated by $\sim$130$\degree$ in longitude.  The size of the leading cool starspot determines if a ``forward'' or ``reverse'' asymmetric light curve is seen.

2MASS J16272533-2506211, hereafter designated 'J211', is peculiar due to its unique and difficult to interpret brightness and color variations.  Fig 18 contains \emph{K$_{s}$}, (\emph{J}-\emph{H}) and (\emph{H}-\emph{K$_{s}$}) photometry for J211 folded to P = 0.485143 $\pm$ 0.000050 days.  This is the shortest period star among the periodic variables.  The peak-to-trough \emph{K$_{s}$} amplitude is 0.40 mag and the $\Delta$(\emph{H}-\emph{K$_{s}$}) color amplitude is 0.30 mag.  The (\emph{J}-\emph{H}) color for J211 clearly becomes bluer for a phase duration of $\sim$0.3 ($\sim$3.5 hrs) centered approximately on the times of maximum brightness.  Somewhat surprisingly, however, no similar variation is seen in the (\emph{H}-\emph{K$_{s}$}) color during the same period.  The data is deemed reliable as the \emph{J} and \emph{H} photometry are significantly brighter than the survey completeness limits.  The variability mechanism is not identified for J211 as this \emph{K$_{s}$}-color behavior is inconsistent with any criteria discussed in $\S$5.  The shape of the light curve coupled with the short period suggests J211 might be a RR Lyrae variable.  However the peculiar color behavior is not expected in these stars. 

\begin{figure}
  \plotone{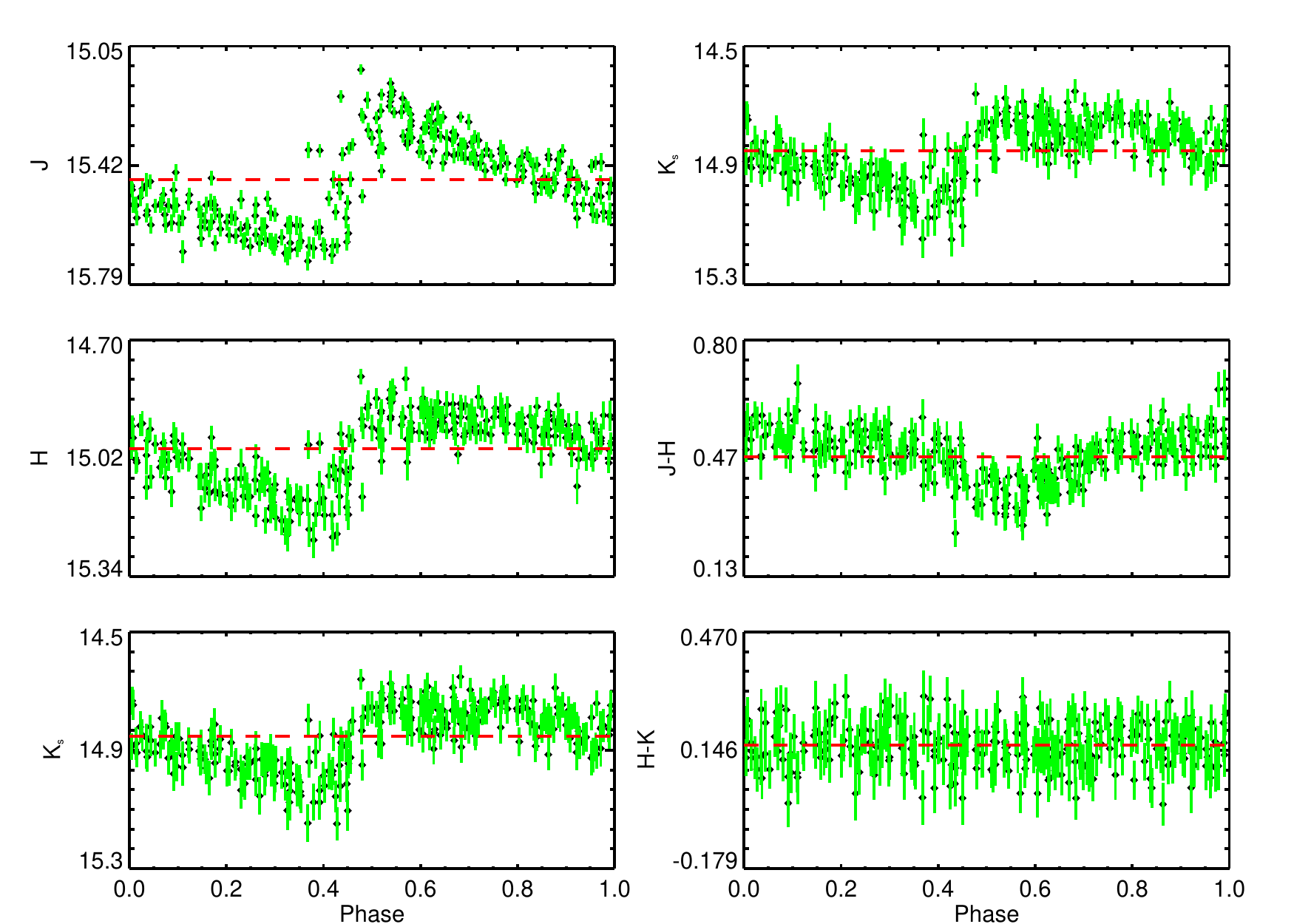}
  \caption{\emph{Left}: The \emph{J}, \emph{H} and \emph{K$_{s}$} light curves of 2MASS J16272533-2506211 folded to a period of 0.485143 $\pm$ 0.000050 days.  \emph{Right}: The \emph{K$_{s}$}, (\emph{J}-\emph{H}) and (\emph{H}-\emph{K$_{s}$}) light and color curves folded to the above period.  The significant difference in form of the \emph{J} folded light curve to the other two suggests the cause for the variability in the (\emph{J}-\emph{H}) color arises mainly from the \emph{J} band.  The red line in each plot indicates the mean value in each case.}
\end{figure}

\subsubsection{Eclipse- and Inverse Eclipse-like Periodic Variables}
Eclipse-like periodic variables possess photometry containing sharp periodic drops in source flux.  The duration of these drops, or possibly ``eclipses'', is in all cases less than a phase of 0.3 when the photometry is folded to the most significant period.  Fig 19 contains the folded light curves for the 6 eclipse-like periodic variables.  These eclipse-like periodic variables are assigned to this subclass by visual inspection; it is not possible to confidently state that these sharp changes in photometry are \emph{bona fide} occultations of star light (i.e. true eclipses).

The periods for the eclipse-like periodic variables range from 2.95 to 8.00 days.  The durations of these eclipses range from 5.8 to 12.7 hours.  The $\Delta$\emph{K$_{s}$} amplitudes range from 0.21 to 0.51, with a median value of 0.31 mag.  The $\Delta$(\emph{H}-\emph{K$_{s}$}) color amplitude range from 0.10 to 0.25 mag, with a median value of 0.11 mag.  These amplitudes in both magnitude and color represent the total change in stellar flux and they do not necessarily represent eclipse depths since there is considerable scatter in the out-of-eclipse photometry.  The eclipse depths and how they are determined are described below.  The variability mechanism for the eclipse is determined in the same manner as with the sinusoidal-like periodic variables.  Correlations between the \emph{K$_{s}$} photometry and stellar colors (\emph{J}-\emph{H}) and (\emph{H}-\emph{K$_{s}$}) for the eclipse event are assessed visually and compared with the criteria discussed in $\S$5.  For 4 (66$\%$) eclipse-like periodic variables, the (\emph{J}-\emph{H}) and (\emph{H}-\emph{K$_{s}$}) colors become redder as the star dims.  The color correlation coupled with the short, periodic behavior favor extinction, possibly by the inner region of a circumstellar disk, as the dominant variability mechanism causing the eclipse.  The variability in the colors during the eclipse for ISO-Oph 102 are not correlated with the \emph{K$_{s}$} photometry consistent with variability caused by rotational modulation of cool starspots.  Unfortunately, both the \emph{J} and \emph{H} photometry for YLW 10C are dimmer than the survey completeness limits and the lack of color information prevents confident identification of a variability mechanism.  All 6 eclipse-like periodic variables are classified as YSO Class II (15$\%$)\footnote{The percentage indicates the percentage of variable Class II stars that are eclipse-like periodic variables}.  All stars in this sub-category are located ``on cloud'' except 2MASS J16271513-2451388.

\begin{figure}
  \plotone{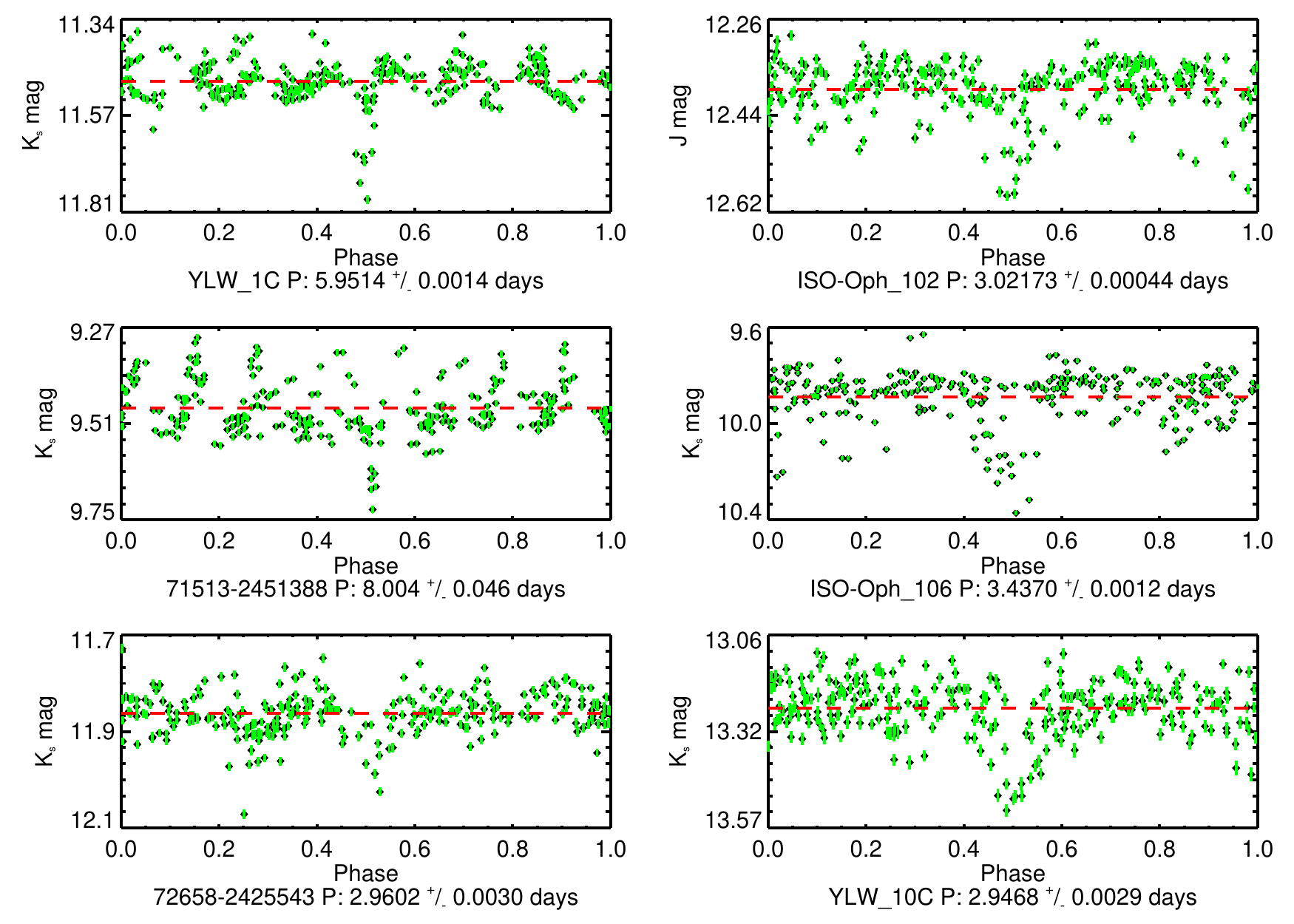}
  \caption{The folded \emph{J} and \emph{K$_{s}$} light curves for 6 eclipse-like periodic variables.  The catalog name for stars labeled with a 2MASS designation have been truncated by 2MASS J162.  The red line indicates the star's mean magnitude}
\end{figure}

\citet[YSOVAR]{morales11} qualitatively identified a number of similar eclipse-like variables in their mid-IR variability survey of YSOs within the Orion Nebula Cluster.  The survey found 38 stars exhibiting brief, sharp drops in stellar flux.  These stars are identified as AA Tau or ``dipper'' variables.  The variability mechanism is believed to be due to high latitude warps in the inner accretion disk periodically occulting the star \citep{bertout00,bouvier03}.  The 4 eclipse-like variables that show evidence of extinction could be considered AA Tau variables, and possibly all 6 eclipse-like systems.

Under the assumption all 6 eclipse-like periodic variables are AA Tau variables, constraints on the spatial location within the circumstellar disk and size of the hypothetical occulter are investigated.  \citet{marsh10} performed a deep mid-IR imaging survey of the $\rho$ Oph 2MASS Calibration field used in this work.  They computed the T$_{eff}$, A$_{V}$ and mass for 5 of the 6 eclipse-like periodic varibles.  This was done by fitting model spectra to observed SEDs.  The SEDs were computed from photometry in the \emph{J}, \emph{H}, \emph{K}, [3.5] and [4.5] bands.  The model spectra were obtained using the COND, DUSTY and NextGen models.  T$_{eff}$ and A$_{V}$ are found by minimizing the following equation:

\begin{equation}
  \phi(T_{eff},\alpha,A_{V}) = \sum_{\lambda=1}^{5}\frac{1}{\sigma_{\lambda}^{2}}[\emph{f}_{\lambda}^{obs}-\alpha10^{-0.4r_{\lambda}A_{V}}\emph{f}_{\lambda}^{mod}(T_{eff})]^{2}-A_{V}\beta
\end{equation}

\noindent where $\alpha$ is a flux scaling factor, \emph{f}$_{\lambda}^{obs}$ and \emph{f}$_{\lambda}^{obs}$(T$_{eff}$) are the respective observed and model fluxes at wavelength $\lambda$, $\sigma_{\lambda}$ is the flux uncertainty, r$_{\lambda}$ is the absorption at wavelength $\lambda$ relative to A$_{V}$ and $\beta$ is a constant penalty parameter.  In the Rayleigh-Jeans regime a degeneracy exists between T$_{eff}$ and A$_{V}$ such that a high temperature star seen through low extinction will have a SED similar to that of a low temperature star seen through high extinction.  The parameter $\beta$ is used to break this degeneracy by penalizing solutions with low values of A$_{V}$.  This parameter was optimized by using the published photometry of a large sample of spectroscopically confirmed brown drawfs.  Values of A$_{V}$ have errors between 1 to 2.7 while T$_{eff}$ is accurate to within 860 K.  The COND and DUSTY models then yield a model-unique mass for each star.  They used a mass-temperature relationship to derive the mass for hotter stars fit by the NextGen models.  They conclude an accuracy in the mass estimate to within a factor of $\sim$2-3.

For the case of ISO-Oph 106, a 1 Myr isochrone from \citet{siess00} is used to determine the T$_{eff}$ and radius for ISO-Oph 106 by assuming a mass of 0.5 M$_{\sun}$.  The stellar radius for the remaining eclipse-like periodic variables is computed by $R_{\star} = (\frac{L_{\star}}{T_{eff}^{4}})^{\frac{1}{2}}$, where each quantity is in solar units.  The stellar luminosity, L$_{\star}$, is computed using the prescription outlined in \citet{natta06} except in the case of YLW 10C.  The prescription relates the stellar luminosity as a function of \emph{J} magnitude and extinction A$_{J}$.   The extinction is computed using the \emph{(J-H)} and \emph{(H-K)} colors corrected into the CIT system using the $\rho$ Oph extinction law by \citet{kenyon98} and the CTTS locus defined by \citet{meyer97}.  The \emph{J} band photometry for YLW 10C is below the survey completeness limit.  The luminosity for this star is found using the 1 Myr isochrone mentioned above and the mass determined by \citet{marsh10}.  The estimated mass, T$_{eff}$, A$_{V}$, and R$_{\sun}$ are in Table 5.

\begin{deluxetable}{c c c c c c c c}
  \tablecolumns{8}
  \tablewidth{0pc}
  \tablecaption{Summary of Eclipse-Like Periodic Variable Characteristics}
  \tabletypesize{\scriptsize}
  \tablehead{
    \colhead{Star} & \colhead{L} & \colhead{M} & \colhead{T$_{eff}$} & \colhead{R$_{\star}$} & \colhead{a} & \colhead{D} & \colhead{$\Delta$K$_{s}$}\\
    \colhead{} & \colhead{(L$_{\sun}$)} & \colhead{(M$_{\sun}$)} & \colhead{(K)} & \colhead{(R$_{\sun}$)} & \colhead{(R$_{\star}$)} & \colhead{(R$_{\star}$)} & \colhead{(mag)}
  }
  \startdata
  YLW 1c&1.81&1.16&4738&2.01&7.21&2.04&0.29\\
  2MASS J16271513-2451388&0.52&0.11&3033&2.63&3.09&0.58&0.26\\
  2MASS J16272658-2425543&0.062&0.043&2783&1.07&2.84&2.05&0.17\\
  ISO-Oph 102&0.22&0.059&2888&1.88&1.82&1.37&0.12\\
  ISO-Oph 106&0.88&0.50&3769&2.06&3.69&3.48&0.51\\
  YLW 10c&0.746&0.35&3901&1.90&3.21&3.63&0.28\\
  \enddata
\end{deluxetable}

Assuming the occulter has negligible mass and orbits under Keplerian rotation, the occulter's distance from the host star can be computed for each eclipse-like periodic variable.  The diameter of the occulter is computed from the duration of the eclipse event and its location in the circumstellar disk.  Strictly speaking, the computed diameters are along the orbital path.  No assumption concerning the occulter geometry (i.e. spherical, ellipsiod) is made.  This diameter is considered a strict lower bound as the disk geometry, occulter impact parameter and occulter opacity are unknown.  The eclipse depth, however, is determined by first removing the eclipse-event photometry from the time-series data (see $\S$4.1).  The eclipse depth is then determined to be the difference between the maximum magnitude in the eclipse feature relative to a median non-eclipse mean magnitude.  This calculation results in occulter distances that range between 1.83 to 7.21 R$_{\star}$, with a median value of 3.20 R$_{\star}$.  The occulter size ranges from 0.58 to 5.07 R$_{\star}$, with a median value of 2.05 R$_{\star}$.  The range in $\Delta$K$_{s}$ eclipse depth is 0.12 to 0.51 mag, with a median value of 0.27 mag.  Table 5 summarizes the results of this investigation.

Inverse eclipse-like periodic variables are similar to eclipse-like periodic variables, but the ``eclipse'' is an increase in source flux rather than a decrease.  Fig 20 contains the folded \emph{K$_{s}$} and color curves for WL 4 and YLW 16A, the 2 inverse eclipse-like periodic variables in the variable catalog.  WL 4 is a Class II YSO whose period of variability is 65.61 $\pm$ 0.40 days.  The peak-to-trough $\Delta$K$_{s}$ amplitude id 0.67 mag and the peak-to-trough $\Delta$(\emph{H}-\emph{K$_{s}$}) color amplitude is 0.19 mag.  The (\emph{J}-\emph{H}) color for WL 4 becomes redder as the star brightens during the inverse eclipse event.  However, the (\emph{J}-\emph{H}) color change starts just prior and ends just after the inverse eclipse event.  The difference in each case is $\sim$0.1 in phase or $\sim$6.6 days.  The (\emph{H}-\emph{K$_{s}$}) color is not correlated with the \emph{K$_{s}$} variability.  

The period of variability for YLW 16A is longer than WL 4 at 92.28 $\pm$ 0.84 days.  The amplitudes of variability for YLW 16A, a Class I YSO, are also larger with peak-to-trough $\Delta$K$_{s}$ and $\Delta$(\emph{H}-\emph{K$_{s}$}) amplitudes of 0.95 and 0.34 mag, respectively.  As the \emph{J} band photometry is dimmer than the survey completeness limits, no reliable (\emph{J}-\emph{H}) color information is available.  The (\emph{H}-\emph{K$_{s}$}) color variability is sinusoidal-like, but is not aligned with the \emph{K$_{s}$} variability.  Both stars reside ``on cloud''.

\begin{figure}
  \plotone{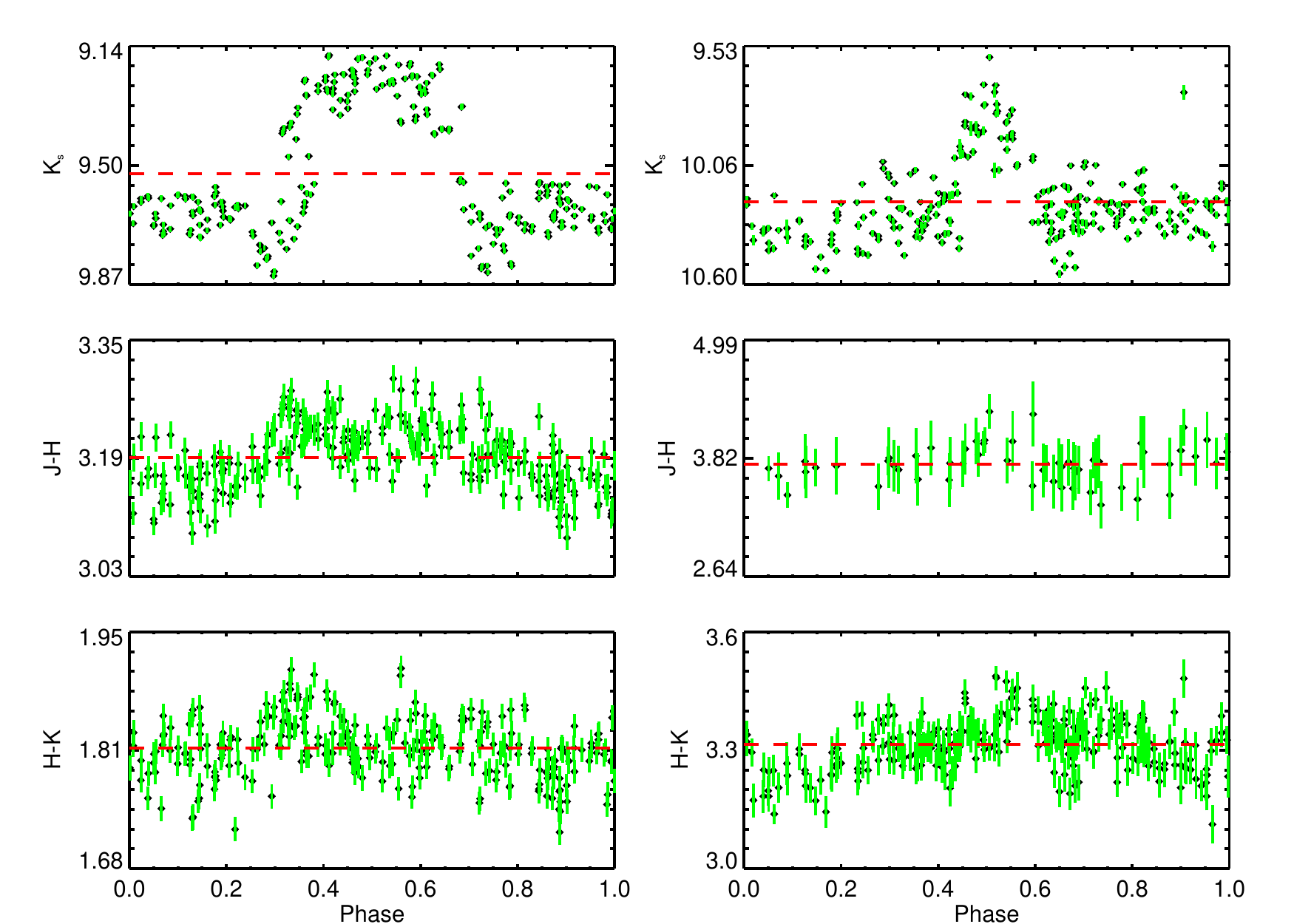}
  \caption{The folded \emph{K$_{s}$} and color curves for the inverse eclipse-like periodic variables WL 4 (P = 65.6 days) and YLW 16A (P = 92.3 days).  The red line in each plot indicates the mean value in each case.}
\end{figure}
 
The variability mechanism for both WL 4 and YLW 16A is believed to be related, and similar to the interpretations proposed in separate letters \citep{plavchan08a,plavchan13}.  Here the proposed variability mechanism is summarized.  Both systems contain a visual binary companion detected through high resolution direct imaging \citep{ratzka05,plavchan13}.  The two visible components for WL 4 are separated by 0.176$\arcsec$ and separated by 0.3$\arcsec$ in the case of YLW 16A.  This corresponds to a projected linear separations of 23 AU and 39 AU, respectively, given a mean distance of 129 pc (see $\S$3.4).  In each system, the large amplitude variability is believed to be intrinsic to one of the visible pair.  This component is, in turn, hypothesized to be a close binary surrounded by a circumbinary disk; this system is thus a triple system.  The influence of the wide companion has caused the plane of the circumbinary disk to be inclined to the orbital plane of the inner binary.  The variability results when each component of the inner binary is periodically obscured by the circumbinary disk as the binary orbits around the barycenter.  \citet{kusakabe05} proposed a similar model to explain the variability for KH-15D. 

\subsection{Long Time-Scale Variables}
The largest amplitude variability in long time-scale variables is not observed to be periodic, but show consistent trends, unlike irregular variables.  The \emph{K$_{s}$} light curves are shown in Fig 21 to 25; all 31 LTVs (31$\%$ of the variable catalog) are listed in Table 7.

\begin{deluxetable}{c c c c c c c}
  \tablecolumns{7}
  \tablewidth{0pc}
  \tablecaption{Time-Scale Variables}
  \tabletypesize{\scriptsize}
  \tablehead{
    \colhead{Catalog ID\tablenotemark{a}} & \colhead{Time-Scale} & \colhead{$\Delta$K$_{s}$} & \colhead{$\Delta$(\emph{J}-\emph{H})} & \colhead{$\Delta$(\emph{H}-\emph{K$_{s}$})}  & \colhead{YSO Class} & \colhead{Var. Mech.}\\
    \colhead{} & \colhead{(days)} &\colhead{(mag)} & \colhead{(mag)} & \colhead{(mag)} & \colhead{} & \colhead{}
  }
  \startdata
  ISO-Oph 88&310&0.500&0.415&0.270&II&Extinction\\
  WL 14&384&0.218&0.185&0.380&II&Extinction\\
  70072-2446272&543&0.061&0.091&0.084&---&Extinction\\
  ISO-Oph 91&143&0.078&0.074&0.180&III&Extinction\\
  70266-2446345&313&0.049&0.061&0.122&---&Extinction\\
  ISO-Oph 94&64&0.894&---&0.707&II&Extinction\\
  WL 1\tablenotemark{b}&226&0.294&0.275&0.523&II&Accretion?\\
  WL 17\tablenotemark{b}&578&1.125&---&1.065&I&Unknown\\
  ISO-Oph 107&355&0.195&---&0.119&II&Accretion\\
  YLW 8A&355&0.198&1.272&0.098&II&Accretion\\
  ISO-Oph 112\tablenotemark{b}&207&0.984&---&1.123&II&Extinction\\
  ISO-Oph 113&120&0.058&0.067&0.122&III&Unknown\\
  WL 19\tablenotemark{b}&589&1.012&---&0.784&II&Accretion?\\
  ISO-Oph 117&530&0.560&0.249&0.133&II&Accretion\\
  YLW 10B\tablenotemark{b}&578&0.562&---&0.850&II&Unknown\\
  ISO-Oph 119&76&0.445&---&---&II&Unknown\\
  WL 20E&81&0.299&0.262&0.297&II&Accretion\\
  WL 20W&122&0.213&0.258&0.305&II&Extinction\\
  71726-2422283&132&0.163&---&1.137&---&Unknown\\
  YLW 12A&92&0.728&---&0.813&I&Accretion\\
  ISO-Oph 126&349&0.135&---&0.371&III&Extinction\\
  WL 6&172&1.199&---&1.256&I&Accretion?\\
  72297-2448071&327&0.134&0.097&0.062&---&Accretion?\\
  72357-2412288&354&0.052&0.066&0.080&---&Extinction\\
  ISO-Oph 137&89&0.749&---&---&I&Unknown\\
  72514-2446335&750&0.065&0.089&0.173&---&Unknown\\
  YLW 15A&478&0.393&---&0.927&I&Extinction\\
  YLW 16B&140&2.312&---&1.318&I&Extinction\\
  YLW 17B\tablenotemark{b}&516&0.155&0.110&0.233&II&Accretion\\
  ISO-Oph 151&398&0.282&0.106&0.160&II&Accretion\\
  ISO-Oph 150&239&0.926&---&---&I&Unknown\\
  \enddata
  \tablenotetext{a}{The catalog ID has been truncated by 2MASS J162 for 2MASS catalog stars}
  \tablenotetext{b}{Candidate sinusoidal-like periodic LTV}
\end{deluxetable}

\begin{figure}
  \plotone{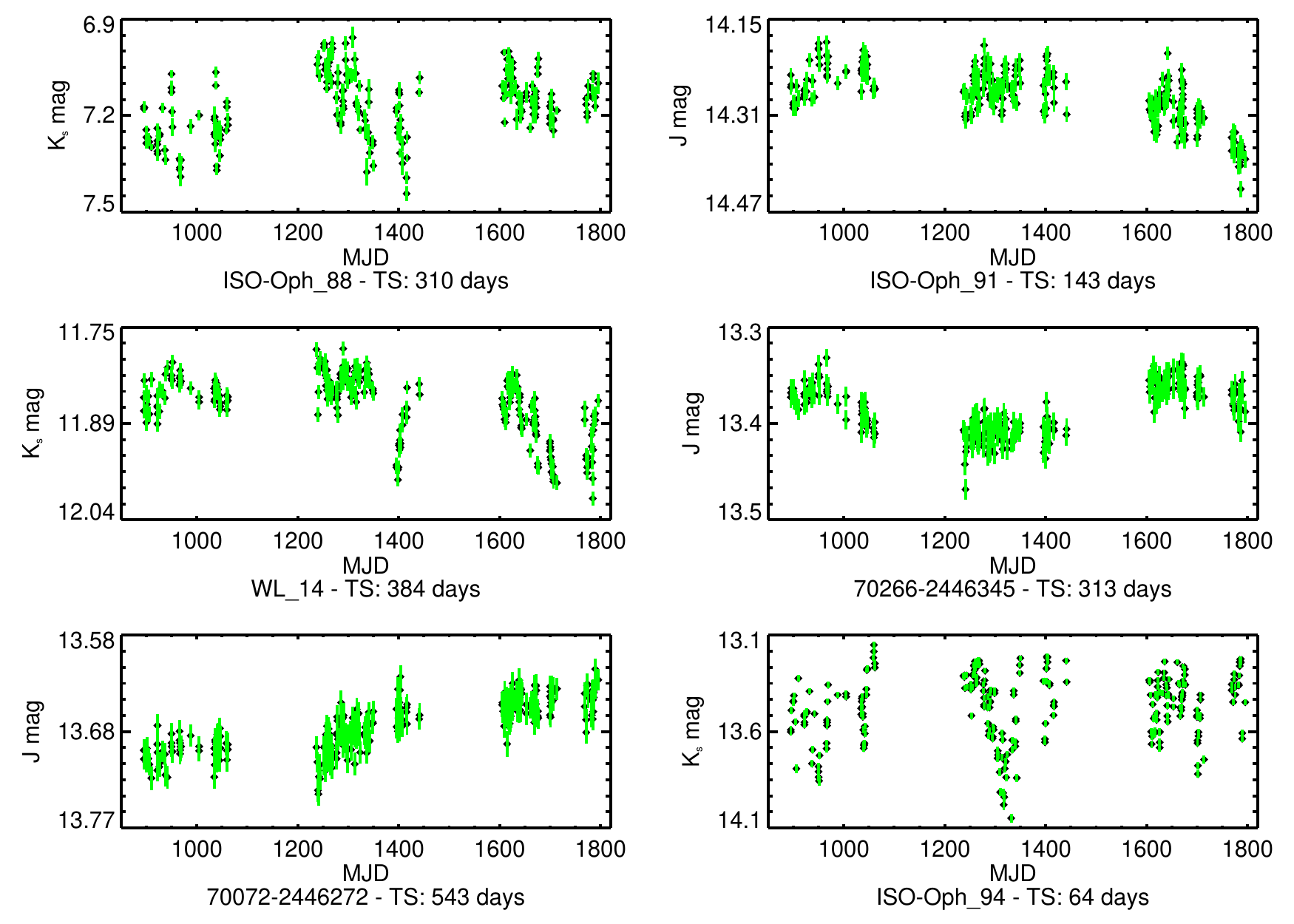}
  \caption{The \emph{J} or \emph{K$_{s}$} light curves for 6 long time-scale variables.  The the highest signal-to-noise light curve of these 2 is illustrated.}
\end{figure}

\begin{figure}
  \plotone{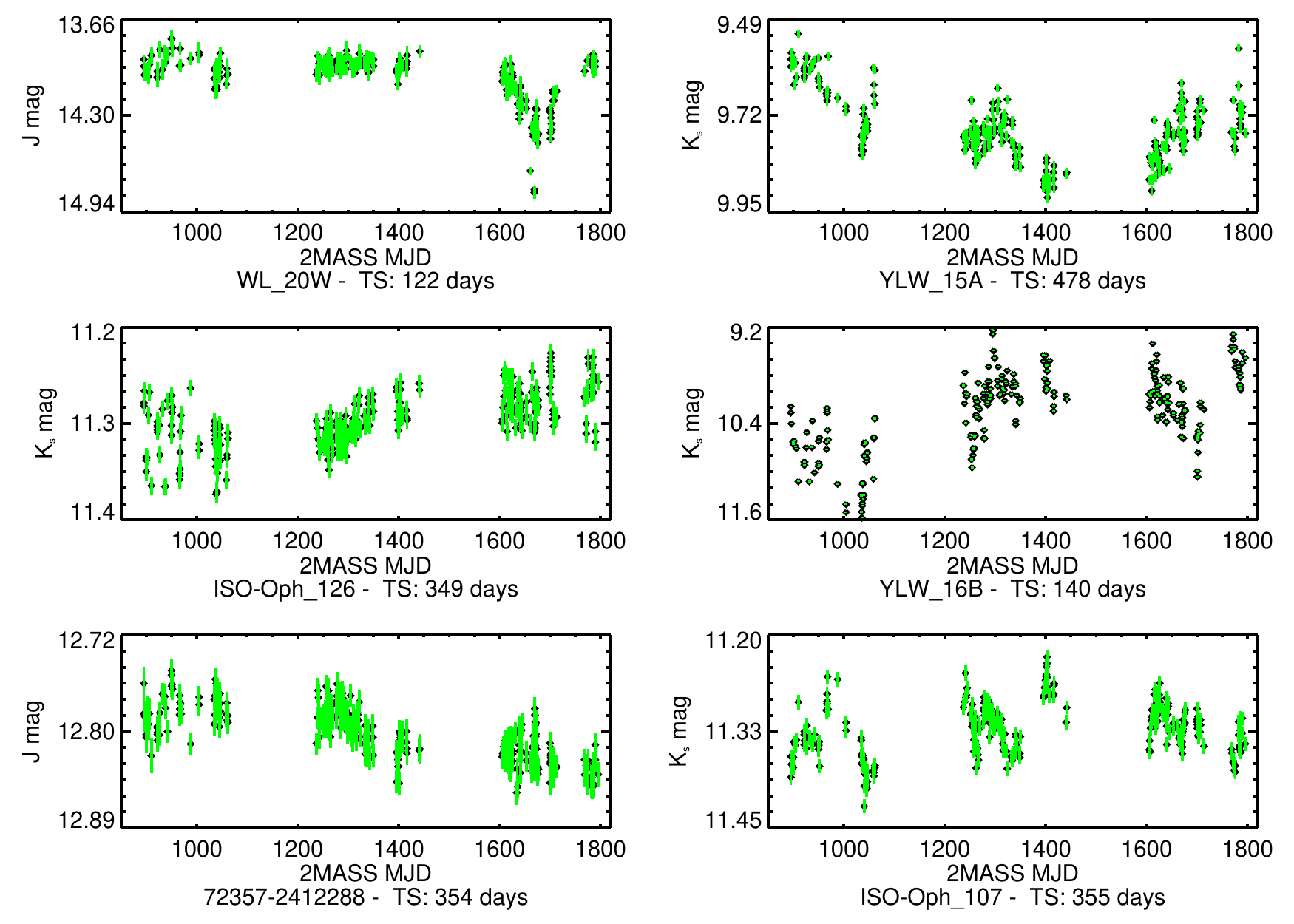}
  \caption{The \emph{J} or \emph{K$_{s}$} light curves for 6 long time-scale variables.  The the highest signal-to-noise light curve of these 2 is illustrated.}
\end{figure}

\begin{figure}
  \plotone{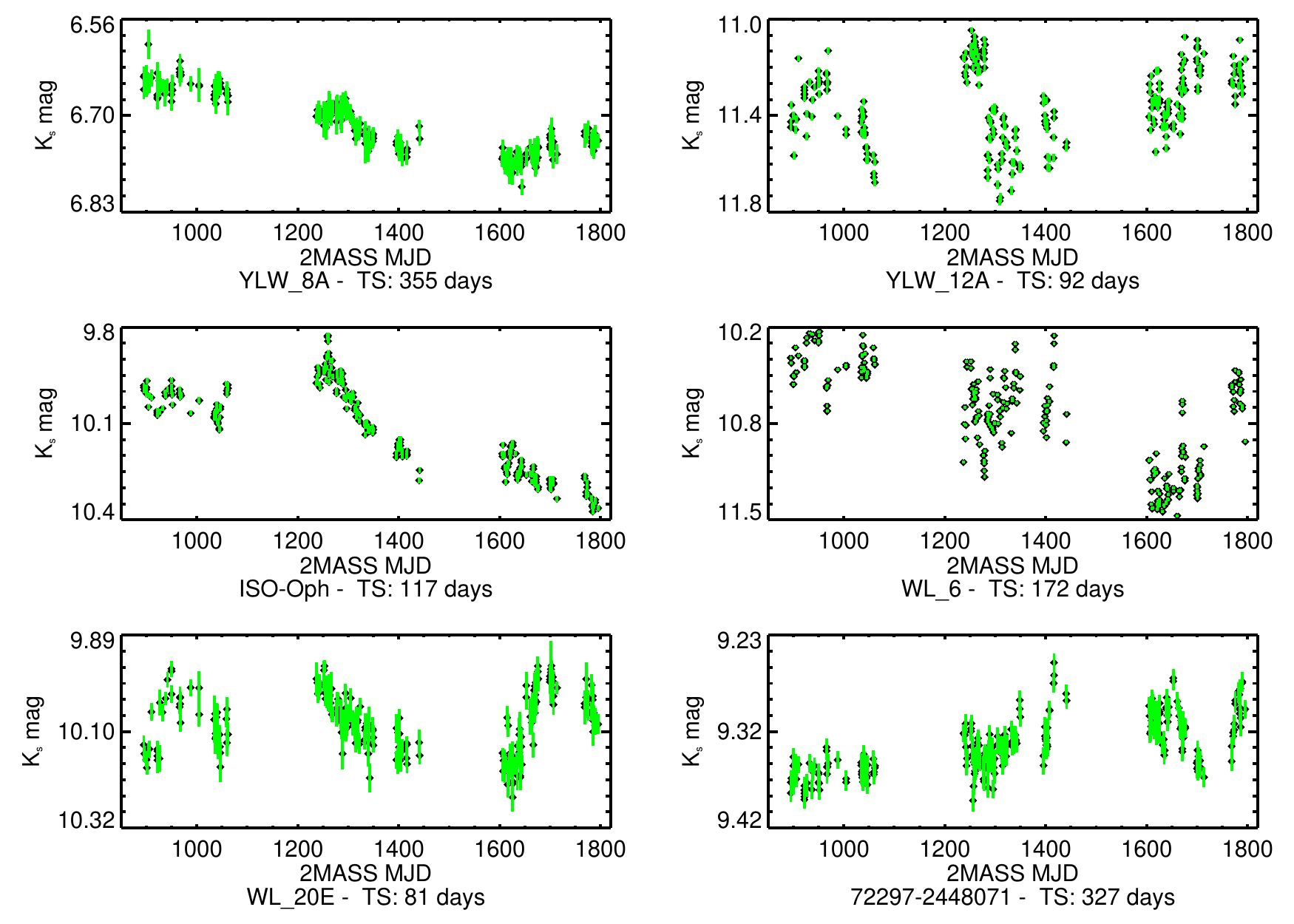}
  \caption{The  \emph{K$_{s}$} light curves for 6 long time-scale variables.}
\end{figure}

\begin{figure}
  \plotone{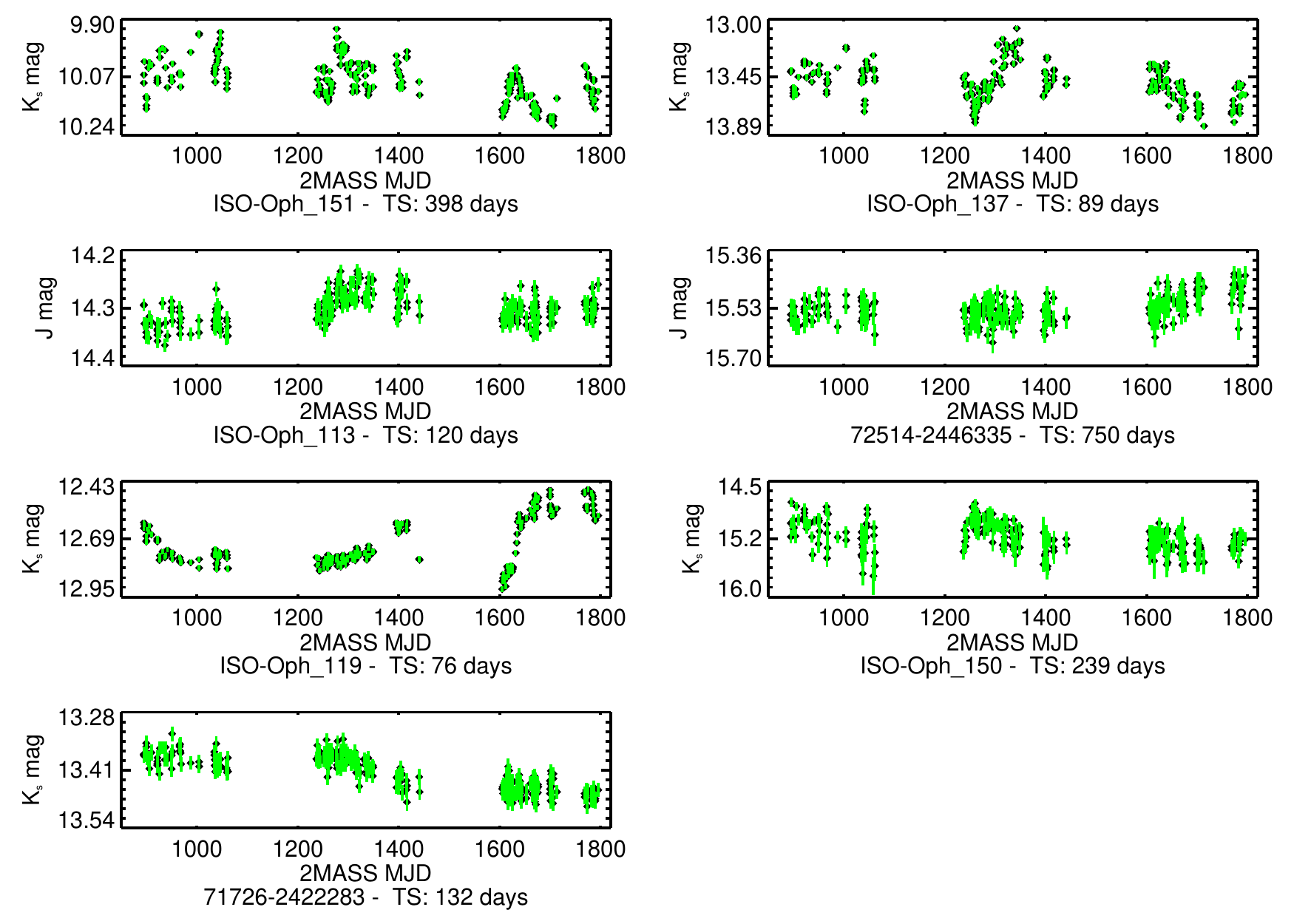}
  \caption{The \emph{J} or \emph{K$_{s}$} light curves for 7 long time-scale variables.  The the highest signal-to-noise light curve of these 2 is illustrated.}
\end{figure}

In $\S$4.2, a method for quantifying the time-scale of the extended brightness variations is detailed.  These time-scales range from 64 to 790 days, the latter being near the full duration of the observing campaign.  The peak-to-trough $\Delta$K$_{s}$ amplitudes for all LTVs range from 0.05 to 2.31 mag, with a median value of 0.29 mag.  The peak-to-trough $\Delta$(\emph{H}-\emph{K$_{s}$}) color amplitudes range from 0.06 to 1.32 mag, with a median value of 0.23 mag.  The two most probable mechanisms for these aperiodic variations with time-scales much longer than typical stellar rotation periods are variable extinction and variable mass accretion rates.  Fig 25 shows two examples of the change in \emph{K$_{s}$} brightness and stellar color caused by these two mechanisms.  Extinction causes the star to become redder as the star dims.  Changes in the mass accretion rate cause stars to become bluer as the star dims.  For 12 LTVs (39$\%$), the (\emph{J}-\emph{H}) and (\emph{H}-\emph{K$_{s}$}) colors become redder as the star dims favoring variable extinction as the dominant variability mechanism.  The (\emph{J}-\emph{H}) and (\emph{H}-\emph{K$_{s}$}) colors become bluer as the star dims in 11 LTVs (34$\%$) favoring variable mass accretion rates as the dominant variability mechanism.  The remaining 7 LTVs (23$\%$), either do not have useful color information because the \emph{J} or \emph{H} (or both) photometry is below the survey completeness limits, or the brightness-color correlation does not agree with any of the 4 listed variability criteria.  No dominant variability mechanism is assigned to these stars.

One intriguing scenario to explain why LTVs do not seem to favor one variability mechanism over another is the viewing angle.  For LTVs where variable accretion is the favored mechanism, the system could be more face-on providing a clearer view of the inner disk hole.  Variable extinction due to circumstellar disk asymmetries is more easily seen at higher disk inclinations where the ``puffed up'' outer disk attentuates the light from the inner disk.  As these two mechanisms have opposing brightness-color correlations, systems with no measured correlations may represent intermediate viewing angles.  In this case, the measured effects from variable accretion will ``cancel'' out or confuse the measured effects from variable extinction.

\begin{figure}
  \plotone{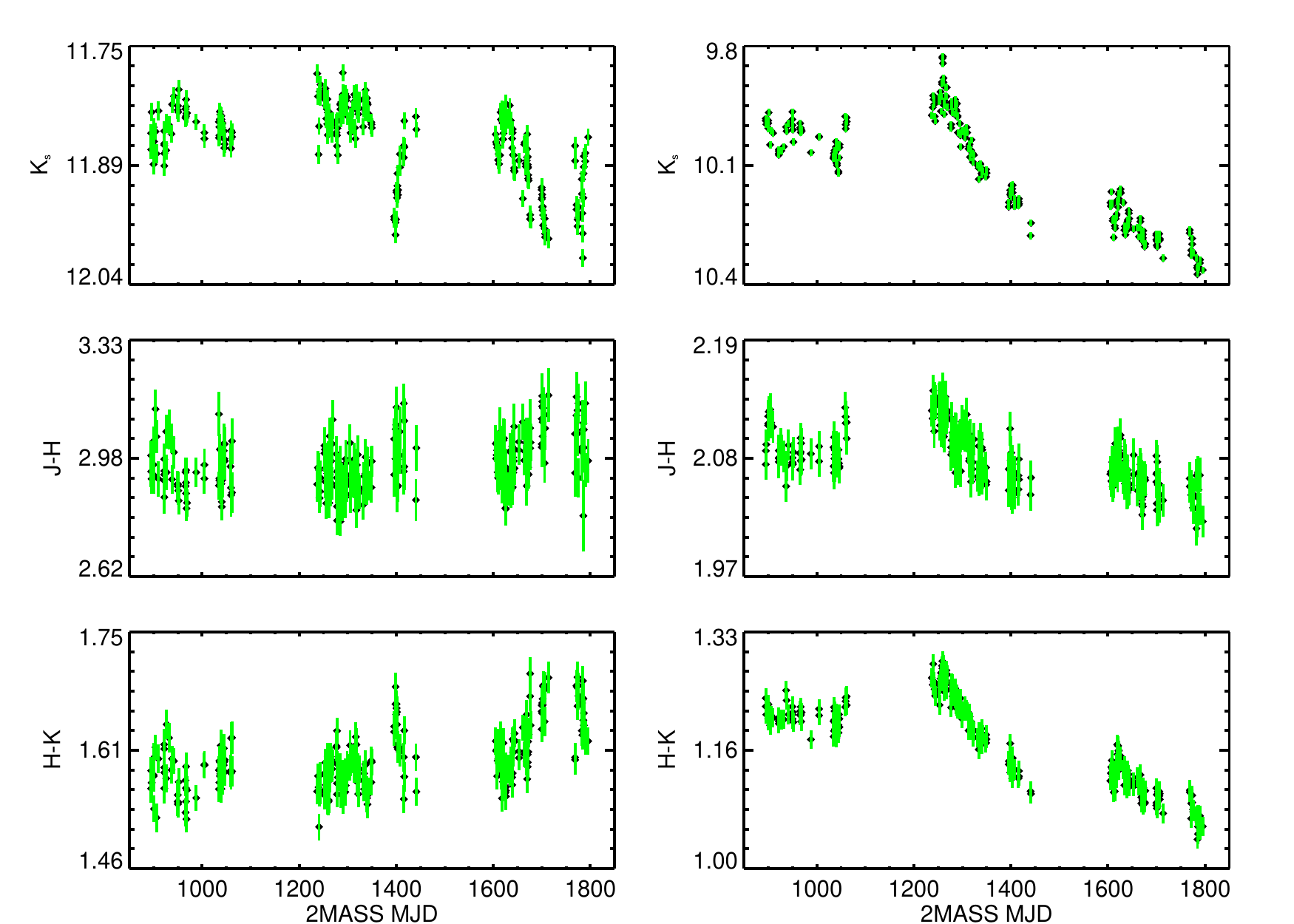}
  \caption{\emph{Left}: The \emph{K$_{s}$} light, (\emph{J}-\emph{H}) color curve, and (\emph{H}-\emph{K$_{s}$}) color curve for the long time-scale variable WL 14. This is an example of variability caused by extinction. As the \emph{K$_{s}$} magnitude drops the colors become redder. \emph{Right}: The same light and color curves for the long time-scale variable ISO-Oph 117.  This is an example of variability caused by variable mass accretion.  As the \emph{K$_{s}$} magnitude drops the colors become bluer.}
\end{figure}

All LTVs are located ``on cloud'' and 25 stars of the 31 LTVs are classified as a YSO: 7 Class I (58$\%$), 15 Class II (44$\%$) and 3 Class III (27$\%$).\footnote{The percentages indicate the percentage of variable stars in each class that are sinusoidal-like periodic variables.}  The favored variability mechanism in the Class I LTVs is variable extinction for 2 stars, variable mass accretion for 2 stars and unidentified for 3 stars.  The variability in the Class II LTVs is consistent with variable extinction in 5 stars, variable mass accretion in 8 stars and is not identified for 2 stars.  The \emph{K$_{s}$}-color correlation in 2 Class III LTVs favors variable extinction as the dominant variability mechanism while the mechanism for variability in the third Class III LTV is not identified.

Based upon visual inspection of folded light curves, 6 LTVs are considered candidate periodic variables.  These candidate periodic LTVs are denoted in Table 6 and Fig 26 contains their folded \emph{K$_{s}$} light curves.  The variability time-scales, ranging from 207 to 589 days, for the candidate periodic LTVs are measured using the Lomb-Scargle algorithm.  The PA does not find the time-scales found by Lomb-Scargle to be significant.  The stars, on average, have higher flux and color amplitude variability than the LTVs taken as a whole.  The peak-to-trough $\Delta$\emph{K$_{s}$} amplitude for these candidate periodic variables range from 0.16 to 1.13 mag, with a median value of 0.77 mag.  The peak-to-trough color amplitude range from 0.23 to 1.12 mag, with a median value of 0.82 mag.  For 3 candidate periodic LTVs, the (\emph{J}-\emph{H}) and (\emph{H}-\emph{K$_{s}$}) colors become bluer as the star dims favoring a variable mass accretion rate as the dominant variability mechanism.  The (\emph{H}-\emph{K$_{s}$}) color of ISO-Oph 112 reddens as the star dims.  This is consistent with variable extinction as the dominant variability mechanism.  A combination of \emph{J} band photometry below the survey completeness limits and the \emph{K$_{s}$}-(\emph{H}-\emph{K$_{s}$}) color correlation not matching any of the 4 criteria precludes the identification of the dominant variability mechanism for 2 candidate periodic LTVs.  All the candidate periodic LTVs are classified as a YSO: 1 Class I star and 5 Class II stars.

\begin{figure}
  \plotone{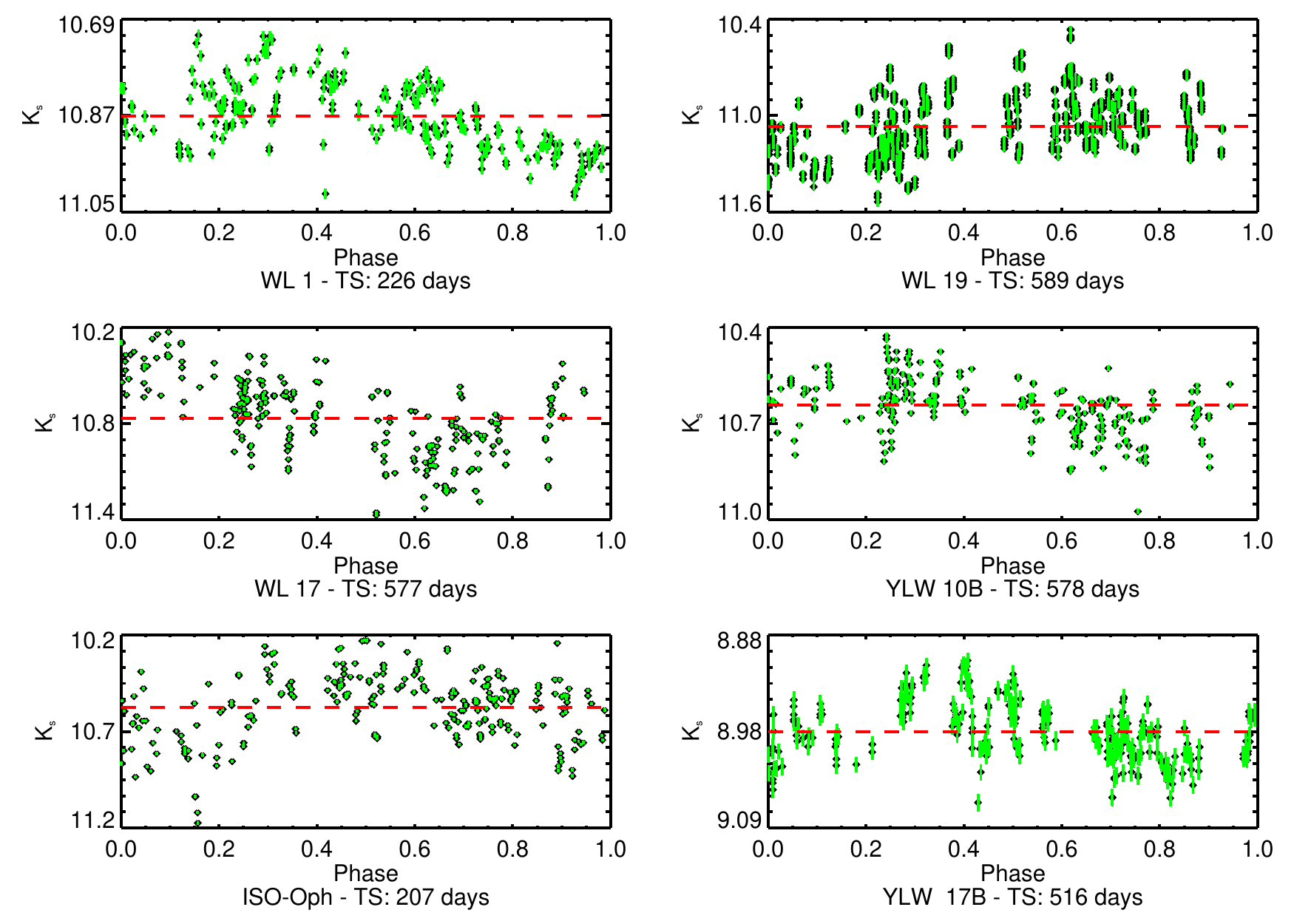}
  \caption{The \emph{K$_{s}$} folded light curves for 6 candidate periodic long time-scale variables.  The red line indicates the star's mean magnitude}
\end{figure}

\subsection{Irregular Variables}
The variable catalog contains 40 stars (40$\%$) that are clearly variable, but the largest amplitude variability is not periodic or coherent on long time-scales.  The \emph{K$_{s}$} light curves are located in Figs 27 to 31.  Table 8 contains the list of irregular variables.

\begin{deluxetable}{c c c c c c}
  \tablecolumns{6}
  \tablewidth{0pc}
  \tablecaption{Irregular Variables}
  \tabletypesize{\scriptsize}
  \tablehead{
    \colhead{Catalog ID\tablenotemark{a}} & \colhead{$\Delta$K$_{s}$} & \colhead{$\Delta$(\emph{J}-\emph{H})\tablenotemark{b}} & \colhead{$\Delta$(\emph{H}-\emph{K$_{s}$})\tablenotemark{c}} & \colhead{YSO Class} & \colhead{Var Mech}\\
    \colhead{} & \colhead{(mag)} & \colhead{(mag)} & \colhead{(mag)} & \colhead{} & \colhead{}
  }
  \startdata
  65576-2508150&0.114&0.150&0.123&---&Unknown\\
  65699-2455192&0.167&0.153&0.180&---&Unknown\\
  WL 21&0.166&---&---&II&Unknown\\
  65744-2452589&0.160&0.105&0.139&---&Unknown\\
  65789-2452371&0.854&---&---&---&Unknown\\
  65789-2457518&0.108&0.093&0.124&---&Unknown\\
  65861-2446029&0.131&0.198&0.168&---&Accretion?\\
  ISO-Oph 87&0.294&0.241&0.180&II&2 ``flare'' events\\
  70054-2446444&1.109&---&---&---&Unknown\\
  WL 22&0.631&---&---&I&Unknown\\
  65967-2415433&0.078&0.124&0.073&---&Unknown\\
  70055-2416255&0.061&0.082&0.066&---&Unknown\\
  WL 16&0.082&0.083&0.176&---&Unknown\\
  70276-2502437&0.053&0.062&0.083&---&Unknown\\
  70285-2418546&0.060&0.086&0.064&---&Unknown\\
  70501-2508484&0.052&0.084&0.077&---&Unknown\\
  70516-2420077&0.049&0.079&0.078&---&Unknown\\
  70591-2459376&0.044&0.078&0.055&---&Unknown\\
  70597-2428363&0.182&---&0.157&II&Unknown\\
  60819-2442286&0.069&0.061&0.195&---&Unknown\\
  71003-2429133&0.334&---&0.366&II&Unknown\\
  71096-2445298&0.056&0.073&0.074&---&Unknown\\
  71173-2447109&0.069&0.079&0.074&---&Unknown\\
  ISO-Oph 116&0.155&0.085&0.090&II&Accretion\\
  71377-2505450&0.059&0.065&0.091&---&Unknown\\
  71384-2415441&0.205&---&0.291&---&Unknown\\
  71404-2415096&0.109&0.200&0.149&---&Unknown\\
  71531-2415515&0.069&0.099&0.088&---&Unknown\\
  71605-2415039&0.352&---&0.534&---&Unknown\\
  71604-2416163&0.391&---&0.517&---&Unknown\\
  71744-2413079&0.155&0.217&0.315&---&Unknown\\
  YLW 13B&0.215&0.276&0.158&II&Accretion\\
  ISO-Oph 131&0.050&0.184&0.069&III&Unknown\\
  72330-2507282&0.700&---&0.751&---&Unknown\\
  72325-2448357&0.074&0.099&0.094&---&Unknown\\
  ISO-Oph 138&0.326&0.432&0.175&II&Unknown\\
  73052-2432347&0.070&0.075&0.082&---&Unknown\\
  73107-2504004&1.057&---&---&---&Unknown\\
  73122-2504172&0.566&0.386&0.473&---&Unknown\\
  73208-2508545&0.613&0.318&0.679&---&Unknown\\
  \enddata
  \tablenotetext{a}{The catalog ID has been truncated by 2MASS J162 for 2MASS catalog stars.}
  \tablenotetext{b}{Entries with --- mark stars with photometry below the survey \emph{J} band completeness limit.}
  \tablenotetext{c}{Entries with --- mark stars with photometry below the survey \emph{H} band completeness limit.}
\end{deluxetable}

\begin{figure}
  \plotone{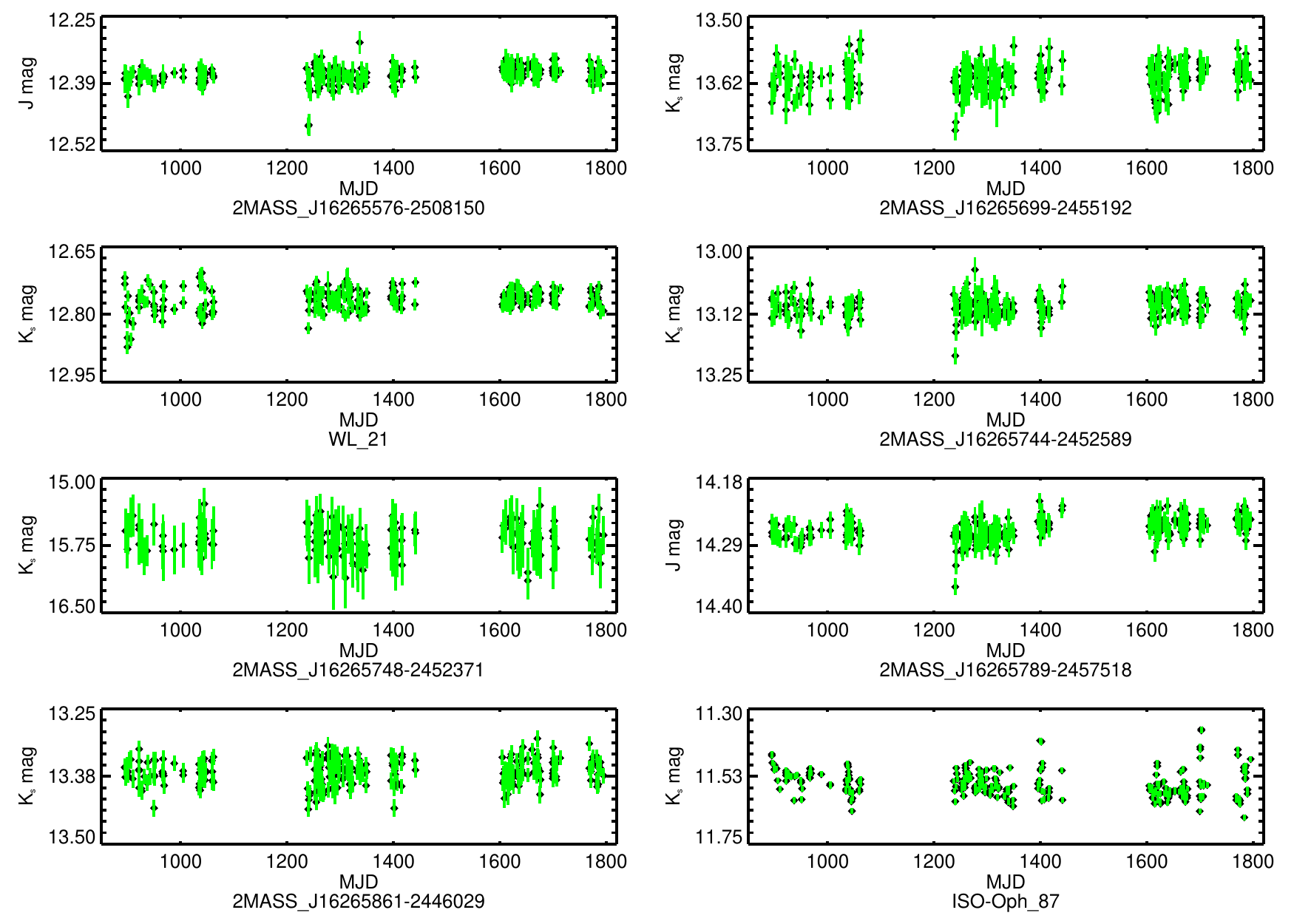}
  \caption{The \emph{J} or \emph{K$_{s}$} light curves for 8 irregular variables.}
\end{figure}

\begin{figure}
  \plotone{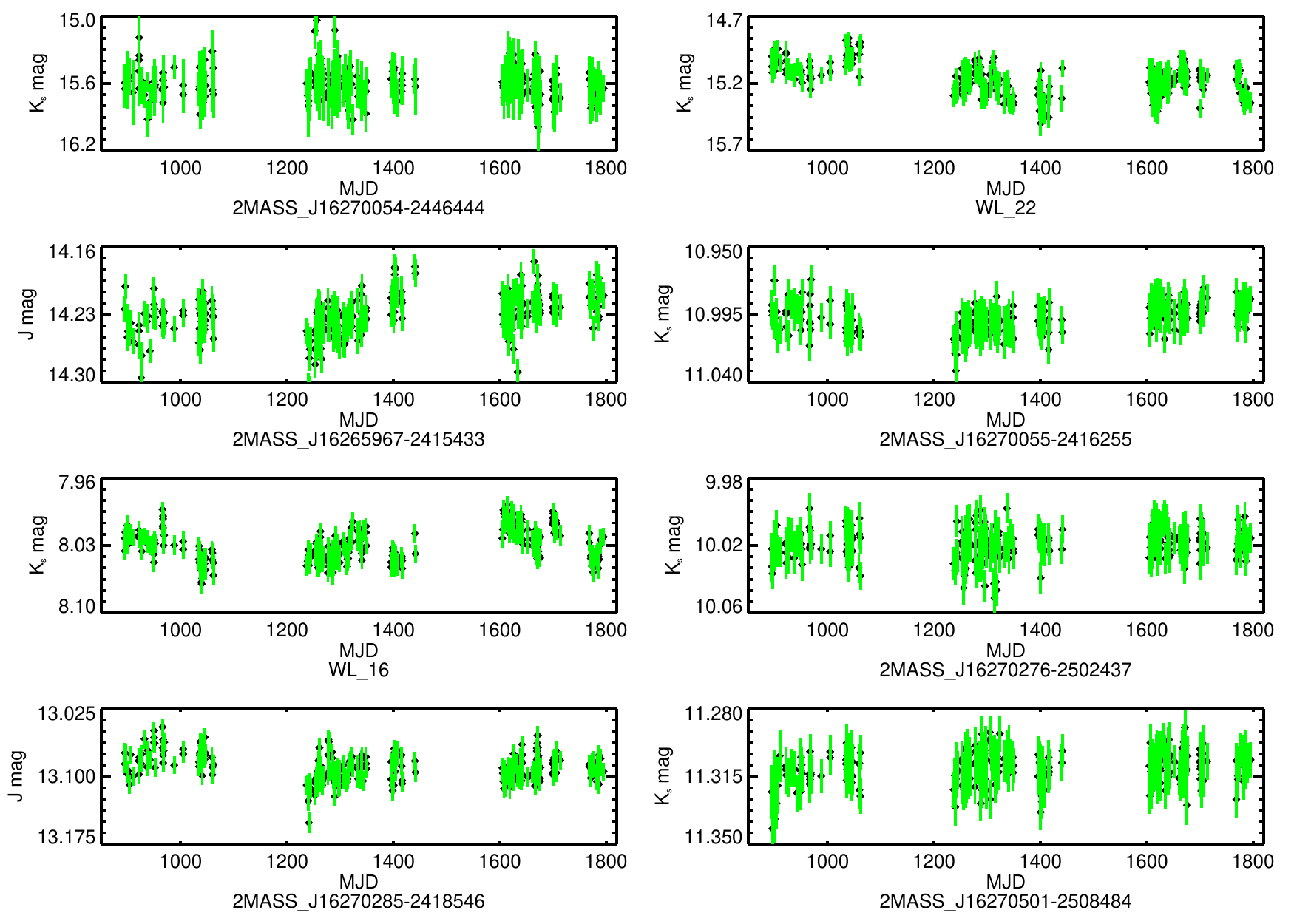}
  \caption{The \emph{J} or \emph{K$_{s}$} light curves for 8 irregular variables.}
\end{figure}

\begin{figure}
  \plotone{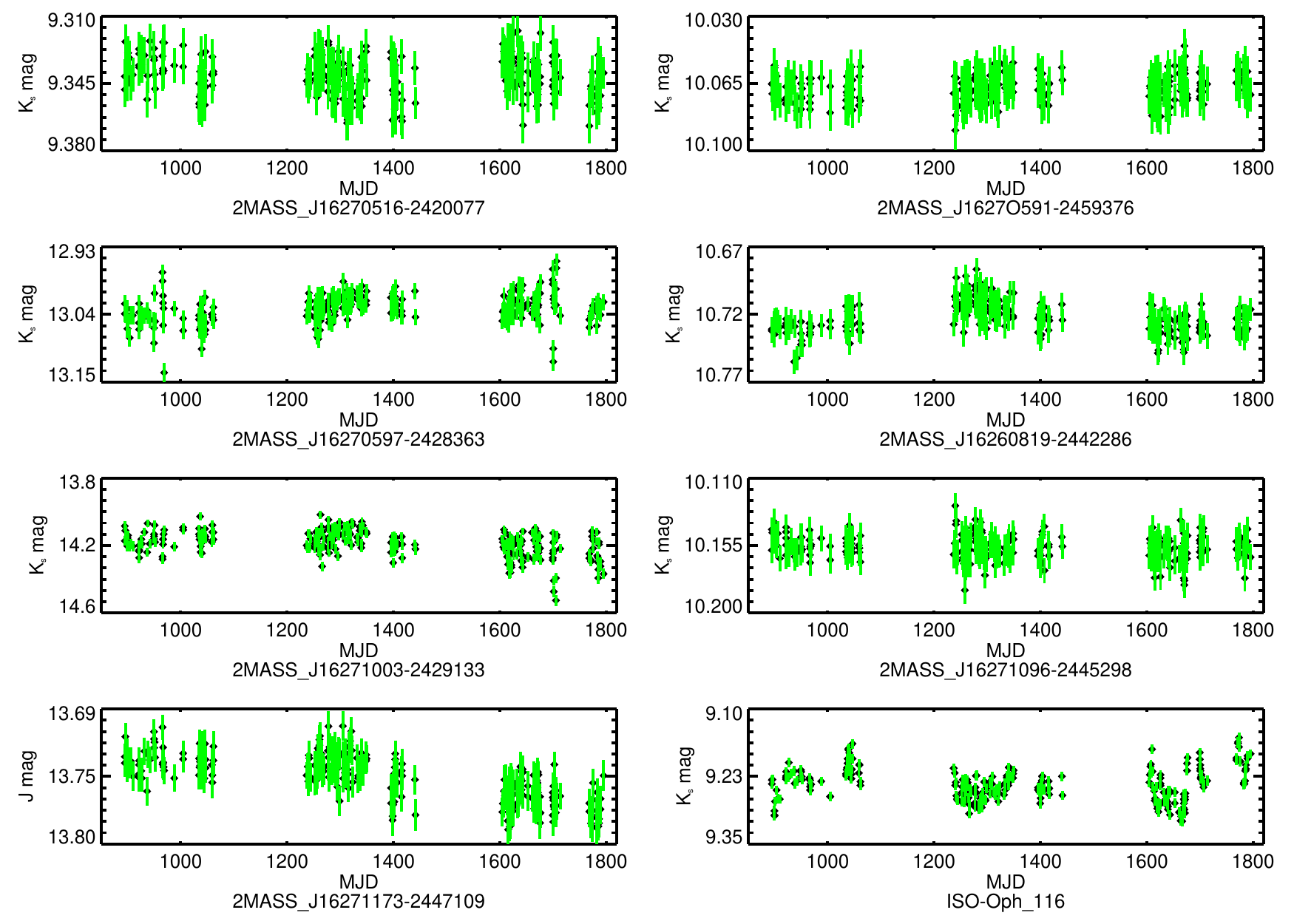}
  \caption{The \emph{J} or \emph{K$_{s}$} light curves for 8 irregular variables.}
\end{figure}

\begin{figure}
  \plotone{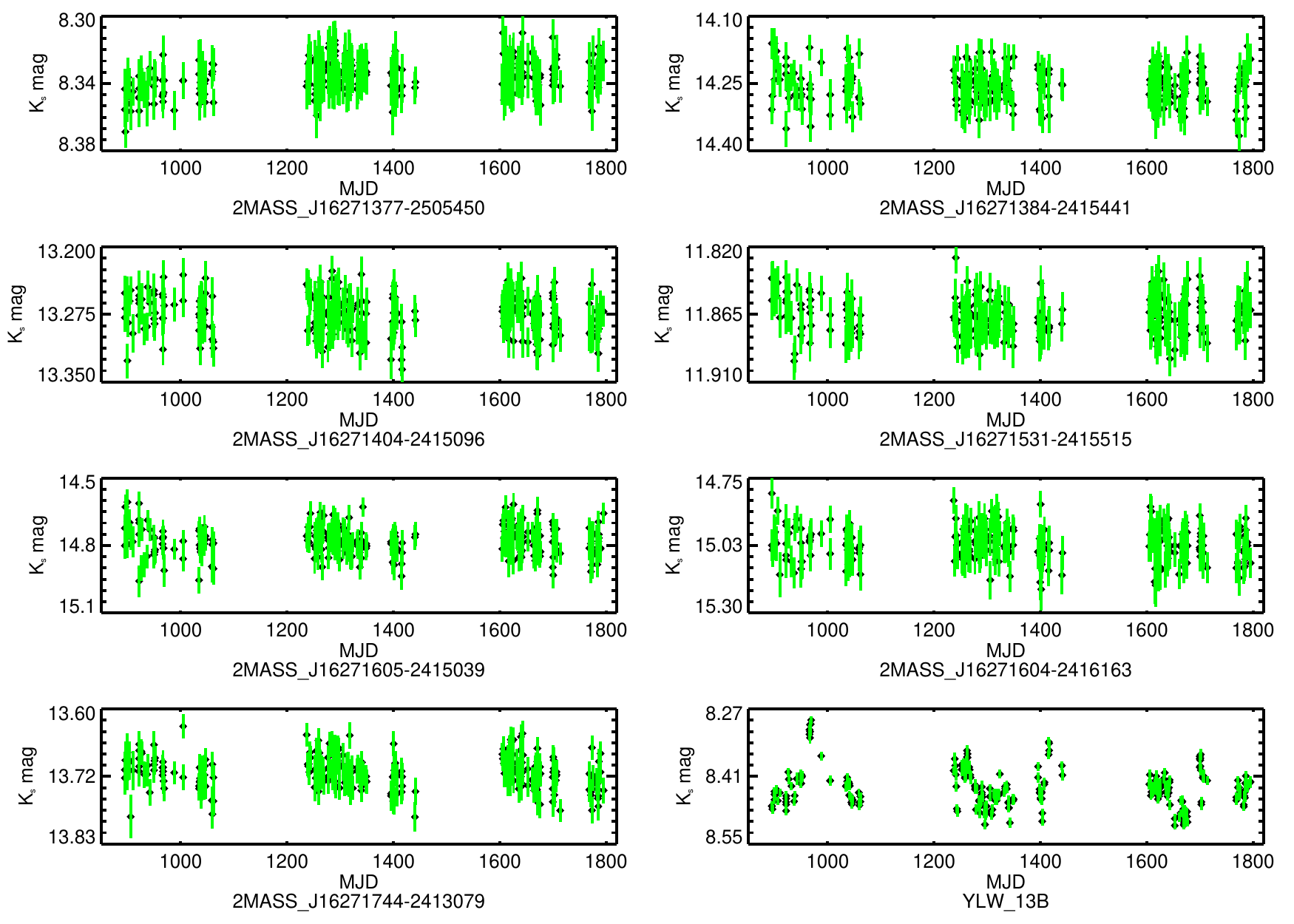}
  \caption{The \emph{K$_{s}$} light curves for 8 irregular variables.}
\end{figure}

\begin{figure}
  \plotone{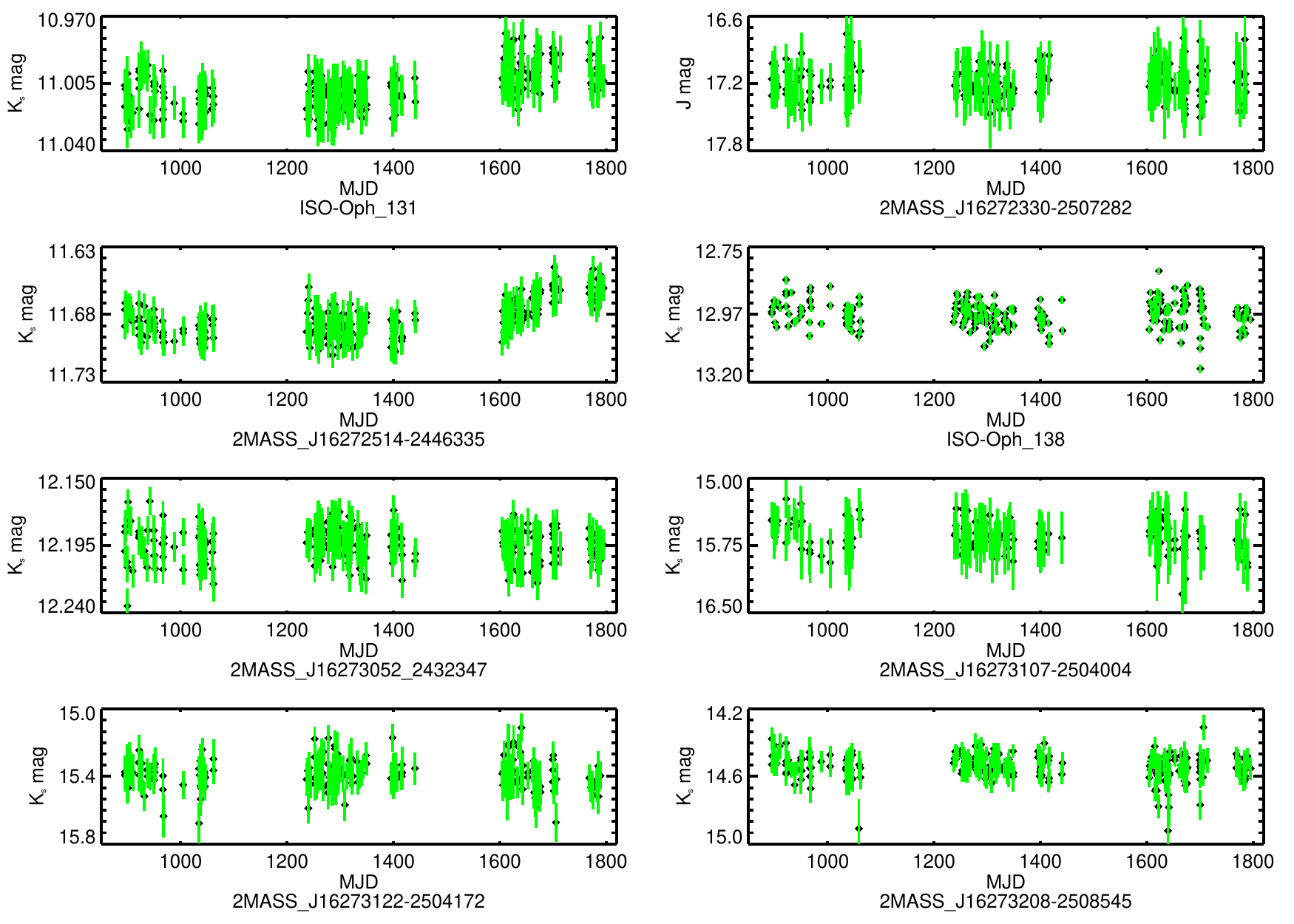}
  \caption{The \emph{J} or \emph{K$_{s}$} light curves for the 8 irregular variables.}
\end{figure}

The $\Delta$K$_{s}$ amplitude range from 0.04 to 1.11 mag, with a median value of 0.14 mag.  The $\Delta$(\emph{H}-\emph{K$_{s}$}) color amplitude range from 0.05 to 0.75 mag with a median value of 0.14 mag.  Using the variability criteria, discussed in $\S$5, the primary variability mechanism is only identified for 4 irregular variables (2MASSJ16265861-2446029, ISO-Oph 116, YLW 13B, ISO-Oph 87).  The first three 3 exhibit a long time-scale variation in at least one observing season where the star becomes bluer as it dims.  This is indicative of variable mass accretion as the variability mechanism.  These stars are not considered LTVs as this variability is not the largest amplitude variability in the time-series.  The variable accretion for YLW 13B is identified to occur for $\sim$ 115 days in the second year with a $\Delta$K$_{s}$ $\sim$0.15 mag and $\Delta$(\emph{H}-\emph{K$_{s}$}) 0.11 mag (see Fig 30).  ISO-Oph 116 varies via variable accretion at least twice (see Fig 28).  The first time occurs for $\sim$170 days in the first year with $\Delta$K$_{s}$ 0.14 mag and $\Delta$(\emph{H}-\emph{K$_{s}$}) 0.07 mag.  The second time occurs for $\sim$70 days in the third year with $\Delta$K$_{s}$ $\sim$0.11 mag and $\Delta$(\emph{H}-\emph{K$_{s}$}) 0.07 mag.  The average error in both \emph{K$_{s}$} and (\emph{H}-\emph{K$_{s}$}) is 0.01 mag for both YLW 13B and ISO-Oph 116.  In the case of 2MASSJ16265861-2446029, only the (\emph{J}-\emph{H}) color becomes bluer making the mechanism identification tentative. The YSO classification for this star is unknown.    The \emph{J} band photometry is too dim in 11 irregular variables to identify the variability mechanism.  Of these 11 variables, the \emph{H} band is also too dim in 5 stars. 

The variability in ISO-Oph 87 is peculiar due to two flare-like events that occur in the \emph{K$_{s}$} photometry on approximately 1400 and 1700 2MASS MJD..  Fig 32 contains the \emph{K$_{s}$}, (\emph{J}-\emph{H}) and (\emph{H}-\emph{K$_{s}$}) photometry for this Class II YSO.  If the events are truly related to an increase in stellar activity, the star is expected to become bluer in both (\emph{J}-\emph{H}) and (\emph{H}-\emph{K$_{s}$}).  However, no change is seen in the (\emph{J}-\emph{H}) color and the star reddens in (\emph{H}-\emph{K$_{s}$}).  The first event lasts for $\sim$10 days with $\Delta$K$_{s}$ 0.20 mag and $\Delta$(\emph{H}-\emph{K$_{s}$}) 0.13 mag.  The second event occurs for $\sim$6 days and $\Delta$K$_{s}$ 0.18 mag and $\Delta$(\emph{H}-\emph{K$_{s}$}) 0.16 mag.  Two additional, lower amplitude spikes in the \emph{K$_{s}$} photometry between 1600 and 1700 2MASS MJD might also be similar flare-like events.

\begin{figure}
  \plotone{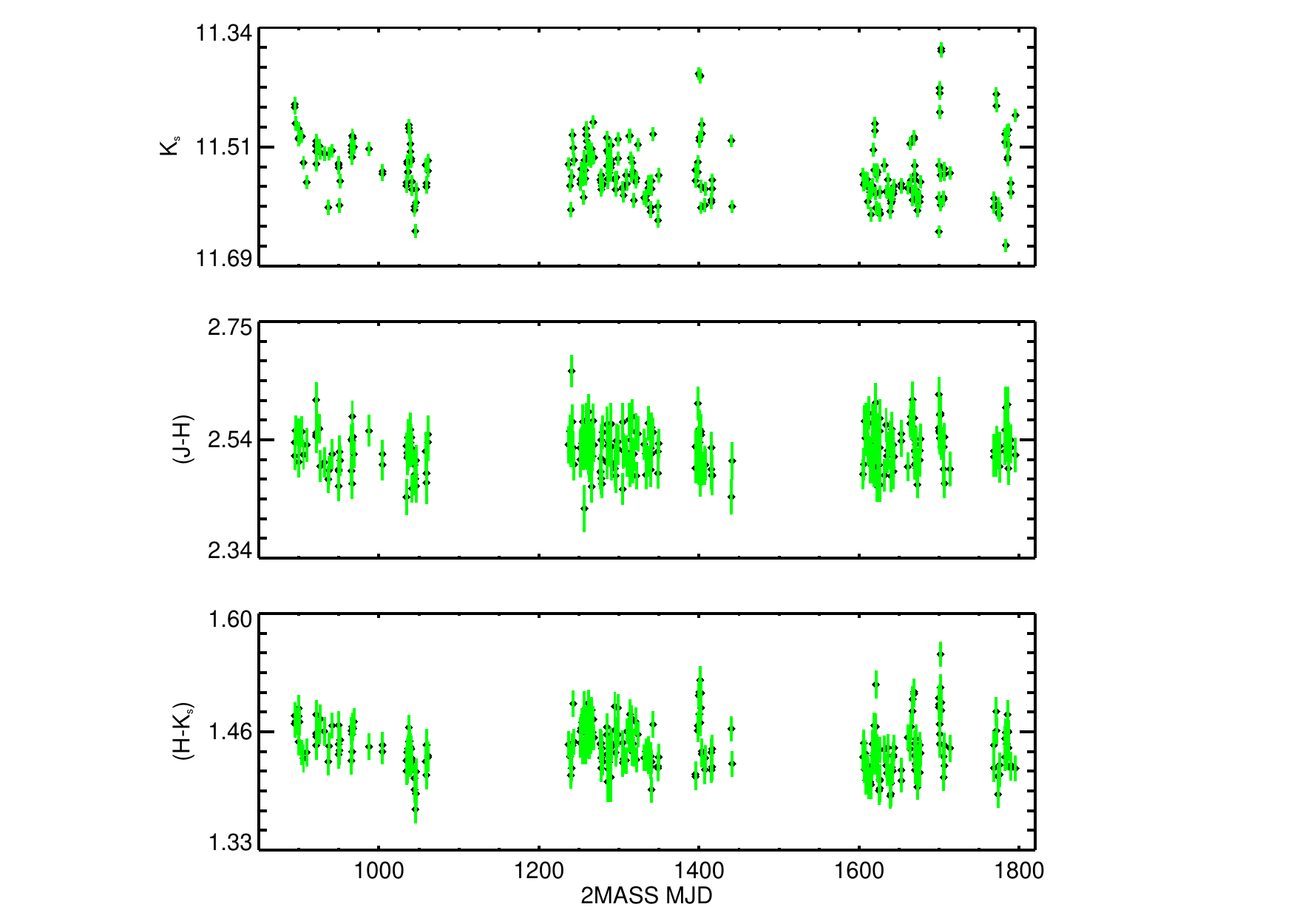}
  \caption{The K$_{s}$, (\emph{J}-\emph{H}) and  (\emph{H}-\emph{K$_{s}$}) photometry for ISO-Oph 87.  The photometry contains at least two ``flare'' events, where the star brightens sharply in \emph{K$_{s}$},  occurring at 1400 and 1700 2MASS MJD. Two other possible ``flare'' events occur between 1600 and 1700 2MASS MJD.  The events become redder as the star brightens rather than becoming bluer as expected for stellar flares.}
\end{figure}

Irregular variables have the smallest percentage (68$\%$) of stars located ``on cloud''.  Only 9 stars in this sub-category are classified as a YSO: 1 Class I (8$\%$), 7 Class II (21$\%$) and 1 Class III (9$\%$).\footnote{The percentages indicate the percentage of variable stars in each class that are irregular variables.}  Most (16 stars) of the 22 candidate $\rho$ Oph members are irregular variables.  If these candidate members are YSOs then the fraction of YSO irregular variables is comparable to the fraction of periodic and LTV YSOs.  

\subsection{Examples of Multiple Variability Mechanisms}
As discussed in the Introduction, variability studies indicate young stars sometimes exhibit complex photometric behavior believed to result from multiple variability mechanisms acting concurrently.  For most stars in this survey, only the highest amplitude variability can be confidently characterized.  However, a lower amplitude, second type of variability is definitely seen in 7 variable stars.  Four of these seven stars (YLW 1C, 2MASS J16272658-2425543, YLW 10C, WL 4) show evidence for two separate, yet statistically distinct periodic variations.  The remaining three stars (WL 20W, ISO-Oph 126, WL 15) are periodically variable underneath a higher amplitude, long time-scale variation.  The methods used to identify both the primary and secondary variabilities in these stars is discussed in $\S$4.1.  The following subsections contain detailed discussions for each of these stars except WL 4 which is described in \citet{plavchan08a}.

\begin{itemize}

\item YLW 1C (ISO-Oph 86): This CTTS exhibits both sinusoidal-like and eclipse-like periodic variability at 2 distinct periods; the periods are distinct from each other to a 20$\sigma$ confidence level.  Fig 33 contains the \emph{K$_{s}$}, (\emph{J}-\emph{H}) and (\emph{H}-\emph{K$_{s}$}) photometry folded to the sinusoidal-like period, P = 5.7792 $\pm$ 0.0085 days.  The peak-to-trough $\Delta$K$_{s}$ amplitude for this variability is 0.14 mag.  The variability in both the (\emph{J}-\emph{H}) color and (\emph{H}-\emph{K$_{s}$}) color is not correlated with the \emph{K$_{s}$} variability.  This favors variability caused by rotational modulation of a cool starspot(s).  Fig 34 contains the \emph{K$_{s}$}, (\emph{J}-\emph{H}) and (\emph{H}-\emph{K$_{s}$}) photometry folded to the eclipse-like period, P = 5.9514 $\pm$ 0.0014 days.  The $\Delta$K$_{s}$ eclipse depth is 0.29 mag.  The (\emph{H}-\emph{K$_{s}$}) color reddens during the eclipse event consistent with variability caused by extinction.  This behavior is not seen in the (\emph{J}-\emph{H}) color.  Since the \emph{J} band photometry is near the survey completeness limits, the absence of a clear reddening trend may be due to low signal-to-noise in the color curve.

\begin{figure}
  \plotone{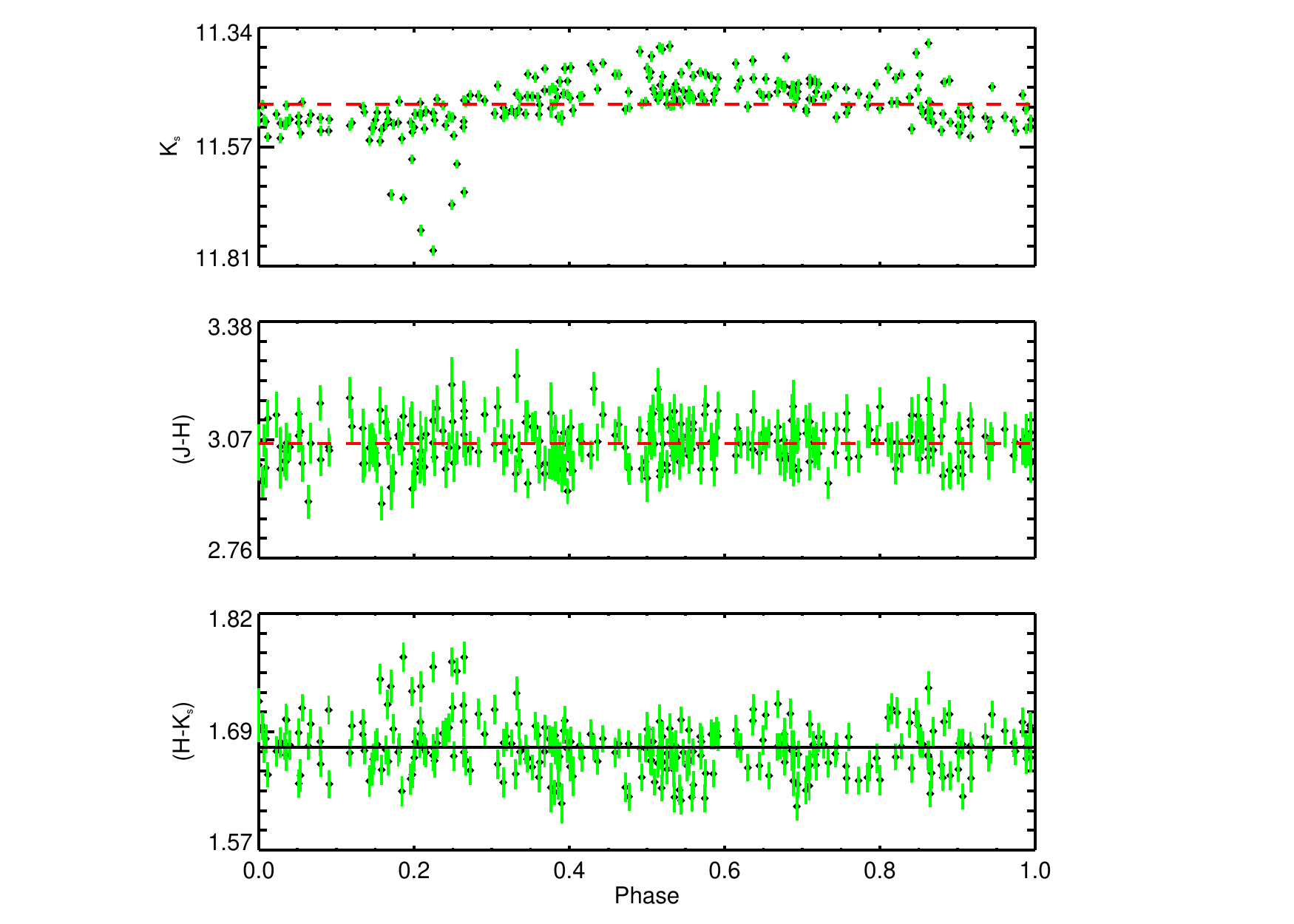}
  \caption{The \emph{K$_{s}$}, (\emph{J}-\emph{H}) and (\emph{H}-\emph{K$_{s}$}) photometry for YLW 1C phased to the 5.7752 $\pm$ 0.0085 day sinusoidal-like period.  The red line indicates the mean value in each panel.  The lack of color correlation with $\Delta$\emph{K$_{s}$} points to rotational modulation of cool starspots as the variability mechanism.}
\end{figure}

\begin{figure}
  \plotone{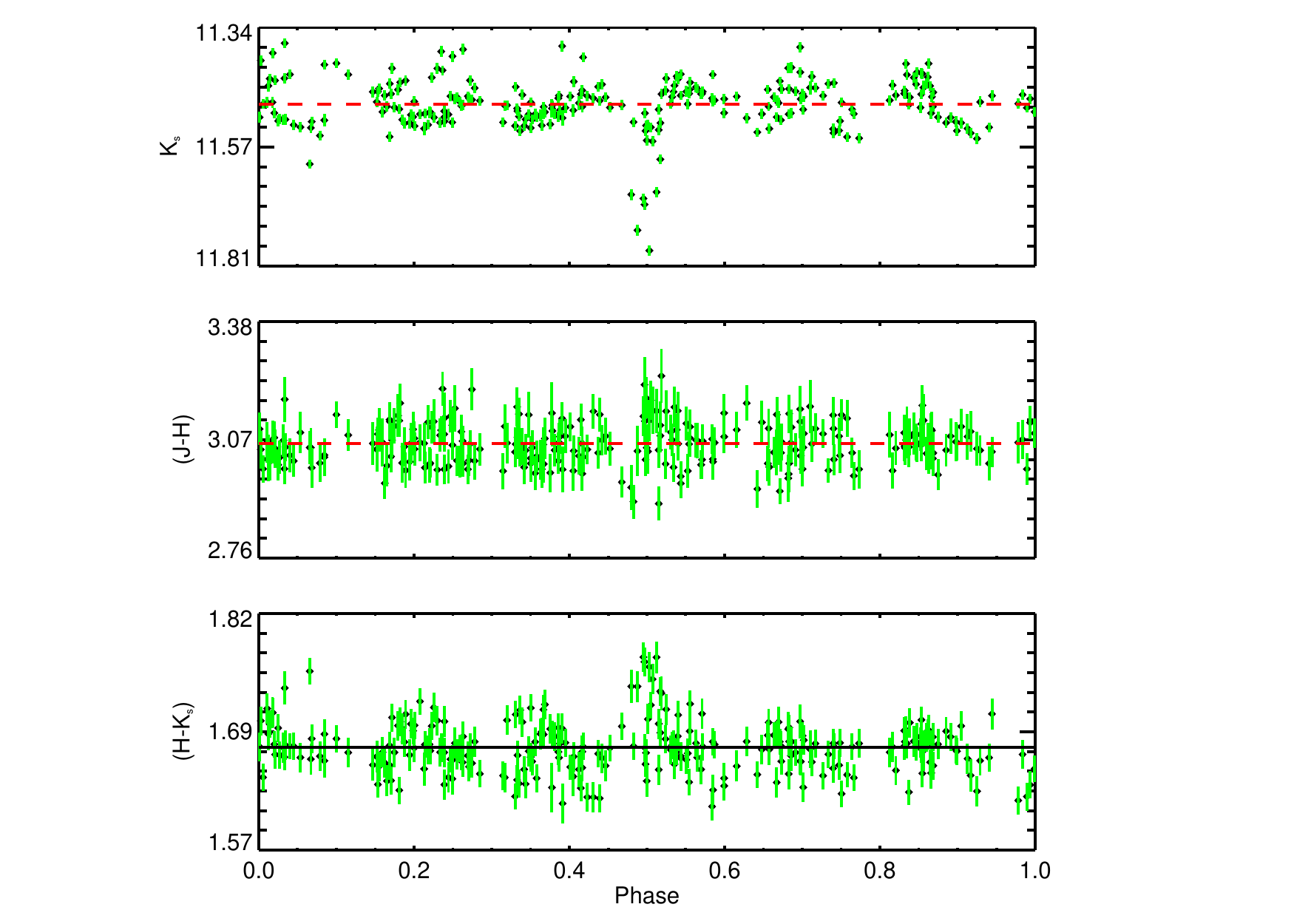}
  \caption{The \emph{K$_{s}$}, (\emph{J}-\emph{H}) and (\emph{H}-\emph{K$_{s}$}) photometry for YLW 1C phased to the 5.9514 $\pm$ 0.0014 day eclipse-like period.  The red line indicates the mean value in each panel.  Both colors become redder as \emph{K$_{s}$} dims indicating variable extinction as the likely variability mechanism.}
\end{figure}

The two periods for YLW 1C is consistent with the interpretation that these events are true occultation events, as proposed for AA Tau (see $\S$5.1.2).  The short period, arising from a stellar surface feature(s), traces the stellar rotation rate.  The longer period suggests the occultation of the star by an obscuration located just beyond the circumstellar disk co-rotation radius.  Following the anaylsis described in $\S$5.1.2, the occulter of YLW 1C is located 7.2 R$_{\star}$ from the host star and the duration of the occultation is $\sim$6.4 hours corresponding to a minimum occulter diameter of $\sim$2.0 R$_{\star}$.  The reader is reminded this diameter represents the extent within the orbital path and makes no claim on any preferential occulter shape.  The eclipse depth is $\Delta$K$_{s}$ = 0.29 mag.  

The large occulter diameter argues against the direct detection of a hot protoplanet.  However recent imaging results suggest that gas giant planets maybe considerably extended in the mass accretion phase \citep{quanz13,kraus12}. If true in this case, this would demonstrate the existance of a hot protoplanet with a period of 6 days very near the peak in the period distribution for exoplanets \citep{wright12}.  Alternatively, the event could be caused by an occultation of a warped portion of a circumstellar disk.  This scenario has been proposed to explain the near- to mid-IR variability in LRLL 31 \citep{flaherty10,flaherty12}.  While most YSO disk models invoke axisymmetry, objects such as YLW 1C are prompting the creation of more complex models.

\item 2MASS J16272658-2425543: This CTTS, designated 'J543' hereafter, is another star exhibiting both sinusoidal-like and eclipse-like variability with two distinctly different periods.  Fig 35 contains the \emph{K$_{s}$}, (\emph{J}-\emph{H}) and (\emph{H}-\emph{K$_{s}$}) photometry folded to the sinusoidal-like period, P = 1.52921 $\pm$ 0.00065 days.  The peak-to-trough $\Delta$K$_{s}$ amplitude for this variability is 0.20 mag.  The variability in the (\emph{J}-\emph{H}) and (\emph{H}-\emph{K$_{s}$}) colors are not correlated with the K$_{s}$ photometry, which favors variability caused by rotational modulation of a cool starspot.  Fig 36 contains the \emph{K$_{s}$}, (\emph{J}-\emph{H}) and (\emph{H}-\emph{K$_{s}$}) photometry folded to the eclipse-like period, P = 2.9602 $\pm$ 0.0013 days.  The $\Delta$K$_{s}$ eclipse depth is 0.17 mag.  Both the (\emph{J}-\emph{H}) and (\emph{H}-\emph{K$_{s}$}) colors redden during in the eclipse event consistent with variable extinction.

\begin{figure}
  \plotone{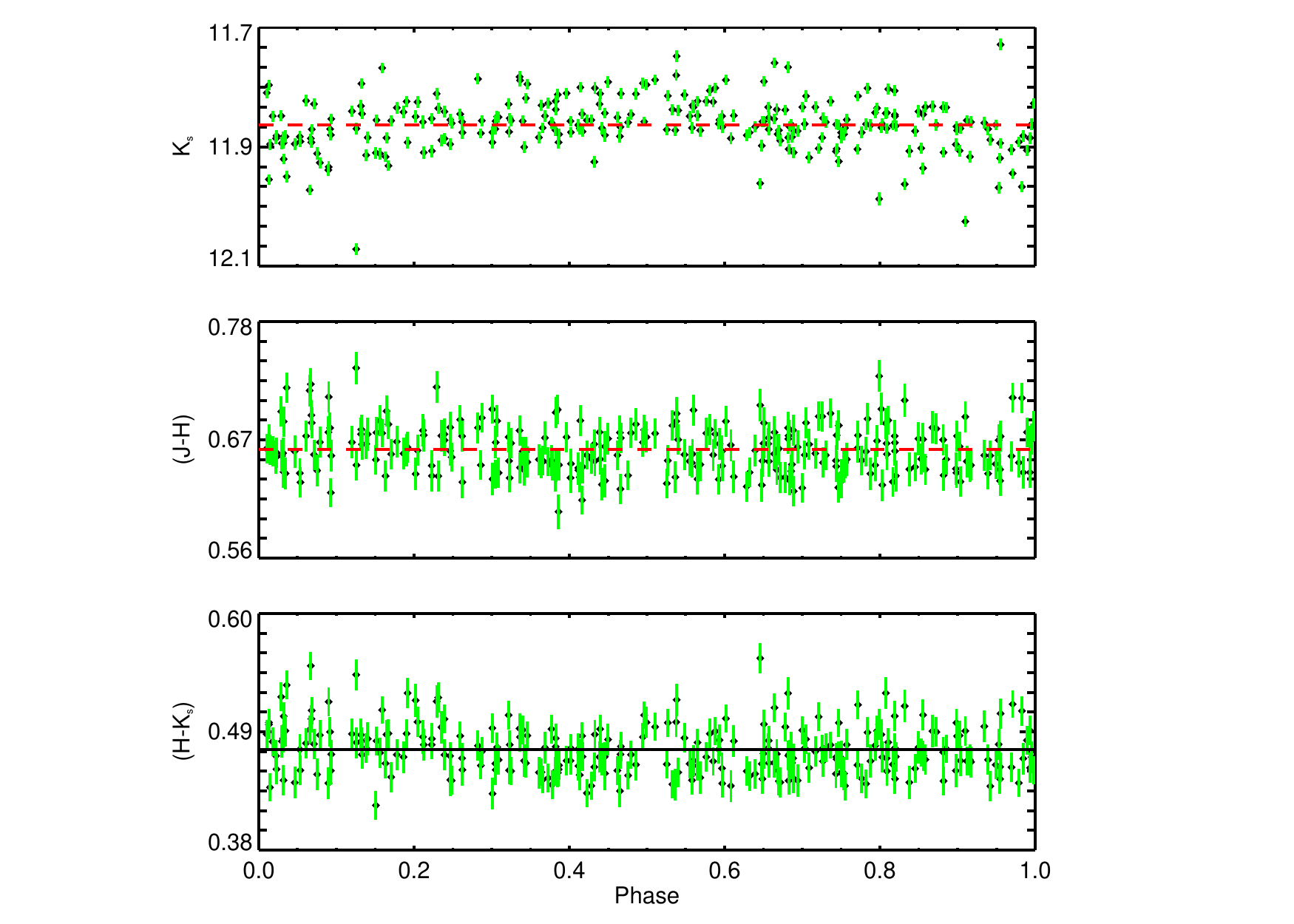}
  \caption{The \emph{K$_{s}$}, (\emph{J}-\emph{H}) and (\emph{H}-\emph{K$_{s}$}) photometry for J543 phased to the 1.52921 $\pm$ 0.00065 day sinusoidal-like period.  The red line indicates the mean value in each panel.  The lack of color correlation with $\Delta$\emph{K$_{s}$} points to rotational modulation of cool starspots as the variability mechanism.}
\end{figure}

\begin{figure}
  \plotone{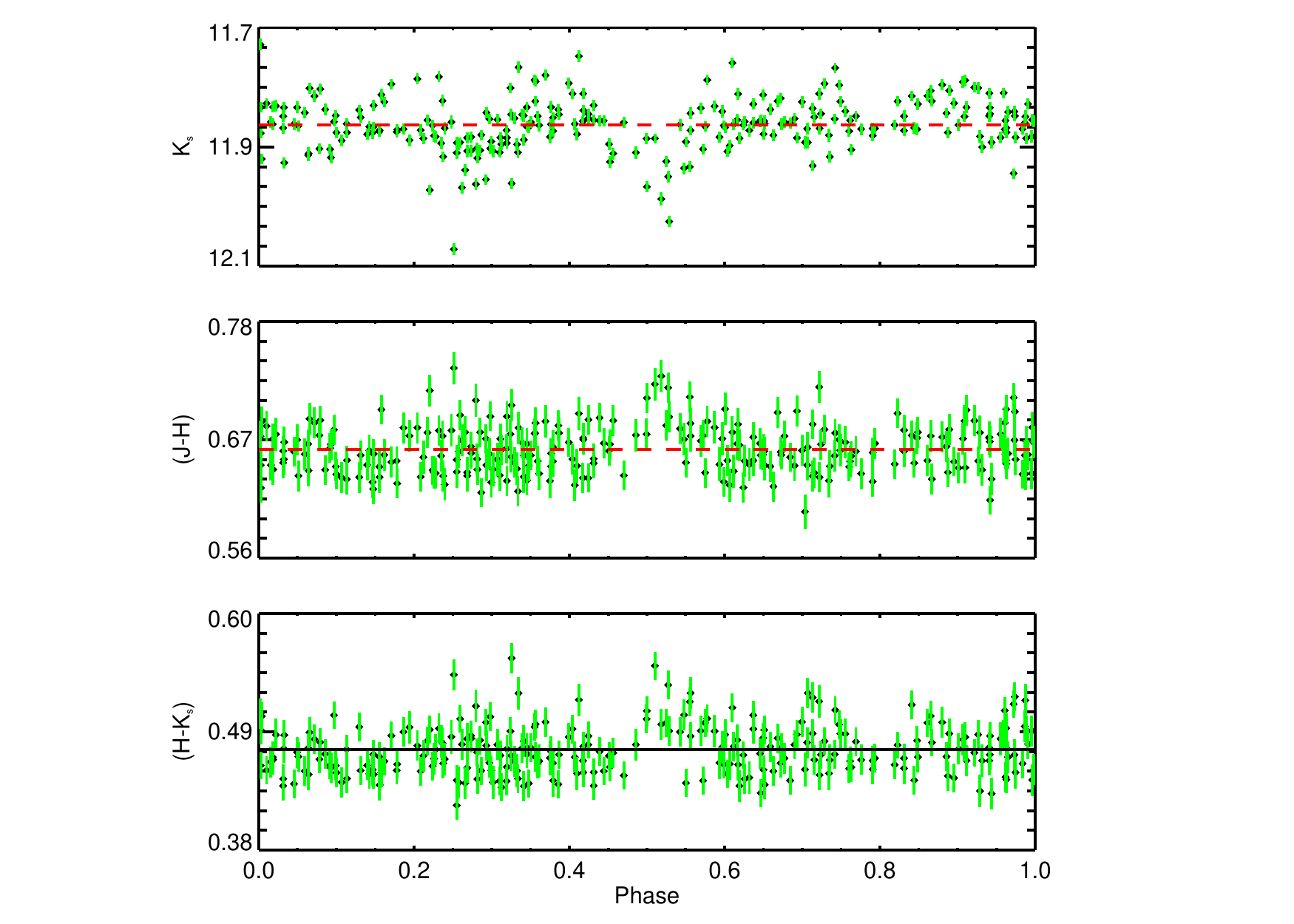}
  \caption{The \emph{K$_{s}$}, (\emph{J}-\emph{H}) and (\emph{H}-\emph{K$_{s}$}) photometry for J543 phased to the 2.9602 $\pm$ 0.0013 day eclipse-like period.  The red line indicates the mean value in each panel.  Both colors become redder as \emph{K$_{s}$} dims indicating variable extinction as the likely variability mechanism.}
\end{figure}

The physical interpretation for the observed variability is identical to that of YLW 1C.  The 1.6 day period corresponds to the stellar rotation rate and the 3.0 day period arises from a periodic occultation by an asymmetry in the inner circumstellar disk.  The occulter size and distance are 2.8 R$_{\star}$ and $\sim$3.0 R$_{\star}$.  Differing from YLW 1C, the occulter for J543 is located approximately a stellar radius beyond the co-rotation radius.  

\item YLW 10C (ISO-Oph 122): YLW 10C is the third CTTS where two distinct periods are identified.  Fig 37 contains the \emph{K$_{s}$}, (\emph{J}-\emph{H}) and (\emph{H}-\emph{K$_{s}$}) photometry folded to the sinusoidal-like period, P = 3.0779 $\pm$ 0.0025 days.  The peak-to-trough $\Delta$K$_{s}$ amplitude for this variability is 0.25 mag.  Fig 38 contains the \emph{K$_{s}$}, (\emph{J}-\emph{H}) and (\emph{H}-\emph{K$_{s}$}) photometry folded to the eclipse-like period, P = 2.9468 $\pm$ 0.0029 days.  The $\Delta$K$_{s}$ eclipse depth is 0.28 mag.  Unfortunately, both the \emph{J} and \emph{H} photometry are below the survey completeness limits preventing the identification of either variability mechanism.

\begin{figure}
  \plotone{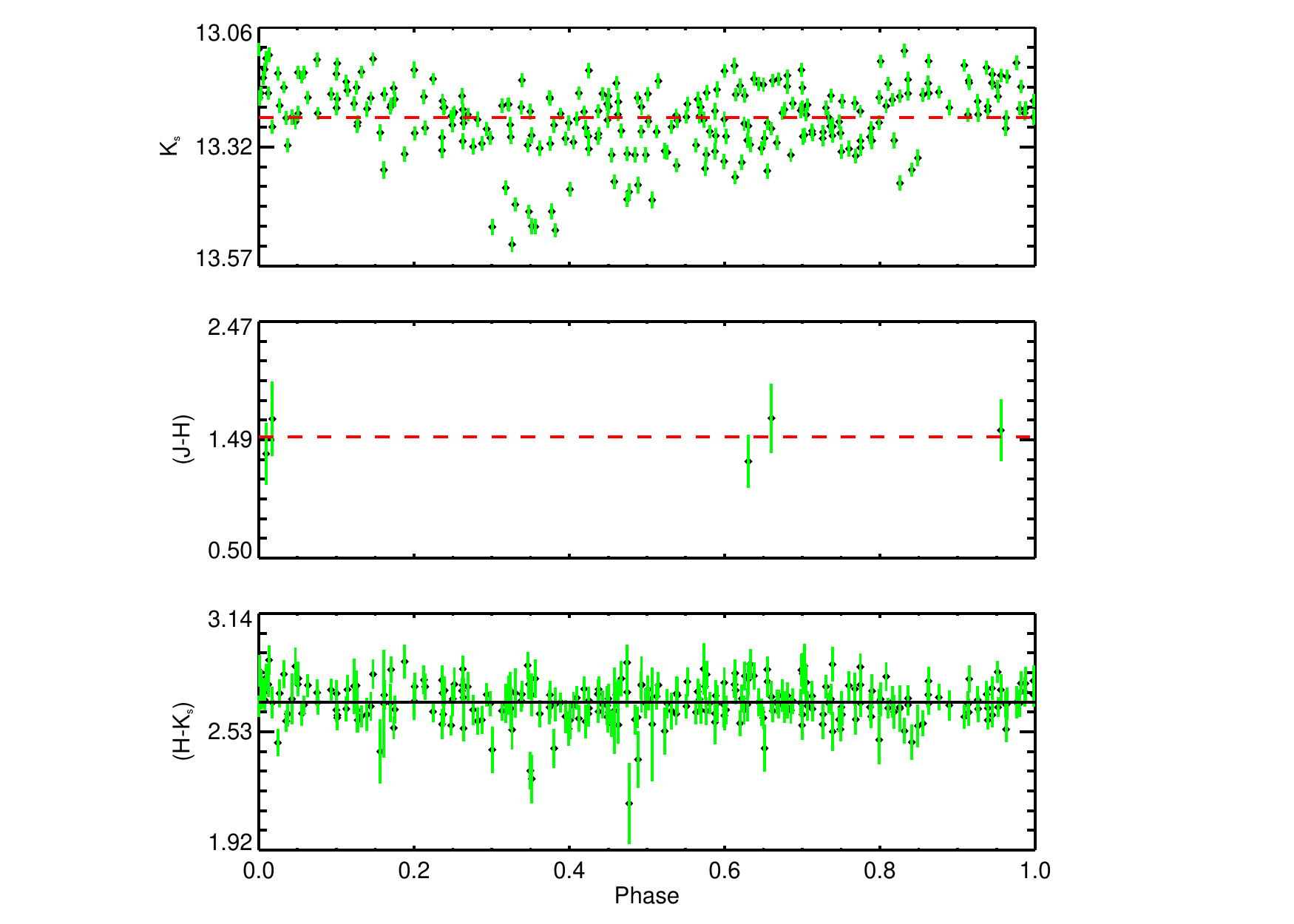}
  \caption{The \emph{K$_{s}$}, (\emph{J}-\emph{H}) and (\emph{H}-\emph{K$_{s}$}) photometry for YLW 10C phased to the 3.0779 $\pm$ 0.0025 day sinusoidal-like period.  The red line indicates the mean value in each panel.  The lack of reliable \emph{J} and \emph{H} photometry prohibits a confident estimate of the variability mechanism.}
\end{figure}

\begin{figure}
  \plotone{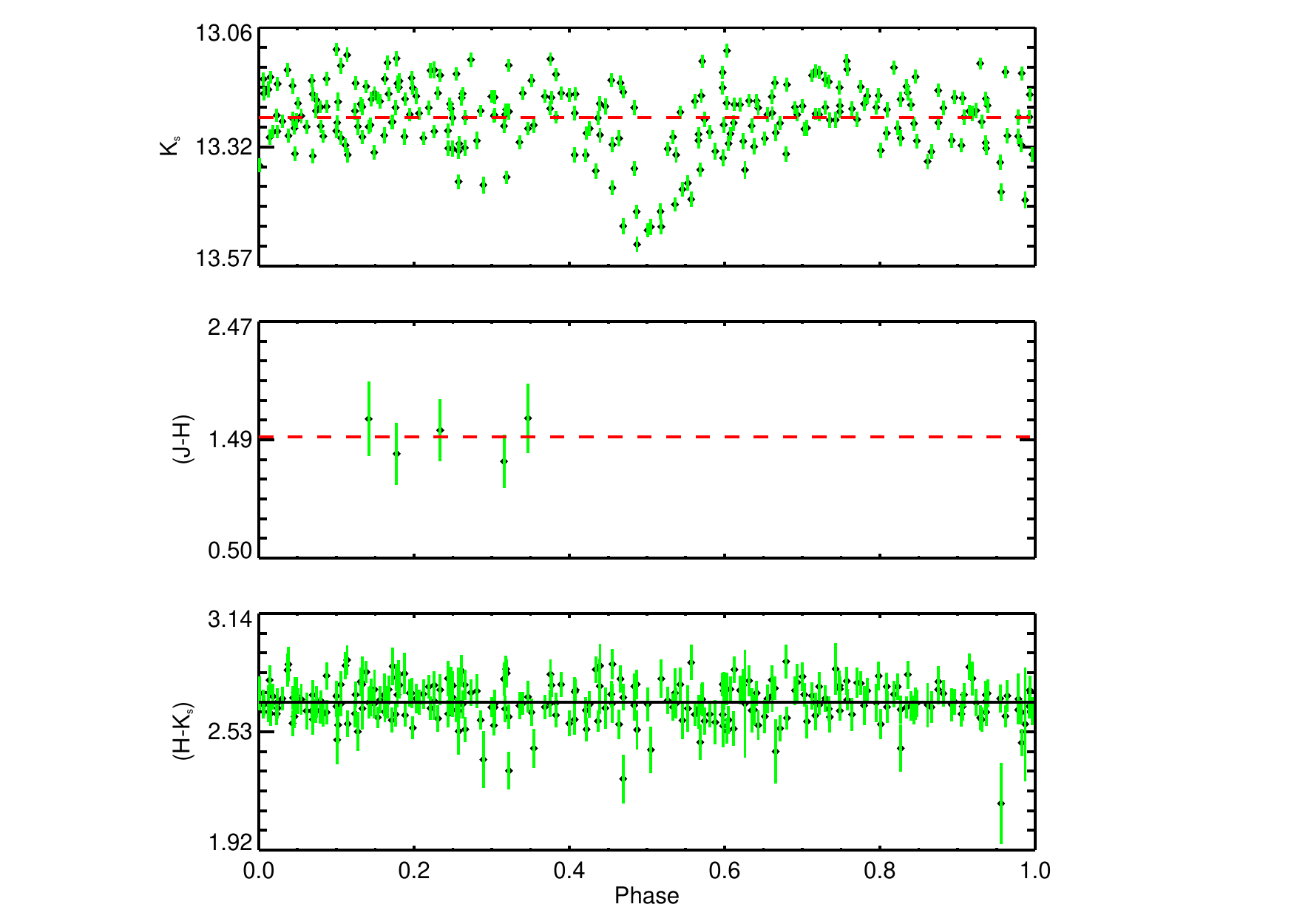}
  \caption{The \emph{K$_{s}$}, (\emph{J}-\emph{H}) and (\emph{H}-\emph{K$_{s}$}) photometry for YLW 10C phased to the 2.9468 $\pm$ 0.0029 day eclipse-like period.  The red line indicates the mean value in each panel.}
\end{figure}

Given the sinusoidal-like and eclipse-like variability is very similar to both YLW 1C and J543, the same physical interpretation is proposed for this star.  However, unlike YLW 1C, the sinusoidal-like variability, presumed to trace the stellar rotation rate, has a longer period by 3.1 hours than the periodic occultations.  This places the hypothetical occulter \emph{within} the co-rotation radius.  The size and distance to the occulter are 3.31 R$_{\star}$ and 3.75 R$_{\star}$.  The occulter is located within the dust sublimation radius as computed using the formulism of \citet{jura98}.  This formulism is only an approximation as it does not take into account dust evaporation and condensation rates, grain size, or grain composition.

\item WL 20W (YLW 11B, ISO-Oph 126): This CTTS is both periodically variable and variable over a long time-scale.  As such, WL 20W is designated both a periodic variable and a LTV.  Fig 39 contains the \emph{K$_{s}$}, (\emph{J}-\emph{H}) and (\emph{H}-\emph{K$_{s}$}) photometry for this star.  The long time-scale variability begins on $\sim$1600 2MASS MJD and has a timescale of 122 days.  The $\Delta$K$_{s}$ depth is 0.26 mag.  Both the (\emph{J}-\emph{H}) and (\emph{H}-\emph{K$_{s}$}) colors become redder during the long time-scale variation, consistent with variable extinction.  The periodic signal is not significant unless the time-series affected by the long time-scale variability is omitted from analysis by the PA (see Fig 11).  Fig 40 contains the \emph{K$_{s}$}, (\emph{J}-\emph{H}) and (\emph{H}-\emph{K$_{s}$}) photometry folded to the sinusoidal-like period, P = 2.1026 $\pm$ 0.0060 days.    The peak-to-trough $\Delta$\emph{K$_{s}$} amplitude for the sinusoidal-like variability is 0.19 mag.  Neither the (\emph{J}-\emph{H}) nor the (\emph{H}-\emph{K$_{s}$}) color is correlated to the \emph{K$_{s}$} variability.  This favors rotational modulation by cool starspots as the variability mechanism.

\begin{figure}
  \plotone{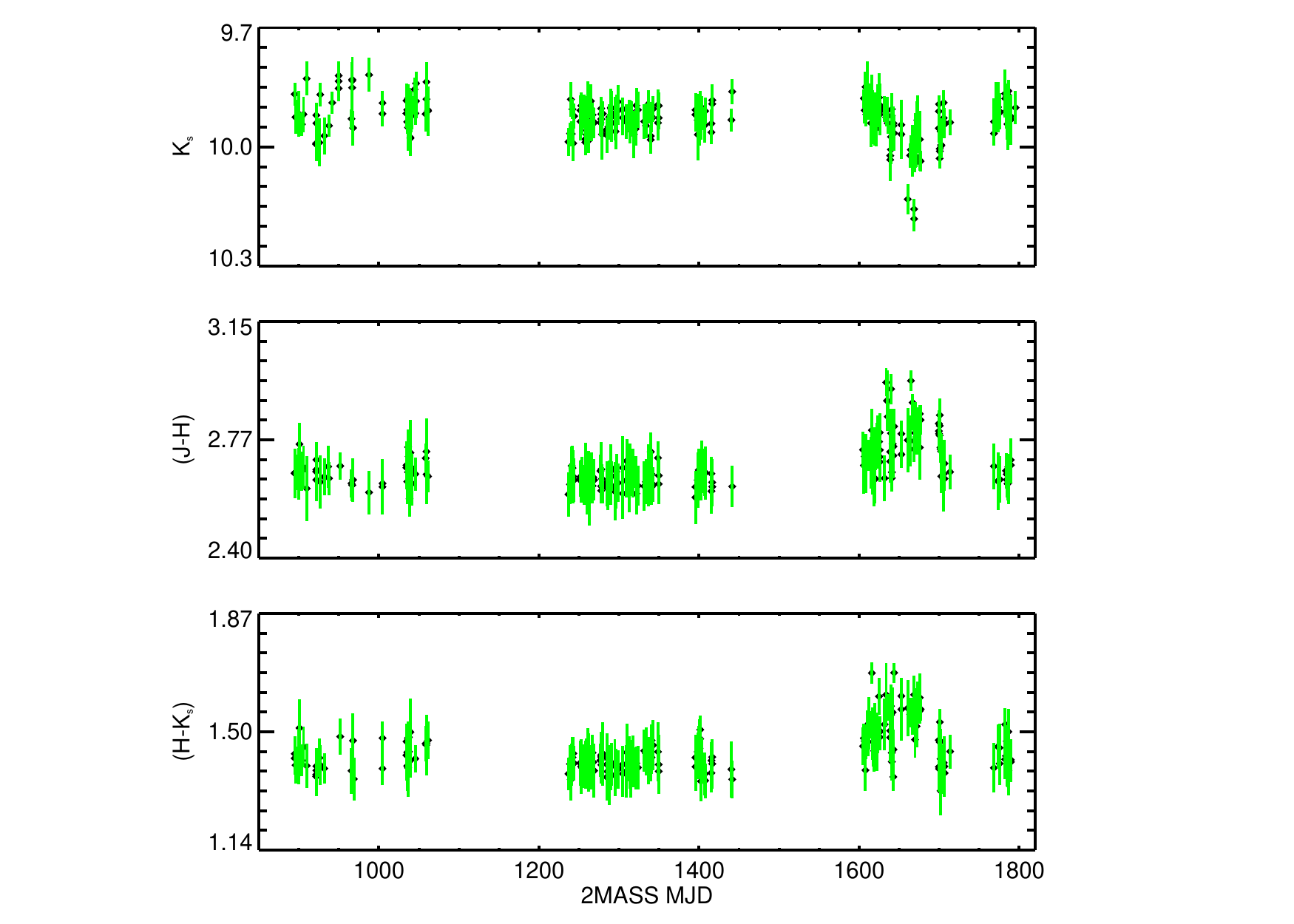}
  \caption{The \emph{K$_{s}$}, (\emph{J}-\emph{H}) and  (\emph{H}-\emph{K$_{s}$}) photometry for WL 20W.  The long time-scale variation is easily seen beginning at $\sim$1600 2MASS MJD.  Both the (\emph{J}-\emph{H}) and  (\emph{H}-\emph{K$_{s}$}) become redder as \emph{K$_{s}$} dims indicating variable extinction as the likely variability mechanism. }
\end{figure}

\begin{figure}
  \plotone{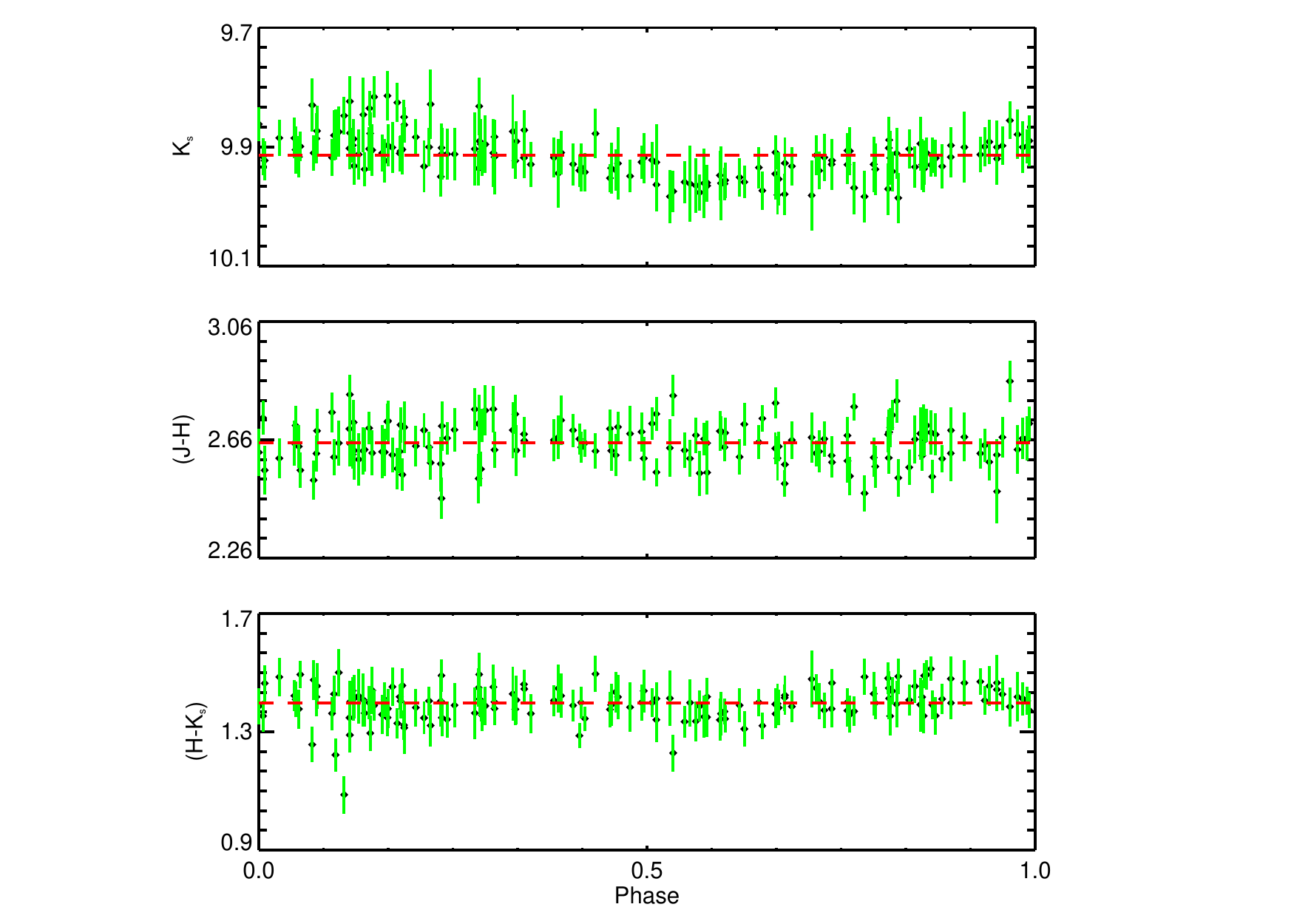}
  \caption{The \emph{K$_{s}$}, (\emph{J}-\emph{H}) and  (\emph{H}-\emph{K$_{s}$}) photometry for WL 20W folded to the period 2.1026 days.  The red line indicates the mean value in each panel. No correlation between $\Delta$\emph{K$_{s}$} and the change in colors points to rotational modulation of cool starspots as the likely variability mechanism. Only the photometry before the time-scale variation is plotted.}
\end{figure}

The variability of AA Tau has been cited as an explanation for the periodic eclipsing variables in this survey.  However recently it has been discovered that AA Tau is exhibiting a long time-scale dimming on the order of 2-3 mag in the V band \citep{bouvier13}.  This long time-scale variation is superimposed on top of the periodic variability.  Additionally the system appears to become bluer in this dim state; a phenomenon seen in UX Ori-type variables \citep{grinin91,herbst94}.  The physical interpretation for UX Ori-type variability is that the star dims due to an asymmetric optically thick occulter beyond the inner circumstellar disk.  The bluer color represents a larger contribution of scattered starlight off the occulting material.

This scenario is an alternative explanation than variable mass accretion for the long time-scale variation observed in WL 20W.  The eclipse depth corresponds to a 2.9 mag dimming when converted to the V band by using the extinction coefficients given in \citet{cohen81}; this is consistent with the AA Tau long time-scale variation.  The (\emph{J}-\emph{H}) and (\emph{H}-\emph{K$_{s}$}) colors do become bluer during the long time-scale variation also consistent with the observations of AA Tau. Our overall interpretation for the variability in WL 20W is a central star rotating with a 2 day period that is occulted by a pocket of optically thick material located beyond the inner circumstellar disk.

It is worth noting this star belongs to a triple system that is spatially resolved in the mid-IR. \citet{ressler01} show the most variable member is, in fact, WL 20S.  They classify this source as Class I through SED fitting of mid-IR photometry.  They show that WL 20E and WL 20W are nearly constant on decadal timescales whereas the flux of WL 20W increased sixfold in 15 years.  As the largest separation between these components is 3.66$\arcsec$, the large aperture size in our work (4$\arcsec$) includes all three stars and thus cannot rule out the possibility that the measured variability arises from this southern component.

\item ISO-Oph 126: This WTTS is similar to WL 20W in that it exhibits both periodic variability and a long time-scale variation.  ISO-Oph 126 is also designated both a periodic variable and a LTV.  Fig 41 contains the \emph{K$_{s}$}, (\emph{J}-\emph{H}) and (\emph{H}-\emph{K$_{s}$}) photometry for this star.  The \emph{J} band photometry is below the survey completeness limits.  Therefore the (\emph{J}-\emph{H}) color is deemed unreliable for analysis.  The long time-scale variation dominates the photometry prior to 1400 2MASS MJD with a time-scale of 349 days.  The $\Delta$\emph{K$_{s}$} depth of this variation from the continuum brightness is 0.10 mag.  The (\emph{H}-\emph{K$_{s}$}) color becomes redder during the long time-scale variation as the star dims.  This is consistent with extinction as the variability mechanism. The PA identifies a significant periodic signal when only the portion of the time-series after 1400 2MASS MJD is analyzed (see Fig 11).  Fig 42 contains the \emph{K$_{s}$}, (\emph{J}-\emph{H}) and (\emph{H}-\emph{K$_{s}$}) photometry folded to the sinusoidal-like period, P = 9.114 $\pm$ 0.090 days.  The peak-to-trough $\Delta$\emph{K$_{s}$} amplitude for the sinusoidal-like variability is 0.06 mag.  The (\emph{H}-\emph{K$_{s}$}) color becomes bluer as \emph{K$_{s}$} brightens favoring a variability mechanism of rotational modulation by accretion-induced hot starspots.

\begin{figure}
  \plotone{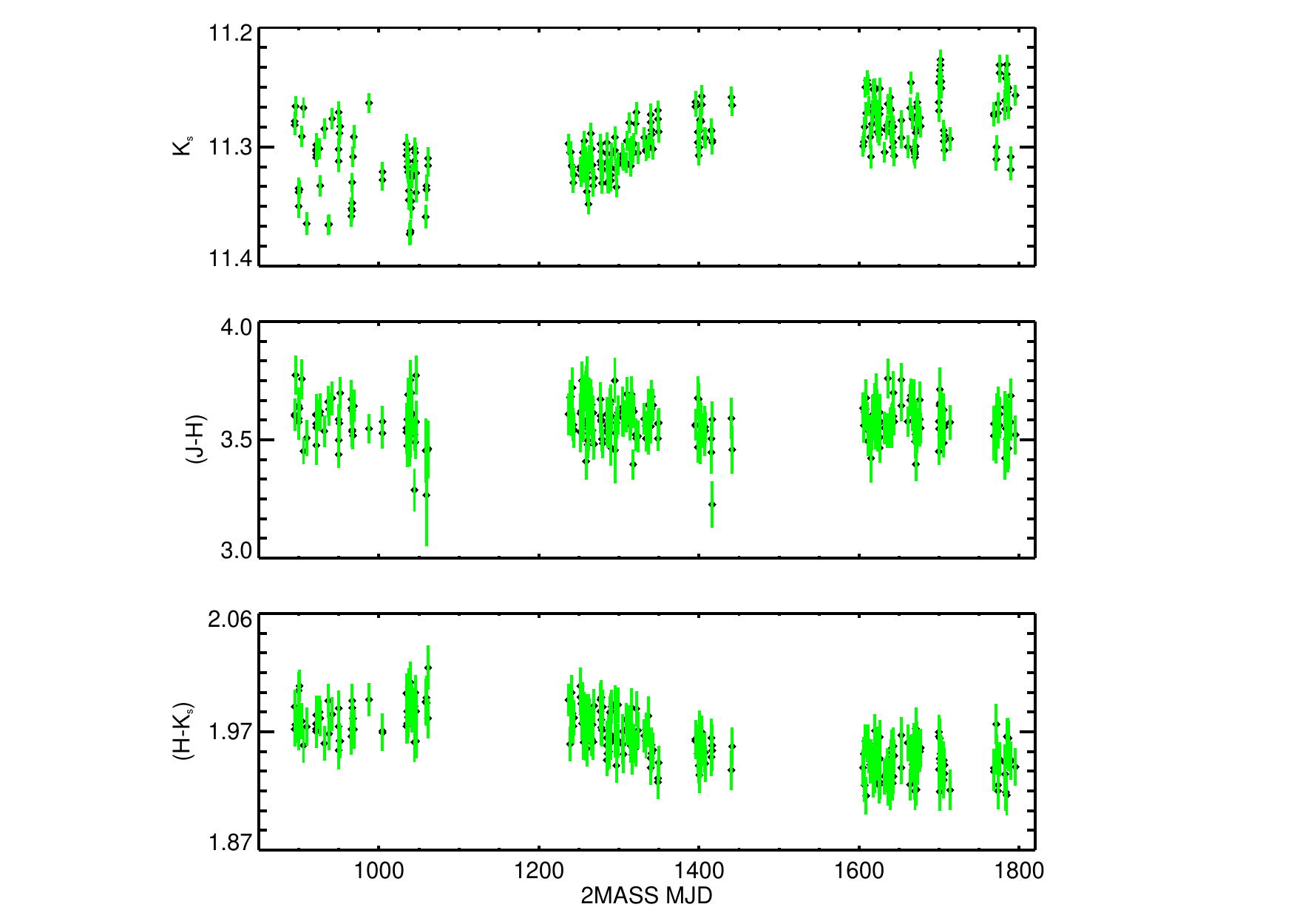}
  \caption{The \emph{K$_{s}$}, (\emph{J}-\emph{H}) and (\emph{H}-\emph{K$_{s}$}) photometry for ISO-Oph 126.  The long time-scale variation is easily seen prior to $\sim$1400 2MASS MJD.  Both the (\emph{J}-\emph{H}) and  (\emph{H}-\emph{K$_{s}$}) become redder as \emph{K$_{s}$} dims indicating variable extinction as the likely variability mechanism.}
\end{figure}

\begin{figure}
  \plotone{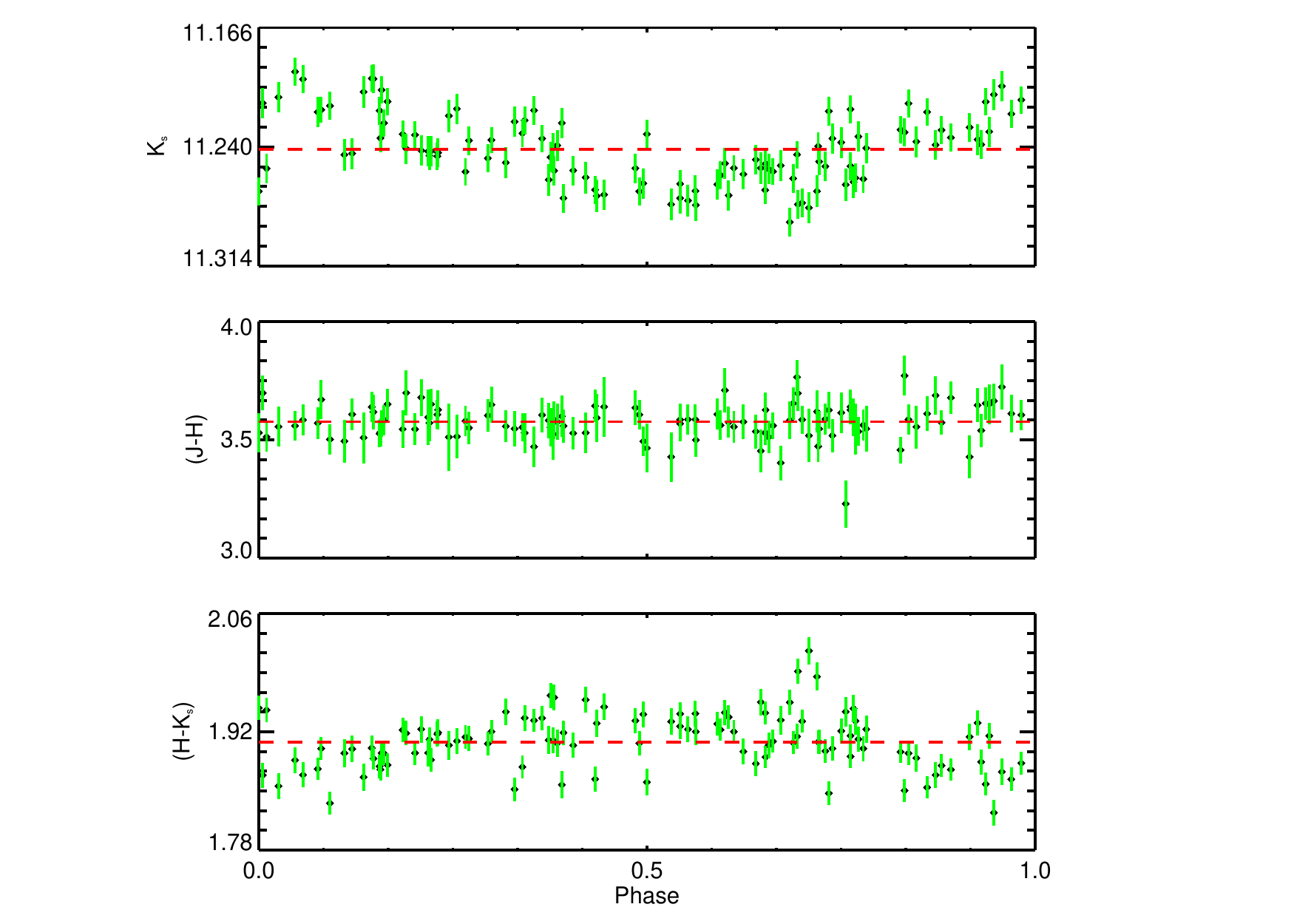}
  \caption{The \emph{K$_{s}$}, (\emph{J}-\emph{H}) and (\emph{H}-\emph{K$_{s}$}) photometry for ISO-Oph 126 folded to the period 9.114 days.  The red line indicates the mean value in each panel. The (\emph{H}-\emph{K$_{s}$}) color becomes bluer as \emph{K$_{s}$} brightens favoring a variability mechanism of rotational modulation by accretion-induced hot starspots.}
\end{figure}

While the origin of the periodic variability can be attributed to stellar surface features, the favored interpretion of extinction in this case potentially challenges the class identification as a diskless WTTS.  While \citet{barsony05} could not provide a YSO classification for ISO-Oph 126, the authors could place an upper limit to the spectral index at $\leq$-0.88.  This allows for the possibility this star is Class II and surrounded by an optically thick accretion disk.

One intriguing option is the occultation by a disk surrounding an orbital companion.  This scenario is invoked to explain long time-scale variability in evolved star systems $\epsilon$ Aur \citep{guinan02,kloppenborg10,stencel11}, EE Cep \citep{mikolajewski99,graczyk03,mikolajewski05,galan10} and most recently in the young star system 1SWASP J140747.93-394542.6 ('J1407') \citep{mamajek12}.  Photometric variations within the long time-scale variations are believed to arise from structure within the occulting disk.  This structure may represent new planets (EE Cep \citep{galan10}), or it may represent planetary moons (J1407 \citep{mamajek12}).  It is noted that there is significant scatter in the \emph{K$_{s}$} time-series during the first half of the long time-scale variation in comparison to the second half of this variation(see Fig 43).  High resolution imaging or radial velocity monitoring may help to confirm the existence of a companion to ISO-Oph 126.

\begin{figure}
  \plotone{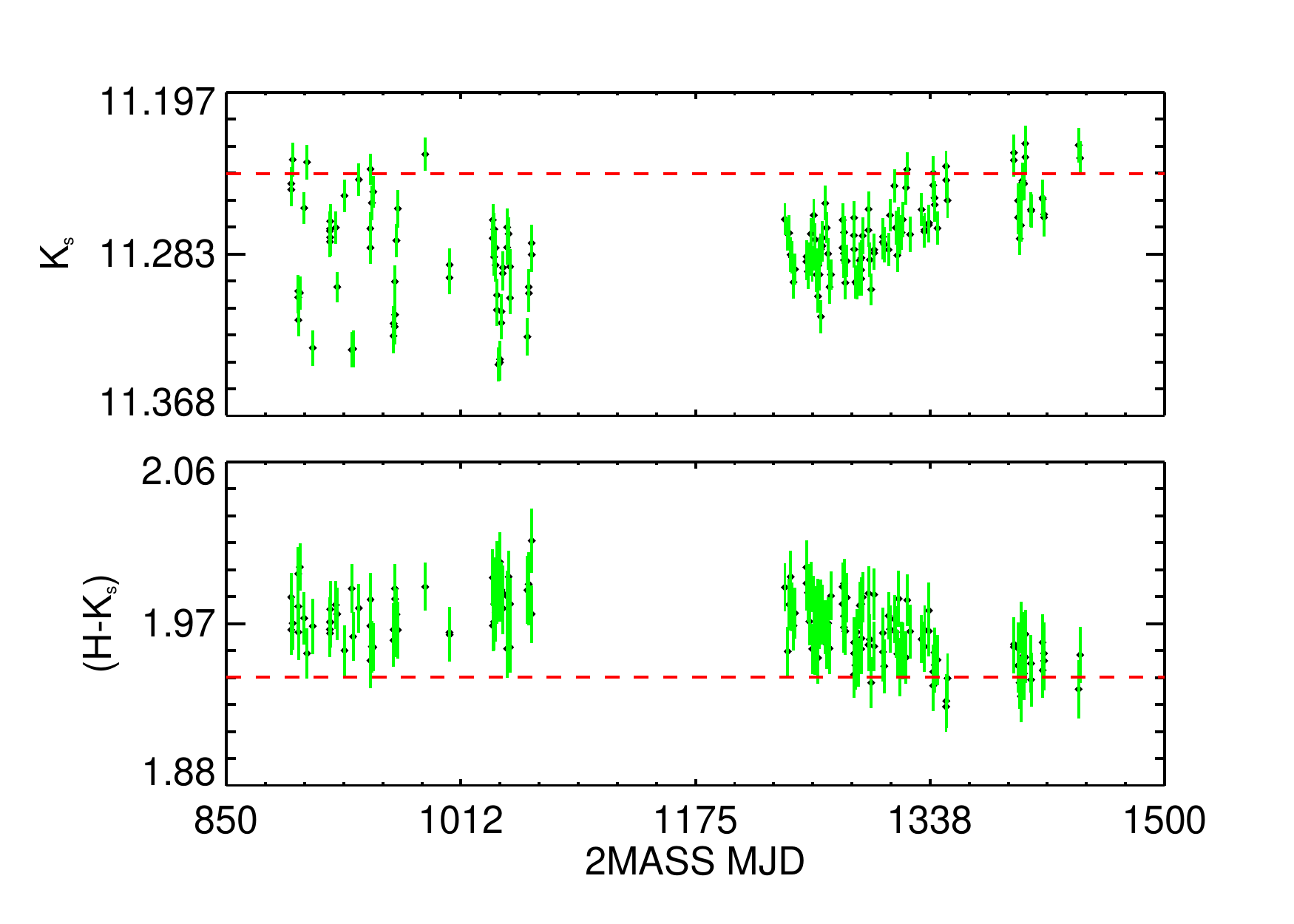}
  \caption{The \emph{K$_{s}$} and (\emph{H}-\emph{K$_{s}$}) photometry for the long time-scale variation in ISO-Oph 126.  The \emph{K$_{s}$} shows considerable structure during ingress, but is nearly stable during the ``eclipse'' egress. The red line indicates the mean magnitude of the continuum photometry.}
\end{figure}

\item WL 15 (YLW 7A, ISO-Oph 108): This star is one of the brightest at K$_{s}$ ($\overline{K_{s}}$ = 7.05 mag) and the reddest (($\overline{\emph{H}-K_{s}}$) = 4.01 mag) in the variable catalog.  WL 15 is a Class I YSO.  Similar to WL 20W and ISO-Oph 126, this star exhibits a large amplitude, long time-scale variation over-top of a smaller amplitude periodic variability.  Unlike WL 20W and ISO-Oph 126, the photometry during the long time-scale variation is too sparse to confidently identify a time-scale.  Therefore WL 15 is only designated a periodic variable.  Fig 44 contains the \emph{K$_{s}$}, (\emph{J}-\emph{H}) and (\emph{H}-\emph{K$_{s}$}) photometry for WL 15.  The J band photometry is below the survey completeness limit and is deemed unreliable for analysis.  The long time-scale variation is observed between 1396 and 1443 2MASS MJD with a $\Delta$\emph{K$_{s}$} amplitude of $\sim$1 mag.  The mean (\emph{H}-\emph{K$_{s}$}) color does not change as the star dims during this event.  Even including the long time-scale variation, the PA found a significant periodic signal.  Fig 45 contains the \emph{K$_{s}$}, (\emph{J}-\emph{H}) and (\emph{H}-\emph{K$_{s}$}) photometry for WL 15 folded to the sinusoidal-like period, P = 19.412 $\pm$ 0.085 days.  This variability has a peak-to-trough $\Delta$\emph{K$_{s}$} amplitude of 0.90 mag.  The (\emph{H}-\emph{K$_{s}$}) color is not correlated to the \emph{K$_{s}$} photometry.  

\begin{figure}
  \plotone{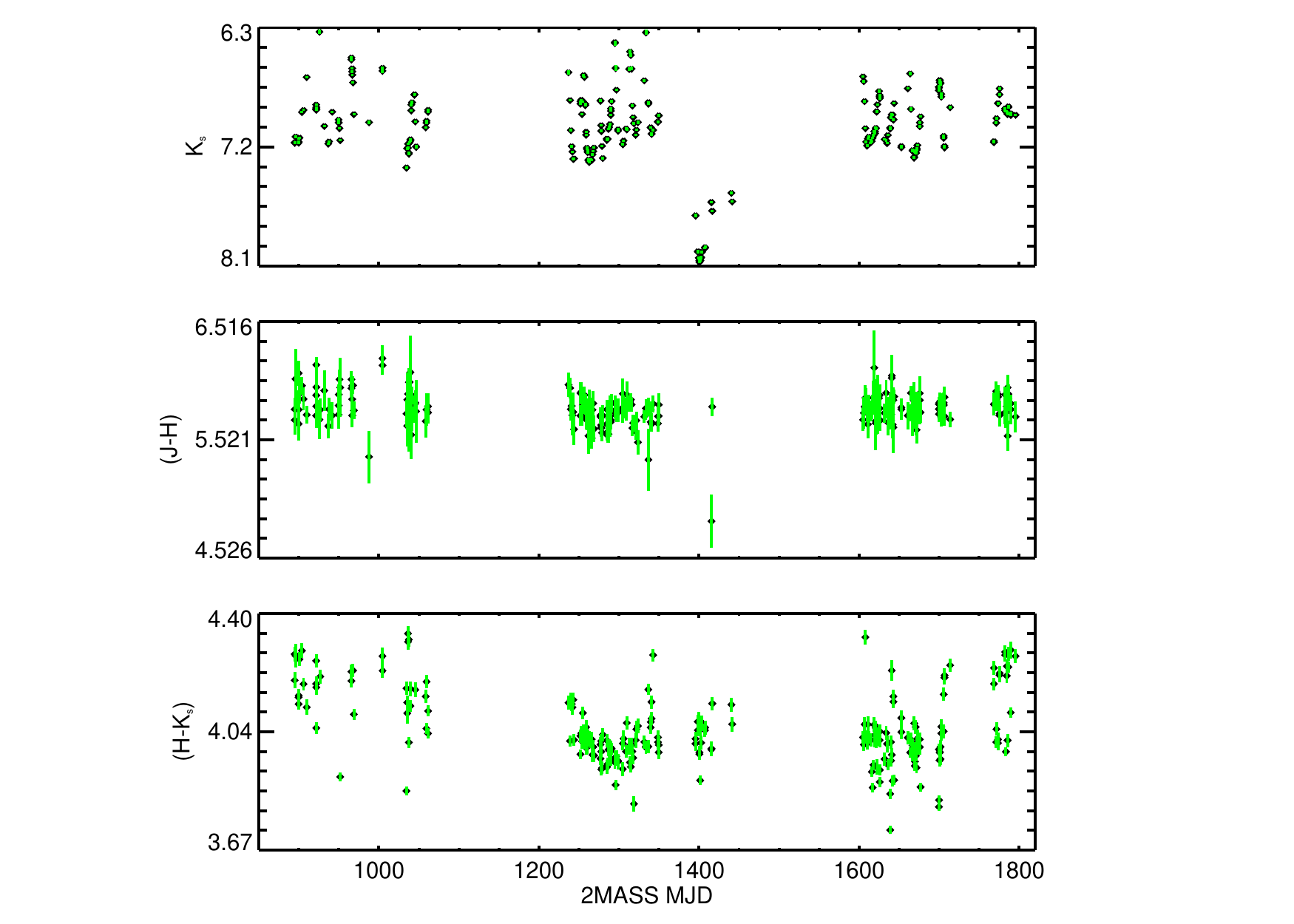}
  \caption{The \emph{K$_{s}$}, (\emph{J}-\emph{H}) and (\emph{H}-\emph{K$_{s}$}) photometry for WL 15.  The $\sim$19 day period is clearly evident and appears to continue even through a $\sim$1 mag drop in \emph{K$_{s}$} band flux.  The photometry during the larger amplitude flux decrease is too sparse to confidently determine a time-scale.  A lack of a trend in the (\emph{H}-\emph{K$_{s}$}) color during this event highly suggests against extinction except by an opaque occulter.}
\end{figure}

\begin{figure}
  \plotone{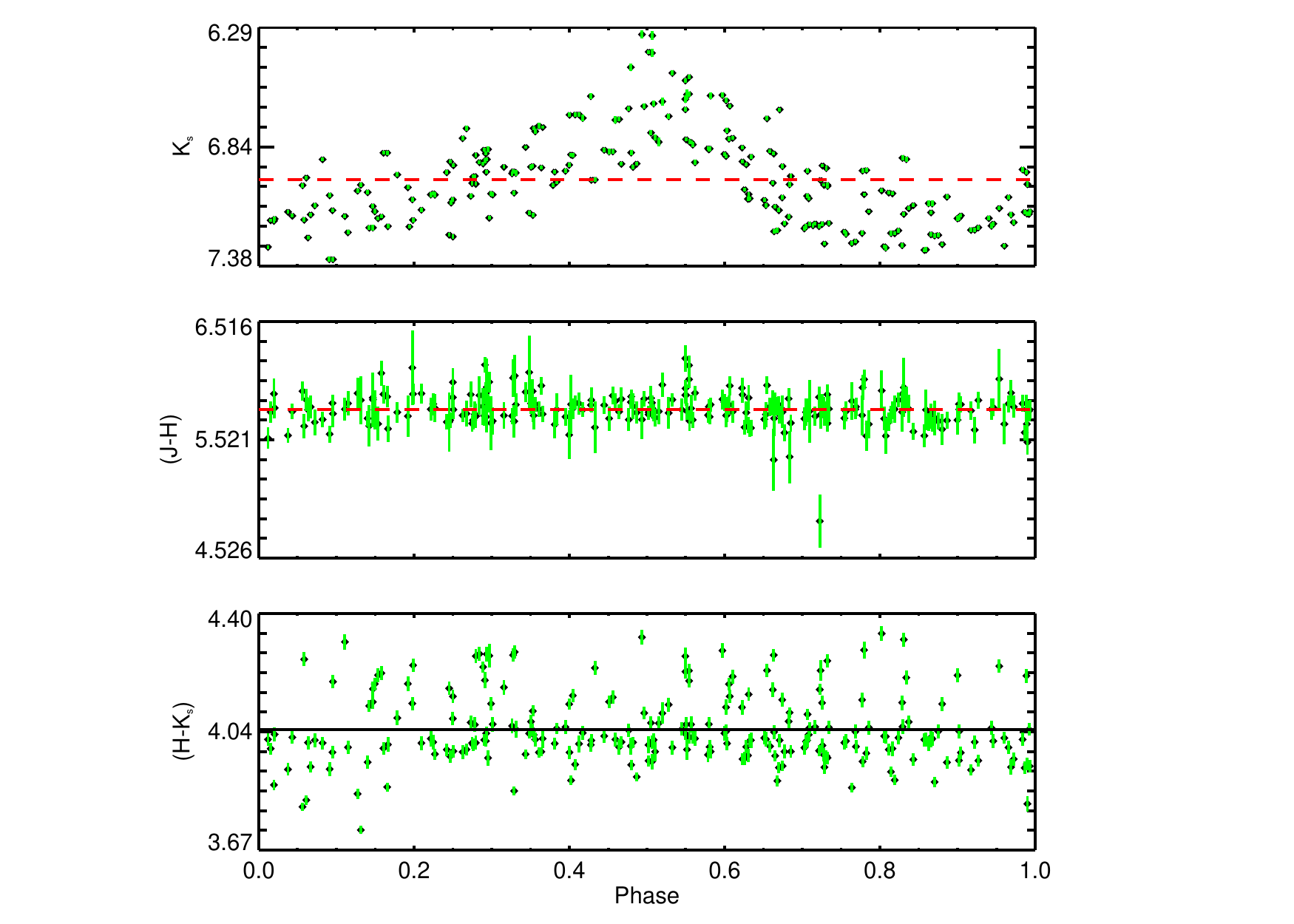}
  \caption{The \emph{K$_{s}$}, (\emph{J}-\emph{H}) and  (\emph{H}-\emph{K$_{s}$}) photometry for WL 15 folded to the period 19.412 days.  The red line indicates the mean value in each panel.}
\end{figure}

While the colorless periodic variability favors rotational modulation by cool starspots, the amplitude of variability does not.  The highest amplitude variability confirmed to as caused by cool starspots, to date, is $\Delta$V = 0.63 mag \citep{strassmeier97}.  This amplitude is nearly 0.3 mag lower than that observed for WL 15.  In addition, the contrast between the starspot and surrounding photosphere increases toward bluer wavelengths.  Therefore the amplitude will be even larger in the optical.  The origin behind the long time-scale variation is equally peculiar.  As the (\emph{H}-\emph{K$_{s}$}) color does not become redder as the star dims, this seems to eliminate extinction as variability mechanism.  However, variability by rotational modulation of surface features seems inplausible given the timescale and amplitude of the variation.

\end{itemize}

\section{SUMMARY}
High precision, high cadence \emph{J}, \emph{H}, \emph{K$_{s}$} photometry is obtained for 7815 stars in the direction of the $\rho$ Oph molecular cloud with a temporal baseline of $\sim$ 2.5 yrs.  Spurious detections, partially resolved doubles or galactic contamination are eliminated from the photometry.  The target sample meeting the specifications for time-series variability analysis includes 1678 stars.  A seven point variability test is used to identify 101 variable stars, which is 6$\%$ of the parent sample.  These tests are sensitive to variability on a variety of different time-scales and forms (e.g. sinusoidal, 'eclipse-like', etc.).

Of the 101 stars in the variable catalog, 15$\%$ are located ``on cloud'' while only 1$\%$ lie within the ``field''.  Location ``on cloud'', variability and (\emph{H}-\emph{K$_{s}$}) colors redder than a 3 Myr isochrone are used to estimate membership in the $\rho$ Oph star forming region.  This method has identified 22 stars as previously unknown candidate $\rho$ Oph members.

The effects of observing strategy on variability detection and measured amplitudes is investigated by comparing this work to the $\rho$ Oph variability study performed by \citet{oliveira08}.  These two studies have 464 stars in common; AC08 identified 7$\%$ as variable stars and this work identifies 18$\%$.  The increase in detection fraction is not caused by different sensitivities in the separate variability criteria used in each survey.  This work also found both a higher \emph{K$_{s}$} and (\emph{H}-\emph{K$_{s}$}) color amplitude in 25 stars identified as variable in both surveys than measured by AC08.  Therefore a high cadence observational strategy will discover more variables within a given set of stars and it will more accurately characterize intrinsically higher amplitude variability.

The \emph{K$_{s}$} variability and stellar color behaviors are used to estimate the physical mechanism responsible for the variability.  Rotational modulation by long-lived cool starspots is expected to produce colorless, periodic variability.  Rotational modulation of long-lived hot starspots (e.g. accretion) is, also, expected to be periodic, while short-lived starspots (e.g. flares) is not.  The star becomes bluer as it brightens in both cases.  Extinction induced variability is either periodic or exists on long time-scales based on the geometry of the occulter relative to the star.  Changes in the mass accretion rate onto the star is not expected to be periodic, but occur on time-scales ranging from days to years.  As this rate changes, the star becomes bluer as the star dims.

Identifying periodic variability within the variable catalog is done via a newly improved period-searching algorithm, the Plavchan algorithm.  The algorithm tests tens of thousands of periods with uniform frequency sampling between 0.1 to 1000 days.  This is done by comparing the observed light curve to a dynamically generated prior.  The statistical significance of individual periods is computed via two methods: the distribution of power values at other periods in the same periodogram and the distribution of maximum power values for all sources in an ensemble survey.  The Plavchan algorithm finds periodic variability in 32$\%$ of the variable catalog with periods ranging from 0.49 to 92 days.

The periodic variables are split into two sub-categories: sinusoidal-like and eclipse/inverse eclipse-like.  Sinusoidal-like periodic variability describes a sinusoidal-like change in the observed flux when the time-series is folded to the most significant period.  Rotational modulation by cool starspots is believed to be the common variability mechanism in this sub-category.  Sinusoidal-like periodic variables are found in each YSO class (3 Class I, 8 Class II, 8 Class III).  Eclipse-like periodic variability results in discrete drops, or ``dips'', in the observed flux when the time-series is folded to the most significant period.  Periods range from 2 to 8 days with the duration of these dips lasting less than 30$\%$ of one periodic epoch.  This sub-category contains 6 stars with a median peak-to-trough $\Delta$\emph{K$_{s}$} and $\Delta$(\emph{H}-\emph{K$_{s}$}) color amplitudes of 0.31 and 0.11 mag, respectively.  Variable extinction is the likely mechanism for the eclipse-like variations.  All stars in this sub-category are Class II YSOs.  The inverse eclipse-like variables, WL 4 and YLW 16A have periods of 65.61 and 92.3 days, respectively.  The variability mechanism proposed in both cases is the periodic obscuration of one component in a close binary by a warped circumbinary disk.

In half of the eclipse-like variables (YLW 1C, 2MASS J16272658-2425543, YLW 10C) an additional statistically significant period is identified.  This sinusoidal-like periodic variability coupled with the presence of ``dips'' suggests a rapidly rotating spotted star occulted by a clump of optically thick material in the inner accretion disk.  These stars strengthen the interpolation posed to explain the variability of other YSO AA Tau-like variables \citep{morales11}.  The periods corresponding to periodic occultations in YLW 1C and YLW 10C are located near thier respective co-rotation radii.  The mechanism driving these occulations could arise from a warped inner circumstellar disk caused by an inclined magnetic dipole, or could be the pre-natal cloud of a forming hot Jupiter.

Long time-scale variables is a variability subclass, containing 31 stars, where the measured flux increases or decreases consistently over months or years.  The variability time-scale is measured using a differencing technique and approximates the time between maximum and minimum brightness.  The measured time-scales range from 64 to 790 days.  The peak-to-trough $\Delta$\emph{K$_{s}$} amplitudes range from 0.05 to 2.31 mag and the peak-to-trough $\Delta$(\emph{H}-\emph{K$_{s}$}) color amplitudes range from 0.06 to 1.32 mag.  Variable extinction and variable accretion rates are both equally likely to cause long time-scale variability.  This subclass contains 25 known YSOs with 7 Class I, 15 Class II and 3 Class III stars. 

The time-series of irregular variables are aperiodic and do not vary over any discernible time-scale.  This subclass contains more members (40) than either the periodic or long time-scale subclasses.  The peak-to-trough $\Delta$\emph{K$_{s}$} amplitudes range from 0.04 to 1.11 mag and peak-to-trough $\Delta$(\emph{H}-\emph{K$_{s}$}) color amplitudes range from 0.05 to 0.75 mag.  No single dominant variability mechanism explains irregular variability.  Only 9 known YSOs (1 Class I, 7 Class II, 1 Class III) are irregular variables.

The CTTS WL 20W and the WTTS ISO-Oph 126 are similar in both have a long time-scale variation superimposed onto a periodic signal.  In both cases, the physical mechanisms for the variability is consistent with an occultation of a rapidly rotating spotted star by optically thick material outside the inner accretion disk.  For WL 20W, the sinusoidal-like variability has a period of 2.1026 days and a peak-to-trough $\Delta$\emph{K$_{s}$} amplitude of 0.19 mag.  The long time-scale variability has a duration of 122 days with a $\Delta$\emph{K$_{s}$} eclipse depth of 0.26 mag.  For ISO-Oph 126, the sinusoidal-like variability has a period of 9.114 days and a peak-to-trough $\Delta$\emph{K$_{s}$} amplitude of 0.06 mag.  The long time-scale variability has a duration of 349 days with a $\Delta$\emph{K$_{s}$} eclipse depth of 0.10 mag. 

The very high amplitude periodic variability measured in the Class I star WL 15 is not consistent with any proposed mechanism.  The observed 47 day colorless decrease in \emph{K$_{s}$} band brightness of $\sim$1 mag is also not easily estimated.

From cross-referencing the target sample with two previous surveys, 72 stars have been assigned a YSO classification (13 Class I, 47 Class II, 12 Class III).  The variability fraction of these YSOs is 79$\%$.  The variability fraction differs according to YSO class with 92$\%$ of Class I and Class III stars identified as variable; this fraction drops to 72$\%$ for Class II stars.  The amplitude of both brightness and color variability are decreasing functions of YSO class.  The median peak-to-trough $\Delta$\emph{K$_{s}$} amplitude for Class I, II and III stars are 0.77, 0.31 and 0.08 mag, respectively.  In addition, the median peak-to-trough $\Delta$(\emph{H}-\emph{K$_{s}$}) color amplitudes are 0.81, 0.21 and 0.07 mag for each class respectively.

\acknowledgments
We will now express our thanks to our anonymous referee for their conversation and comments.  We would also like to express our thanks to Mary Barsony for her many insightful questions and suggestions.  This publication makes use of data products from the Two Micron All Sky Survey, which is a joint project of the University of Massachusetts and the Infrared Processing and Analysis Center/California Institute of Technology, funded by the National Aeronautics and Space Administration and the National Science Foundation.  This research has made use of the NASA Exoplanet Archive, which is operated by the California Institute of Technology, under contract with the National Aeronautics and Space Administration under the Exoplanet Exploration Program.  The authors would like to acknowledge the Infrared Processing and Analysis Center's Visiting Graduate Student program for providing the opportunity and funding for this publication.

\renewcommand{\theequation}{A-\arabic{equation}}
\renewcommand{\thefigure}{A-\arabic{figure}}
\setcounter{equation}{0}  
\setcounter{figure}{0} 
\section*{APPENDIX. Empirical Determination of $\chi_{n_0}^2$ Dependence on Parameters}  
In this Appendix an analysis of the dependence of the PA periodogram $\chi$$_{n_{0}}^2$ power values on the number of observations and periodogram parameters \emph{n$_{0}$} and \emph{p} is presented.  The ensemble survey of mostly non-variable stars is used to carry out this analysis and to present an alternative approach to evaluate the statistical significance of periodogram power values.

First, a random subset of 180 stars is chosen from the survey collection of 1678 stars.  These stars are evenly distributed in N$_{obs}$ and \emph{K$_{s}$} magnitude.  For a given set of parameters \emph{p} and \emph{n$_{0}$}, the maximum $\chi$$_{n_{0}}^2$ periodogram power value is computed for the 180 stars.  Since other algorithms exist that specialize in finding periodic sources with low numbers of detection \citep[N$_{obs}$$\sim$20]{dworetsky83}, test cases are limited to 0.04 $<$ \emph{p} and $\leq$ 0.5 and 12 $<$ \emph{n$_{0}$} $\leq$ 250.  Fig A-1 shows the dependence of $\chi$$_{n_{0}}^2$ on N$_{obs}$ for the 180 stars with a particular set of \emph{p} and \emph{n$_{0}$}.  This dependence is somewhat expected -- a smaller number of observations can result in an increase in the likelihood for false-positive periodogram peaks.

The distribution of $\chi$$_{n_{0}}^2$ values as a function of N$_{obs}$ is well-described by the functional form:

\begin{equation}
  F(N_{obs}) = (\frac{a}{N_{obs} - b})^{1.5} + c
\end{equation}

\noindent where \emph{a}, \emph{b} and \emph{c} represent real numbers that differ for a given \emph{p} and \emph{n$_{0}$}.  As the power law index decreases below 1.5 for \emph{p} $<$ 0.04 and/or \emph{n$_{0}$} $<$ 12, these ranges are excluded from the analysis.  Eqn A-1 is found via trial and error to minimize the residuals when compared to a variety of functional forms tested, rather than from an analytic derivation based upon first principles.

\begin{figure}
  \plotone{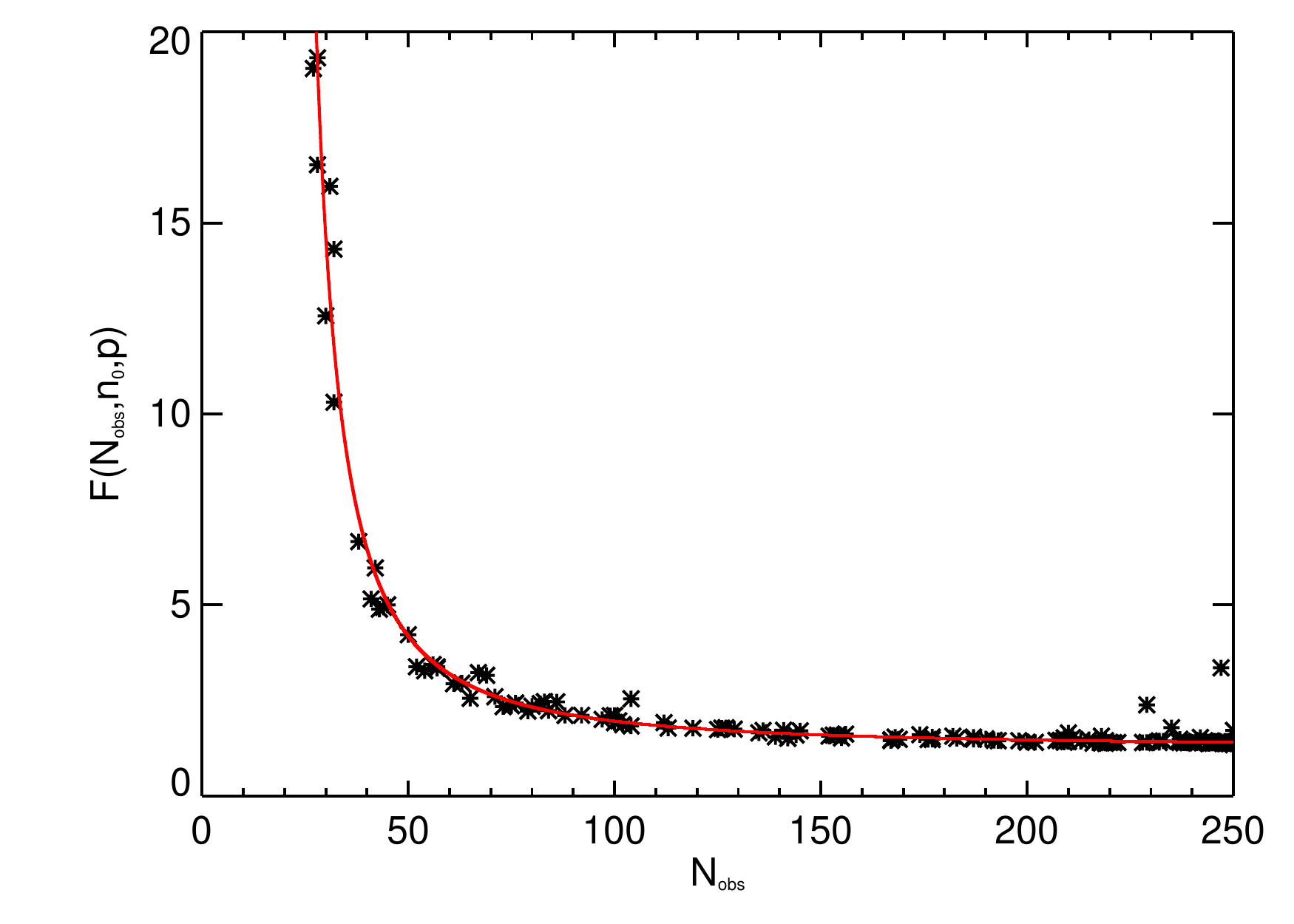}
  \caption{Graph of the $\chi$$_{n_0}^2$ value as a function of detection size, N$_{obs}$ , for the test case with the parameters: n0 = 25 and p = 0.06. The red line represents the functional fit of Eqn A-1 with values of (a,b,c) = (62.5495,18.4963,1.2523) where the residuals are minimized.}
\end{figure}

The next step is to determine how the constants \emph{a}, \emph{b} and \emph{c} vary as functions of the parameters \emph{p} and \emph{n$_{0}$}.  Fixing \emph{n$_{0}$}, the maximum $\chi$$_{n_{0}}^2$ as a function of N$_{obs}$ is empirically fit to 10 chosen \emph{p} values resulting in 10 different values of \emph{a}, \emph{b} and \emph{c}.  The same process is repeated except \emph{p} is fixed and \emph{n$_{0}$} is varied.  Fig A-2 displays the dependence of \emph{a} on the parameters \emph{p} and \emph{n$_{0}$}.  Six fits to the dependence of \emph{a}, \emph{b} and \emph{c} on parameters \emph{p} and \emph{n$_{0}$} are determined empirically through trial and error to be:

\begin{subequations}
  \begin{equation}
    f_a(n_0) = -0.3491(n_0-17.0796) e^{-0.0451n_0} +  63.4573
  \end{equation}
  \begin{equation}
    f_b(n_0) = 1.3023(1-\frac{23.3762}{n_0})  e^{-0.0283n_0} + 18.4347
  \end{equation}
  \begin{equation}
    f_c(n_0) = 0.2796e^{-0.0381n_0} + 1.1467
  \end{equation}
\end{subequations}
\begin{subequations}
  \begin {equation}
    f_a(p) = 82.6288  e^{-12.1989 p} + 25.4356
  \end{equation}
  \begin{equation}
    f_b(p) = 3.0791p^{-0.6377}
  \end{equation}
  \begin{equation}
    f_c{p} = (p-0.0305)^{-0.0395} + 0.0905
  \end{equation}
\end{subequations} 

\begin{figure}
  \plotone{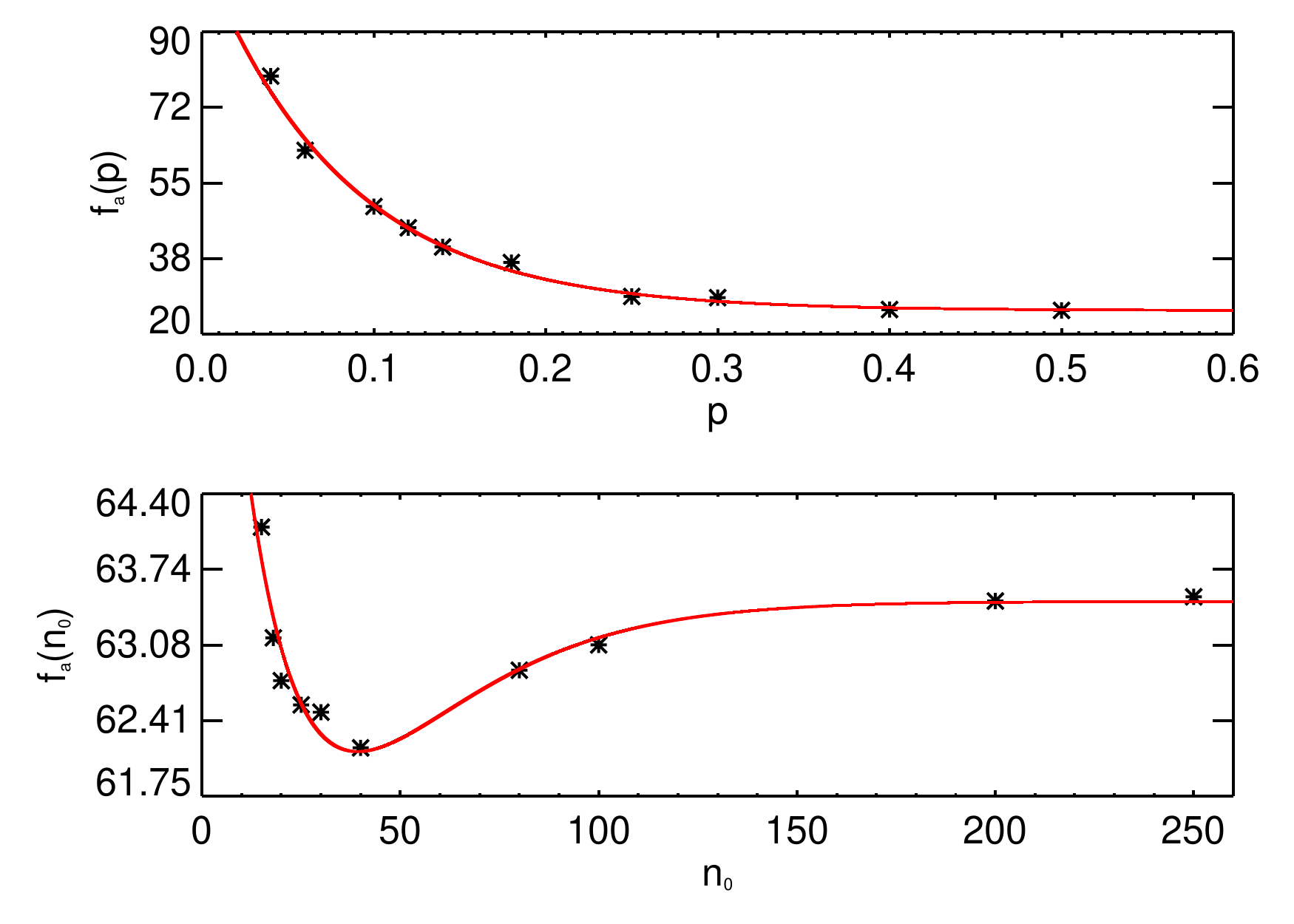}
  \caption{\emph{Top}: Graph of the constant \emph{a} as a function of parameter \emph{p} (see Eqn A-3a).  \emph{Bottom}: Graph of the constant \emph{a} as a function of parameter \emph{n$_{0}$} (see Eqn A-2a).} 
\end{figure}

\noindent The particular functional forms of A-2 and A-3 are again not analytically motivated, but instead minimize the residuals from a variety of functional forms tested.

These six functions of one parameter are combined into three functions of both parameters \emph{p} and \emph{n$_{0}$}.  This is accomplished by replacing the constant term in the \emph{n$_{0}$} function by the entire corresponding \emph{p} function.  The constant term from \emph{f$_{c}$(p)} function is also dropped.  Thus, the following functions are found to adequately describe the dependence of \emph{a}, \emph{b} and \emph{c} on parameters \emph{p} and \emph{n$_{0}$} for this survey:

\begin{subequations}
  \begin{equation}
    f_a(n_0, p) = -0.3491(n_0-17.0796) e^{-0.0451n_0} + 82.6288  e^{-12.1989 p} + 25.4356 
  \end{equation}
  \begin{equation}
    f_b(n_0, p) = 1.3023(1-\frac{23.3762}{n_0})  e^{-0.0283n_0}+3.0791p^{-0.6377}
  \end{equation}
  \begin{equation}
    f_c(n_0, p) = 0.2796e^{-0.0381n_0}+(p-0.0305)^{-0.0395}
  \end{equation}
\end{subequations}

\noindent and therefore Eqn A-1 can be rewritten as:

\begin{equation}
  F(N_{obs},n_0,p) = (\frac{f_a(n_0,p)}{N_{obs}-f_b(n_0,p)})^{1.5} + f_c(n_0,p)
\end{equation}

The values \emph{p} and \emph{n$_{0}$} = 40 -- used throughout this paper to identify periodic variables -- are an optimal choice of parameters for this survey.  They yield the smallest residuals when the 180 test cases are fit in Eqn A-5.  Thus, for this survey the maximum peak power in the periodogram for a non-variable star is approximately given by the expression:

\begin{equation}
  F(N_{obs}, 40, 0.06) = (\frac{63.8629}{N_{obs}-18.6927})^{1.5} + 1.2100
\end{equation}

The validity of the numerical fits (Eqns A-4(a-c)) is verified using an additional 18 test cases with randomly selected values of \emph{p} and \emph{n$_{0}$}.  The $\chi$$_{n_0}^2$ versus N$_{obs}$ distributions are again fit using Eqn A-1, yielding 18 ``observed'' \emph{a}, \emph{b} and \emph{c} values for each pair of \emph{p} and \emph{n$_{0}$}.  Predicted values for \emph{a}, \emph{b} and \emph{c} are found by using Eqns A-4(a-c) and compared to the ``observed'' values.  The mean percent errors between the observed and predicted values for the 18 test cases are -1.1$\pm$3.4$\%$ for \emph{f$_{a}$}, 1.1$\pm$3.5$\%$ for \emph{f$_{b}$} and -0.1$\pm$1.8$\%$ for \emph{f$_{c}$}.  Table A-1 contains the percent errors for each of the individual 18 test cases.  Eqns A-4(a-c) therefore adequately predict the values \emph{a}, \emph{b} and \emph{c} in Eqn A-1 for any set of parameters \emph{p} and \emph{n$_{0}$} within the parameter space explored in this survey.  Thus, this demonstrates the PA algorithm is reasonably ``well-behaved.''  Eqn A-5 could be applied to different surveys and cadences.  However, the particular numerical values in Eqns A-4 and A-6 for \emph{a}, \emph{b} and \emph{c} likely depend on the specific cadence of a survey.

To evaluate the statistical significance of a peak period power value in a periodogram, it does not suffice to identify the expected value for the ensemble survey.  The scatter about the expected value is also necessary.  This scatter, or standard deviation ($\sigma$), of the peak period power values about the expected value depends on N$_{obs}$ in a predictable fashion for this survey (Fig A-3).  To characterize this scatter, the scatter for each survey star is grouped into bins as a function of N$_{obs}$, with a bin size of 25.  An average $\sigma$ is computed for each bin and an empirical fit to this distribution is made, given by Eqn A-7:

\begin{equation}
  \sigma(N_{obs})=\frac{2.3790}{N_{obs}-21.6449}+0.0105
\end{equation}

\begin{figure}
  \plotone{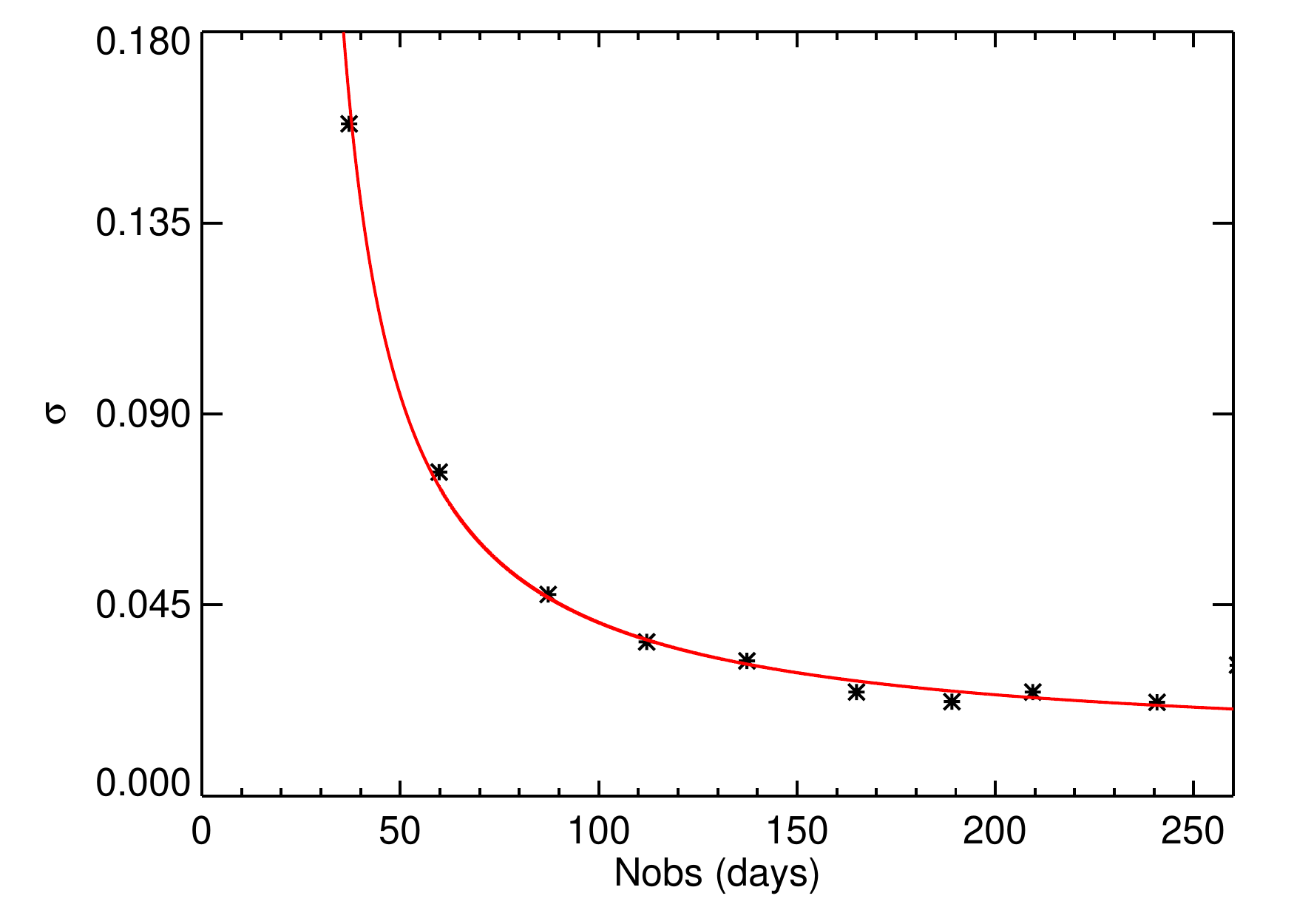}
  \caption{The maximum PA periodogram power value for a star in this survey is found to be well-described by Eqn A-5 (Fig A-1).  The scatter about the value predicted by Eqn A-5, $\sigma$, is also found to be dependent on N$_{obs}$ as shown in this figure.}
\end{figure}

Putting it all together, now an affirmative periodicity condition can be defined.  The statistical significance of a measured period for a star in this survey is simply a function of the number of observations.  Based on visual inspection of star light curves, this condition is defined by the following:

\begin{equation}
  \frac{\chi_{n_0,i}^2}{F(N_{obs_i},n_0,p)} - 1 > 6\sigma(N_{obs_i})
\end{equation}

\noindent Periods where this condition is met can be considered statistically significant for the star investigated.  Periods found using this criteria are generally also found to be statistically significant using the methods outlined in $\S$4.1.

\bibliography{parks_j}

\renewcommand{\thetable}{A-\arabic{table}}
\begin{deluxetable}{c c c c c c c c c c c}
  \rotate
  \tablecolumns{11}
  \tablewidth{0pc}
  \tablecaption{Monte Carlo Simulation: Testing Significance Function}
    \tabletypesize{\scriptsize}
    \tablehead{
      \colhead{n$_0$} & \colhead{p} & \colhead{Obs. Param. A} & \colhead{Pred. Param. A} & \colhead{$\%$ Error\tablenotemark{a}} & \colhead{Obs. Param. B} & \colhead{Pred. Param. B} & \colhead{$\%$ Error\tablenotemark{a}} & \colhead{Obs. Param. C} & \colhead{Pred. Param. C} & \colhead{$\%$ Error\tablenotemark{a}}
    }
    \startdata
    12&0.25&31.338&30.382&-3.054&6.796&6.574&-3.259&1.205&1.239&2.815\\
    13&0.04&80.893&76.952&-4.872&22.235&23.263&4.622&1.393&1.372&-1.500\\
    15&0.06&64.109&65.547&2.243&17.776&18.042&1.500&1.308&1.307&-0.042\\
    18&0.05&69.756&70.192&0.625&19.958&20.568&3.054&1.310&1.309&-0.054\\
    18&0.06&63.138&65.036&3.006&18.173&18.284&0.610&1.287&1.290&0.208\\
    25&0.06&62.550&64.284&2.772&18.496&18.560&0.343&1.253&1.257&0.337\\
    25&0.40&25.656&25.169&-1.898&5.833&5.565&-4.602&1.135&1.148&1.136\\
    25&0.10&49.530&48.938&-1.194&13.192&13.411&1.662&1.200&1.219&1.602\\
    25&0.04&79.775&75.265&-5.653&22.927&24.024&4.784&1.290&1.310&1.534\\
    40&0.06&62.172&63.863&2.719&18.604&18.693&0.479&1.208&1.210&0.194\\
    60&0.24&29.562&28.859&-2.378&8.219&7.796&-5.145&1.128&1.092&-3.168\\
    75&0.055&68.359&66.993&-1.998&18.500&19.682&6.390&1.157&1.174&1.423\\
    125&0.30&28.994&27.429&-5.397&6.329&6.666&5.334&1.080&1.055&-2.269\\
    140&0.45&24.910&25.700&3.171&5.127&5.144&0.335&1.067&1.036&-2.884\\
    175&0.14&43.597&40.392&-7.351&10.569&10.796&2.143&1.090&1.092&0.145\\
    180&0.05&71.813&70.318&-2.082&20.056&20.808&3.751&1.147&1.168&1.867\\
    200&0.06&63.461&65.170&2.694&18.439&18.522&0.451&1.145&1.149&0.385\\
    240&0.40&26.132&26.062&-0.268&5.800&5.525&-4.742&1.070&1.040&-2.794\\
    \enddata
    \tablenotetext{a}{Negative $\%$ errors indicates the significance function underestimates the actual value.}
  \end{deluxetable}
\end{document}